\documentclass[11pt]{article}
\usepackage[utf8]{inputenc}
\usepackage[T1]{fontenc}
\pdfoutput = 1

\usepackage{amsmath}
\usepackage{amssymb}
\usepackage{bbm}
\usepackage{bbold}
\usepackage{braket}
\usepackage{caption}
\usepackage{cite}
\usepackage{color}
    \definecolor{darkgreen}{rgb}{0,0.5,0}
    \definecolor{darkred}{rgb}{0.5,0,0}
    \definecolor{darkblue}{rgb}{0,0,0.6}
    \definecolor{purple}{rgb}{0.4,.2,0.7}
\usepackage[mathcal]{eucal}
\usepackage{float}
\usepackage{graphicx}
\usepackage[hyperfootnotes = true, colorlinks = true, linkcolor = darkblue, citecolor = purple]{hyperref}
\usepackage{mathabx}
\usepackage{mathtools}
\usepackage{pdfsync}
\usepackage{slashed}
\usepackage[normalem]{ulem}
\usepackage{upgreek}
\usepackage{url}

\usepackage{subfiles}

\usepackage[margin = 2.2cm]{geometry}
\usepackage[ragged]{footmisc}
\def\be{\begin{equation}}
\def\ee{\end{equation}}

\def\ben{\begin{equation}}
\def\een{\end{equation}}

\let\a=\alpha   \let\d=\delta 
   
\let\l=\lambda     \let\r=v
 \let\t=\tau

\def\be{\begin{equation}}
\def\ee{\end{equation}}
\def\ba{\begin{array}}
\def\ea{\end{array}}

\def\dalemb#1#2{{\vbox{\hrule height .#2pt
       \hbox{\vrule width.#2pt height#1pt \kern#1pt
               \vrule width.#2pt}
       \hrule height.#2pt}}}

\newcommand{\bea}{\begin{eqnarray}}
\newcommand{\eea}{\end{eqnarray}}

\def\d{{\rm d}}

\renewcommand{\d}{\mathrm{d}}
\renewcommand{\i}{\mathrm{i}}
\renewcommand{\S}{\textsf{S}_0}

\numberwithin{equation}{section}

\begin{document}

\thispagestyle{empty}
\begin{center}
    ~\vspace{5mm}

     {\LARGE \bf 
   Semi-classical dilaton gravity and the very blunt defect expansion
   }
    
   \vspace{0.4in}
    
    {\bf Jorrit Kruthoff$\,^1$ and Adam Levine$\,^2$}

    \vspace{0.4in}
    {
    $^1$ School of Natural Sciences, Institute for Advanced Study, Princeton, NJ 08540\\
    $^2$Center for Theoretical Physics, Massachusetts Institute of Technology, Cambridge, MA 02139, USA}
    \vspace{0.1in}
    
    {\tt kruthoff@ias.edu, arlevine@mit.edu}
\end{center}

\vspace{0.4in}

\begin{abstract}
    We explore dilaton gravity with general dilaton potentials in the semi-classical limit viewed both as a gas of blunt defects and also as a semi-classical theory in its own right. We compare the exact defect gas picture with that obtained by naively canonically quantizing the theory in geodesic gauge. We find a subtlety in the canonical approach due to a non-perturbative ambiguity in geodesic gauge. Unlike in JT gravity, this ambiguity arises already at the disk level. This leads to a distinct mechanism from that in JT gravity by which the semi-classical approximation breaks down at low temperatures. Along the way, we propose that new, previously un-studied saddles contribute to the density of states of dilaton gravity. This in particular leads to a re-interpretation of the disk-level density of states in JT gravity in terms of two saddles with fixed energy boundary conditions: the disk, which caps off on the outer horizon, and another, sub-leading complex saddle which caps off on the inner horizon. When the theory is studied using a defect expansion, we show how the smooth classical geometries of dilaton gravity arise from a dense gas of very blunt defects in the $G_N \to 0$ limit. The classical saddle points arise from a balance between the attractive force on the defects toward negative dilaton and a statistical pressure from the entropy of the configuration. We end with speculations on the nature of the space-like singularity present inside black holes described by certain dilaton potentials.

\end{abstract}

\pagebreak
\setcounter{page}{1}
\tableofcontents

\section{Introduction}\label{sec:intro}

Simplified toy models of quantum gravity in two spacetime dimensions have taught us much about the low-energy and long time structure of black holes \cite{KitSuh17, SaaSheSta18, CotGur16, SaaSheSta19, Yan18, Tur23, MerTur23, BlomKruYao22,SaaStaYanYao22, AlmHarMalSha19, PenSheStaYan19}.\footnote{See \cite{MerTur23} for a review of the progress.} Despite this progress, we still lack a complete microscopic understanding of what happens to the notion of spacetime in the strong gravity regime deep inside the black hole interior; in the context of AdS/CFT, a sharp signature of the near-singularity region of the black hole interior in the boundary theory remains obscure.

To move toward a better understanding of this deep interior regime from a microscopic point of view, we will in this work turn again to simplified two-dimensional models of quantum gravity. While Jackiw-Teitelboim (JT) gravity \cite{Tei83, Jac85} has taught us much about the near-extremal limit of higher dimensional black holes, in pure JT the curvature is constant everywhere in the spacetime. We would like a better understanding of asymptotically AdS black holes where the spacetime curvature in the interior becomes large and where there are no subtleties due to the presence of Cauchy horizons, as happens in classical solutions to JT gravity.

A simple way to modify the interior geometry of an asymptotically AdS$_2$ spacetime away from rigid AdS$_2$ is to consider dilaton gravity with a deformed dilaton potential. Consider the spacetime action with bulk term\footnote{The more physical way of thinking about $G_N$ is as being related to the renormalized value of the dilaton at the boundary, $\Phi_b = \Phi_r/\varepsilon$ \cite{HarJaf18}. The classical $G_N \to 0$ limit corresponds to large renormalized values. Viewed as a dimensional reduction, this means the transverse sphere has large renormalized values.}
\begin{align}\label{eqn:bulkaction}
    S_{bulk} = S_0 \chi(M) + \frac{1}{2G_N}\int_M d^2 x \sqrt{g} \left(\Phi R + U(\Phi)\right).
\end{align}
The first term controls topological suppression and is proportional to the Euler character of the spacetime. In this work, we will ignore higher topology effects, effectively setting $S_0 = \infty$. The dilaton equation of motion imposes $R = -U'(\Phi)$. In JT gravity, $U(\Phi) =2\Phi$ and the dilaton becomes a Lagrange multiplier, enforcing $R = -2$ exactly. More generally, we can deform $U(\Phi)$ at negative values of $\Phi$ while preserving the linear behavior at large positive $\Phi$. This ensures that we still have an asymptotically AdS boundary.

Certain classes of such theories were solved exactly at the disk topology level and shown to be dual to matrix models in \cite{MaxTur21, Wit20MM, TurUsaWen21, EbeTur23}. In \cite{MaxTur21, Wit20MM}, these theories were solved when the deformation to $U$ was a general linear combination of exponentials of the form
\begin{align}\label{eqn:potdef}
    U(\Phi)-2\Phi = \sum_{\alpha_i} \lambda_i e^{-2\pi \frac{(1-\alpha_i)}{G_N} \Phi}, \ \ \pi < 2\pi (1-\alpha_i) < 2\pi.
\end{align}
The authors in \cite{MaxTur21, Wit20MM} proceeded by expanding the gravitational path integral in the couplings $\lambda_i$ and viewing the exponential operators as inserting defects in the spacetime. For $\alpha$ in the range above, these defects have a conical deficit between $\pi$ and $2\pi$. We will refer to such defects as \emph{sharp}. In the work of \cite{TurUsaWen21,EbeTur23}, this restriction on $\alpha$ was extended to the range $0<2\pi (1-\alpha)<\pi$. Expanding in a gas of exponential deformations with $\alpha$ in this range creates defects in the spacetime that we will refer to as \emph{blunt}. Various new subtleties arise in this case due to contact term ambiguities in the definition of the exponential of the dilaton operator. Such contact terms are related to the ability of two blunt defects to merge without violating the upper bound in eq. \eqref{eqn:potdef}. When dilaton gravity is viewed as a limit of the (deformed) $(2,p)$ minimal string, these contact terms are related to operator mixing of the vertex operators, as analyzed in \cite{MooSeiSta91}.

In this work, we would like to focus on theories dominated by a smooth classical geometry, which arise in the classical $G_N \to 0$ limit. The sharp-eyed reader may note the presence of a $G_N$ in the exponential in eq. \eqref{eqn:potdef}, which would seemingly not produce good classical geometries. This $G_N$ is due to the overall normalization of the action we have picked in eq. \eqref{eqn:bulkaction}. A simple re-scaling of $\Phi \to \Phi G_N$ brings us back to the more standard normalization used in \cite{Wit20MM, MaxTur21, TurUsaWen21, EbeTur23}. This is related to the fact that sharp defects create large disturbances in the geometry and so do not reproduce smooth saddle point geometries. Instead, to find smooth geometries, one can work in the normalization of eq. \eqref{eqn:bulkaction} but focus on the ``very blunt'' limit, where $G_N \to 0$ with $(1-\alpha)/G_N$ fixed. In this limit, it is more convenient to work with a re-scaled $(1-\alpha) \to (1-\alpha) G_N$. In terms of this re-scaled and shifted $\alpha$, the action takes the form\footnote{The sum over $i$ in the action may be replaced by an integral over $\alpha$ as long as the upper limit on $1-\alpha$ is less than $1/G_N$ and the results of \cite{TurUsaWen21, EbeTur23} will still hold.}
\begin{align}\label{eqn:classaction}
    S_{bulk} = \frac{1}{2G_N} \int d^2 x \sqrt{g} \left( \Phi R + 2\Phi + \sum_i \lambda_i e^{-2\pi (1-\alpha_i) \Phi}\right).
\end{align}

We can now simply apply the saddle-point approximation to this action in the classical limit. In the interest of concreteness, we will often restrict our attention to a single exponential deformation of strength $\alpha$, but we will point out where our discussion is more general. In particular, for much of this paper we will work with the specific action
\begin{align}\label{eqn:specclassaction}
    S_{bulk} = \frac{1}{2G_N} \int d^2 x \sqrt{g} \left( \Phi R + 2\Phi + \lambda 2\pi (1-\alpha) e^{-2\pi (1-\alpha) \Phi}\right).
\end{align}
As we will detail in Sec. \ref{sec:dilgravrev}, the classical solutions to eq. \eqref{eqn:specclassaction} for $\lambda >0$ (but not too large as discussed in Sec. \ref{sec:dilgravrev}) look like black holes with either three horizons or one horizon. When there are three horizons, the Penrose diagram is an infinite grid of universes.\footnote{See \cite{CasMarTol22} for a realization of such diagrams in the context of higher dimensional dS. As far as the authors know, such diagrams have not been realized by higher dimensional AdS constructions. We will not let this fact bother us due to the existence of a well-defined dual matrix model for our potentials.} When there is one horizon, the diagram looks like that of AdS-Schwarzschild in higher dimensions. As we increase the energy, there is a critical energy, $E = E_c(\lambda, \alpha)$, at which the diagram jumps from having three horizons to one. Part of the motivation for this work was to find a signature of this transition in the matrix model.

Given the set-up, we now turn to studying the theory given by eq. \eqref{eqn:specclassaction} (and more generally \eqref{eqn:classaction}) in the $G_N \to 0$ limit. We proceed with three complementary approaches which we now briefly outline. 

\subsection{Analyzing the exact answer}
In the first part of the paper, we do a more detailed analysis of the exact results of \cite{TurUsaWen21} in the very blunt limit than what has previously appeared in the literature. As we will review in Sec. \ref{sec:exactdos}, the density of states for the theory with action eq. \eqref{eqn:classaction} has a nice form derived in \cite{TurUsaWen21} which is
\begin{align}\label{eqn:exactdosintro}
    &\rho^{\rm exact}(E) = \frac{e^{S_0}}{2\pi G_N}\int_{E^{\rm exact}_0}^E du \frac{1}{\sqrt{E-u}}\frac{d\mathcal{F}(u)}{du},\nonumber \\
    & \mathcal{F}(E) = \int_{\mathcal{C}} \frac{d\Phi_h}{2\pi i} \left( \Phi_h - \sqrt{W(\Phi_h) - E} \right) e^{\frac{2\pi \Phi_h}{G_N}},
\end{align}
where the contour $\mathcal{C}$ is an inverse Laplace transform contour and so lies to the right of all non-analyticities. Here the function $W(\Phi_h)$ is the pre-potential of the theory in question so that\footnote{Strictly speaking, the $W(\Phi)$ that appears in \eqref{eqn:exactdosintro} is only the pre-potential in the classical limit where $(1-\alpha) \ll 1$. More generally, for a potential $U(\Phi) = 2\Phi + \sum_i\lambda_i e^{-2\pi (1-\alpha_i) \Phi} $, then the function $W(\Phi)$ that appears in \eqref{eqn:exactdosintro} is $W(\Phi) = \Phi^2 - \sum_i \frac{\gamma(1-\alpha) \lambda_i}{2\pi} e^{-2\pi (1-\a) \Phi}$ with $\gamma(x) = \Gamma(x)/\Gamma(1-x)$. In the limit of $(1-\alpha)$ small, this becomes the classical pre-potential. Since we will mostly be working in this blunt limit here, we treat $W(\Phi)$ in \eqref{eqn:exactdosintro} as the pre-potential in what follows.}
\begin{align}
W'(\Phi) = U(\Phi).
\end{align}
The ground state energy $E^{\rm exact}_0$ is determined by finding the largest root of the equation 
\begin{align}
    \mathcal{F}(E^{\rm exact}_0) = 0.
\end{align}
The equation $\mathcal{F}(u(x)) = x$ is often referred to as the string equation in the context of matrix models. Knowledge of this function completely determines the disk-level density of states, and allows one to solve the matrix model at all orders in the genus expansion.

In these equations the variable $\Phi_h$ is integrated over and so does not have any a priori meaning. We will propose that it should be interpreted as the value of the dilaton at the horizon. To get a more concrete understanding of the density of states in the $G_N \to 0$ limit, we can perform the $\Phi_h$ integral above by saddle point. We find a saddle point for every root of the equation $W(\Phi_h) = E$. Assuming that $\Phi_h$ is the dilaton at the horizon, we will interpret these saddles geometrically by finding new solutions to the gravity path integral with fixed energy boundary conditions. Usually, the density of states is found by working at fixed boundary length/inverse temperature, $\beta$, and then inverse Laplace transforming to fixed energy. We will take a more direct approach where we work with fixed energy boundary conditions. This amounts to fixing the condition on the dilaton at the boundary
\begin{align}\label{eqn:fixedEbc}
    \Phi_b K - \partial_n \Phi_b = E/2.
\end{align}
Such boundary conditions were discussed in \cite{GoeIliKruYan20} in the context of JT gravity. 

\begin{figure}
    \centering
    \includegraphics[scale=.6]{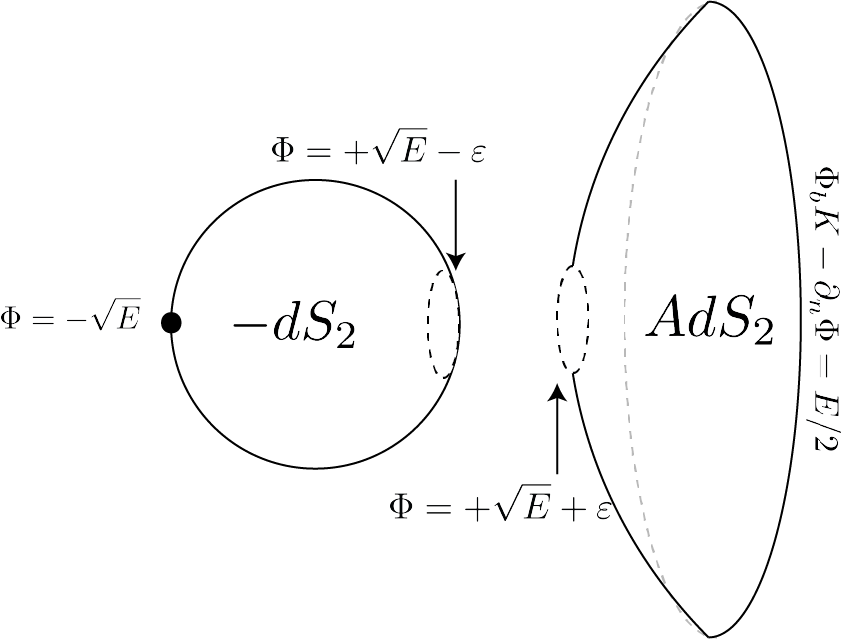}
    \caption{We illustrate what the other saddle point is that contributes to the density of states in JT gravity. This amounts to a complex geometry that is the AdS$_2$ disk. As we will describe in Sec. \ref{sec:exactdos}, the radial coordinate, which can be taken to be the dilaton, follows a complex contour that avoids the pole in the metric at the outer horizon and ends on the negative root of the equation $\Phi^2 = E$. The metric in the range $-\sqrt{E}< \Phi< \sqrt{E}$ looks like that of an anti-sphere, whose metric is Euclidean up to an overall minus sign. This solution has constant curvature $R = -2$.}
    \label{fig:disksphere}
\end{figure}
Using the boundary condition in eq. \eqref{eqn:fixedEbc} allows us to reinterpret the disk level density of states in JT gravity in terms of a contribution from two saddles, one which gives an $e^{2\pi \sqrt{E}/G_N}$ contribution - the standard disk saddle - and another which gives $-e^{-2\pi \sqrt{E}/G_N}$.\footnote{As of now, we do not have a good bulk interpretation of the minus sign for this saddle, except by doing the inverse Laplace transform exactly. Arriving at the minus sign via saddle point methods requires a more careful steepest descent analysis which we discuss in Sec. \ref{sec:exactdos}.} This second saddle is complex and can be described by the disk geometry but with a complex contour for the radial coordinate in Schwarzschild gauge. This contour goes into the complex plane and avoids the pole in the metric at the origin of the disk, ending instead on the pole of the metric associated with the inner horizon. Morally, this solution looks like a disk with an anti-sphere connected onto the horizon. We illustrate this saddle point in Fig. \ref{fig:disksphere}. This saddle can be associated with the presence of an inner horizon in the classical Lorentzian spacetime.\footnote{As we will point out, this saddle violates the Kontsevich-Segal (KS) criterion of \cite{Wit21, KonSeg21}, although several necessary violations of the KS criterion have now been discovered \cite{CheIvoMal23}.}

\subsection{Canonical quantization and classical saddles}

Next, we compare these results to those found by studying the theory defined by eq. \eqref{eqn:classaction} semi-classically. In particular, we will avoid resorting to an expansion in a gas of very blunt defects. We show how to re-derive some of the results in \cite{TurUsaWen21}. In particular, we will show how to canonically quantize general dilaton gravities in geodesic gauge, analogous to what was done in JT \cite{HarJaf18,Yan18}. Just as in JT gravity, we find that the canonically quantized theory is equivalent to a non-relativistic particle propagating in a potential. The position of this particle corresponds to the renormalized length of the Einstein-Rosen bridge connecting the two sides of the black hole.

Comparing with the exact density of states in eq. \eqref{eqn:exactdosintro}, we will find that canonical quantization leads to a (naive) discrepancy with the exact result in eq. \eqref{eqn:exactdosintro}. This discrepancy is most sharply seen by comparing the predicted ground state energies for the matrix model from canonical quantization and from eq. \eqref{eqn:exactdosintro}. We will see that the two answers differ by an amount which is non-perturbatively small in $1/G_N$.

\begin{figure}
    \centering
    \includegraphics[scale=.5]{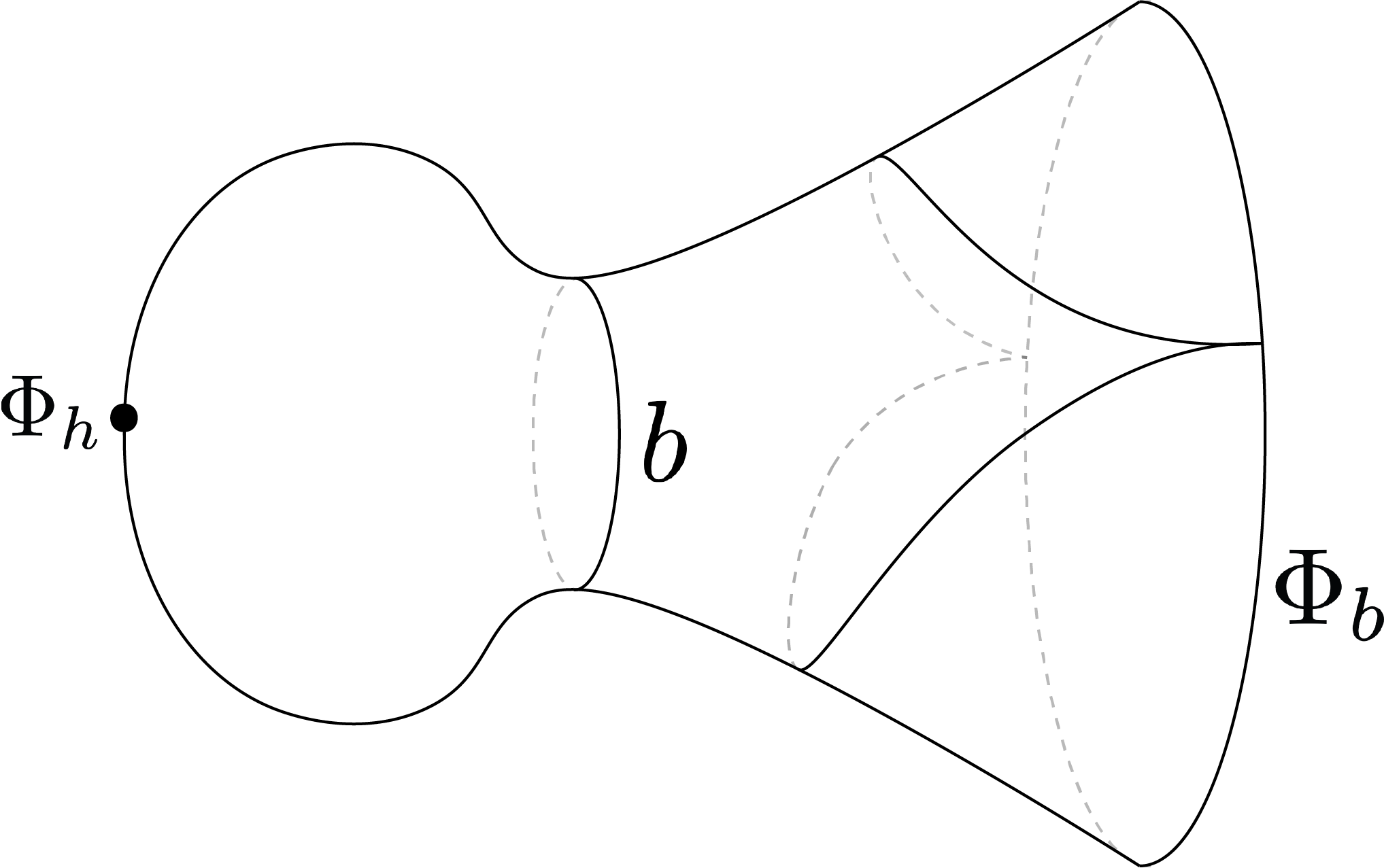}
    \caption{We give an example of what the pacifier spacetimes look like qualitatively, illustrating that there can be multiple geodesics connecting the same pair of boundary points. There exists a closed geodesic of length $b$ and a horizon at $\Phi = \Phi_h$.}
    \label{fig:pacifier}
\end{figure}
We will resolve this discrepancy by pointing out that geodesic gauge is generically ill-defined due to non-perturbative effects in $1/G_N$. These effects are due to the ability of the spacetime to ``nucleate'' a closed geodesic, which leads to multiple geodesics connecting the same pair of boundary points. One should contrast this with JT gravity where geodesic gauge breaks down only non-perturbatively in $S_0$ \cite{IliLevLinMaxMez24}. In JT gravity with a single asymptotic boundary, a closed geodesic can appear only on higher genus geometries, since there are no constant negative curvature Euclidean geometries with a closed geodesic and the topology of the disk.  The difference in our work is that for theories with more general dilaton potentials, the geometry can contain a closed geodesic \emph{at the disk level} since we no longer have the constant negative curvature constraint. We will call such spacetimes which have disk topology together with a closed geodesic \emph{pacifier spacetimes}, illustrated in Fig. \ref{fig:pacifier}.\footnote{Pacifier spacetimes share a qualitative resemblance to the centaur geometries of \cite{AnnHof17, AnnGalHof18}, which connect an approximately constant $R = +2$ sphere to an $R=-2$ exterior along a closed geodesic. We distinguish the pacifier from the centaur because in general the pacifier may not have everywhere positive curvature in the spacetime region bounded by the closed geodesic. } As we will argue, the contribution of pacifier spacetimes to the density of states is suppressed non-perturbatively in $1/G_N$ (but not $S_0$) relative to the leading geometry but becomes enhanced at very low energies. We will argue that this enhancement corresponds to a non-perturbatively small shift in the ground state energy of the matrix model. We find that the leading non-perturbative correction to $E_0$ due to the inclusion of pacifier spacetimes agrees with that predicted by eq. \eqref{eqn:exactdosintro}, up to one-loop factors since we have not done a thorough one-loop analysis of the pacifier geometries. 

The story we describe here is reminiscent of the picture found in \cite{MaxTur21} where instantonic corrections to the 3d gravity path integral for near extremal black holes lead to a non-perturbatively small shift in the ground state energy. The instantons considered in \cite{MaxTur21} create defects in the spacetime, as viewed from the dimensional reduction to 2d gravity. In Sec. \ref{sec:bluntexpansion}, we will argue that the pacifiers arise as a dense gas of these defects in the very blunt limit. We describe this very blunt defect limit shortly, but we hope that the methods of this portion of the paper may shed light on how to treat dilaton gravities more generally when a simple defect gas picture is not present. 

We also comment that already at the level of Lorentzian geometries, the theory specified by the action in eq. \eqref{eqn:specclassaction} with $\lambda >0$ displays a breakdown of geodesic gauge. We point out in Sec. \ref{sec:canquant} that the same theory defined with $\lambda <0$ does not display such a breakdown in Lorentzian signature and yet the ground state energy of the model as predicted by the exact answer in eq. \eqref{eqn:exactdosintro} disagrees with canonical quantization. This discrepancy can instead be interpreted as a modification of the inner product used in the theory.

\subsection{Dense gas of very blunt defects}
\begin{figure}
    \centering
    \includegraphics[scale=.6]{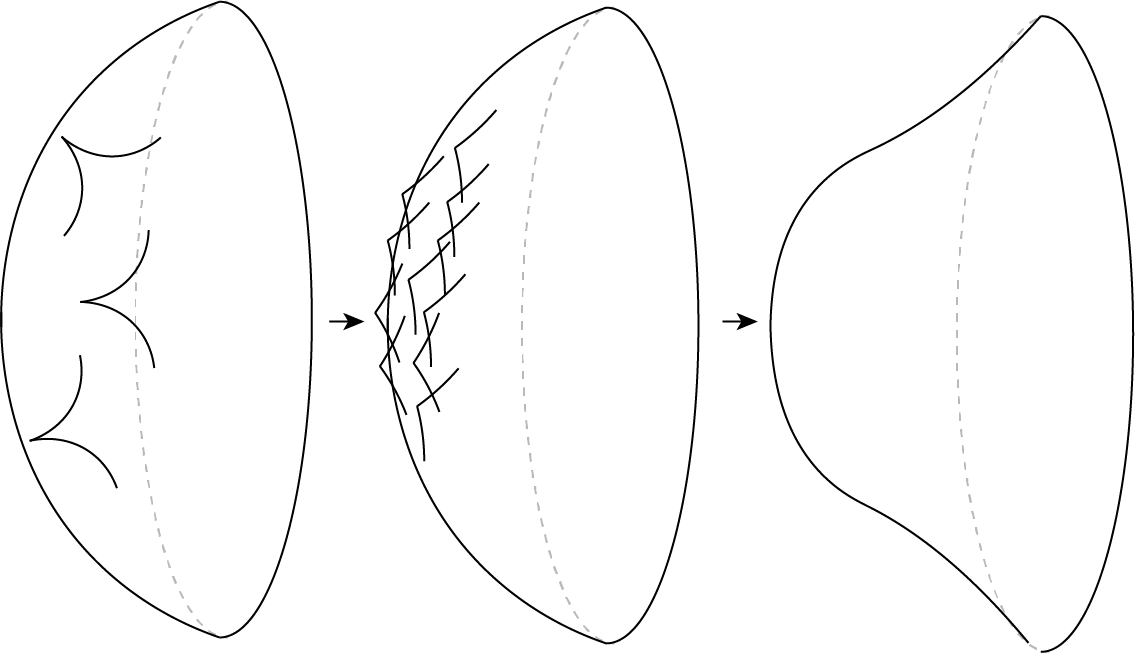}
    \caption{This figure illustrates how a gas of very blunt defects can approximate a smooth spacetime. On the left, we start with a few sharp defects, with the arrows indictating that in the classical limit the defect gas becomes simultaneously denser and blunter.}
    \label{fig:bluntlimit}
\end{figure}

Finally, we return to the defect gas picture and re-analyze the problem in the very blunt, classical limit. We find that in this limit the defect gas expansion simplifies and is dominated by a very large number (scaling like $1/G_N$) of very blunt defects. We show how such a large number of defects builds up a smooth geometry by introducing a new, continuous degree of freedom on the spacetime: the defect density. Given a spacetime with a large number, $L$, of defects of strength $\alpha$, then the spacetime curvature can be written as 
\begin{align}
    R = -2 + 2\pi (1-\alpha) LG_N \rho(x), \ \ \rho(x) \equiv \frac{1}{L} \sum_{i=1}^L \frac{\delta^2 (x-x_i)}{\sqrt{g}},
\end{align}
where the $x_i$ are the positions of each defect. Scaling $L$ with $1/G_N$, we arrive at an action for a continuous density of defects, $\rho(x)$. As we will show, this action has saddles for the density which approximate the semi-classical geometry. These saddles arise due to a balance between the attraction of defects toward more negative values of the dilaton and a statistical pressure from the entropy of the defect configuration. This is how the very blunt limit reproduces the classical geometry. We illustrate this effect in Fig. \ref{fig:bluntlimit}.

We attempt to compute the Weil-Peterson volumes of geometries with a single geodesic boundary of length $b$ and many very blunt defects. We also show that pacifier spacetimes arise due to geometries with a number of defects so large that their ``net'' opening angle is negative. These geometries have a number of defects $L$ with $2\pi (1-\alpha)G_N L > 2\pi$ in terms of the re-scaled $1-\alpha$. This matches the results of \cite{MaxTur21,EbeTur23,Wit20MM}, where it was shown that geometries with $LG_N > \frac{1}{1-\alpha}$ necessarily have a closed geodesic separating the defects from the asymptotic boundary. Our pacifier spacetimes are just the smooth, very blunt approximation to such geometries.

We now briefly summarize the layout of the paper. In Sec. \ref{sec:dilgravrev}, we review the classical solutions to dilaton gravity, going to a gauge where the dilaton is used as a radial coordinate on the geometry. We also review the exact matrix model results. In Sec. \ref{sec:exactdos}, we analyze the exact density of states given by \cite{TurUsaWen21,EbeTur23} in the very blunt limit. We interpret the result in terms of new complex saddles. In Sec. \ref{sec:canquant}, we perform the naive canonical quantization of dilaton gravity in geodesic gauge and find a discrepancy with the exact answer. In Sec. \ref{sec:lengthambiguity}, we resolve this ambiguity by identifying a non-perturbative breakdown in geodesic gauge due to the appearance of pacifier spacetimes. We further show that the inclusion of pacifier spacetimes corresponds to a non-perturbatively small shift in the ground state energy of the model away from that predicted by canonical quantization. In Sec. \ref{sec:bluntexpansion}, we reformulate the very blunt defect expansion in terms of a density of defects. We solve the defect density theory classically and find agreement with the classical limit of dilaton gravity. Finally, in Sec. \ref{sec:discspec}, we end with some speculations about what all of this might imply about the nature of the singularity inside the black hole interior.

\section{Review of Dilaton Gravity and its Matrix Model Dual}\label{sec:dilgravrev}

We now review the basics of classical dilaton gravity. The on-shell Euclidean solutions to dilaton gravity with rotational symmetry are most easily written down in the gauge in which $\Phi$ is the radial coordinate \cite{Wit20PT}. For a general dilaton potential $U(\Phi)$ as in eq. \eqref{eqn:bulkaction}, we introduce the dilaton (pre)-potential $W$ defined by
\begin{equation}
    U(\Phi) = W'(\Phi).
\end{equation}
In terms of $W$, one obtains the solution in Euclidean signature
\be \label{eqn:onshellmetric}
\d s^2 = G(\Phi) \d \t^2 + \frac{\d \Phi^2}{G(\Phi)}
\ee
with 
\be \label{eqn:metricprepot}
G(\Phi) = W(\Phi) - W(\Phi_h).
\ee
Here $\Phi_h$ is the value of the dilaton at the horizon, where $G$ vanishes. For a potential $U$ that becomes linear at large, positive $\Phi$, this geometry approaches AdS$_2$ and does not differ much from the usual JT disk solution. If we want to fix the temperature of the solution, we need to solve the equation
\be \label{eqn:diltemp}
\beta = \frac{4\pi}{U(\Phi_h)}.
\ee
As discussed in the introduction, we will focus on the potential as given in eq. \eqref{eqn:specclassaction} with pre-potential
\begin{align}\label{eqn:specprepot}
    W(\Phi) = W_{\lambda}(\Phi) \equiv \Phi^2 - \lambda e^{-2\pi (1-\alpha) \Phi}.
\end{align}
The exponential behaviour of the potential at large negative $\Phi$ leads to an infinite number of (complex) solutions to eq. \eqref{eqn:diltemp} for a given $\beta$. There is also a real solution that is a smooth deformation of the usual JT disk. This solution corresponds to when $\Phi_h$ has the largest real part of all possible solutions to eq. \eqref{eqn:diltemp}. 

The ADM energy of these solutions is given in terms of $\Phi_h$ by 
\be \label{eqn:E=W}
E_{\rm AdM} = \lim_{\Phi \to \infty} \left(\Phi^2 - G(\Phi)\right) = W(\Phi_h).
\ee
When $\Phi_h$ is real, then so is $E_{\rm AdM}$.  When $\lambda$ is positive and not too large, there is a solution with zero temperature or $\beta \to \infty$. According to eq. \eqref{eqn:diltemp}, this solution occurs when $U(\Phi_h) = 0$. Throughout this work we will refer to this solution as occurring at the ``the outermost mininum of the potential,'' denoted by $\Phi_0$. Furthermore, we will denote the AdM energy of this solution by $E_0 = W(\Phi_0)$. In Sec. \ref{sec:canquant}, we will rederive this ground state energy from canonical quantization.

\begin{figure}
    \centering
    \includegraphics[scale=.4]{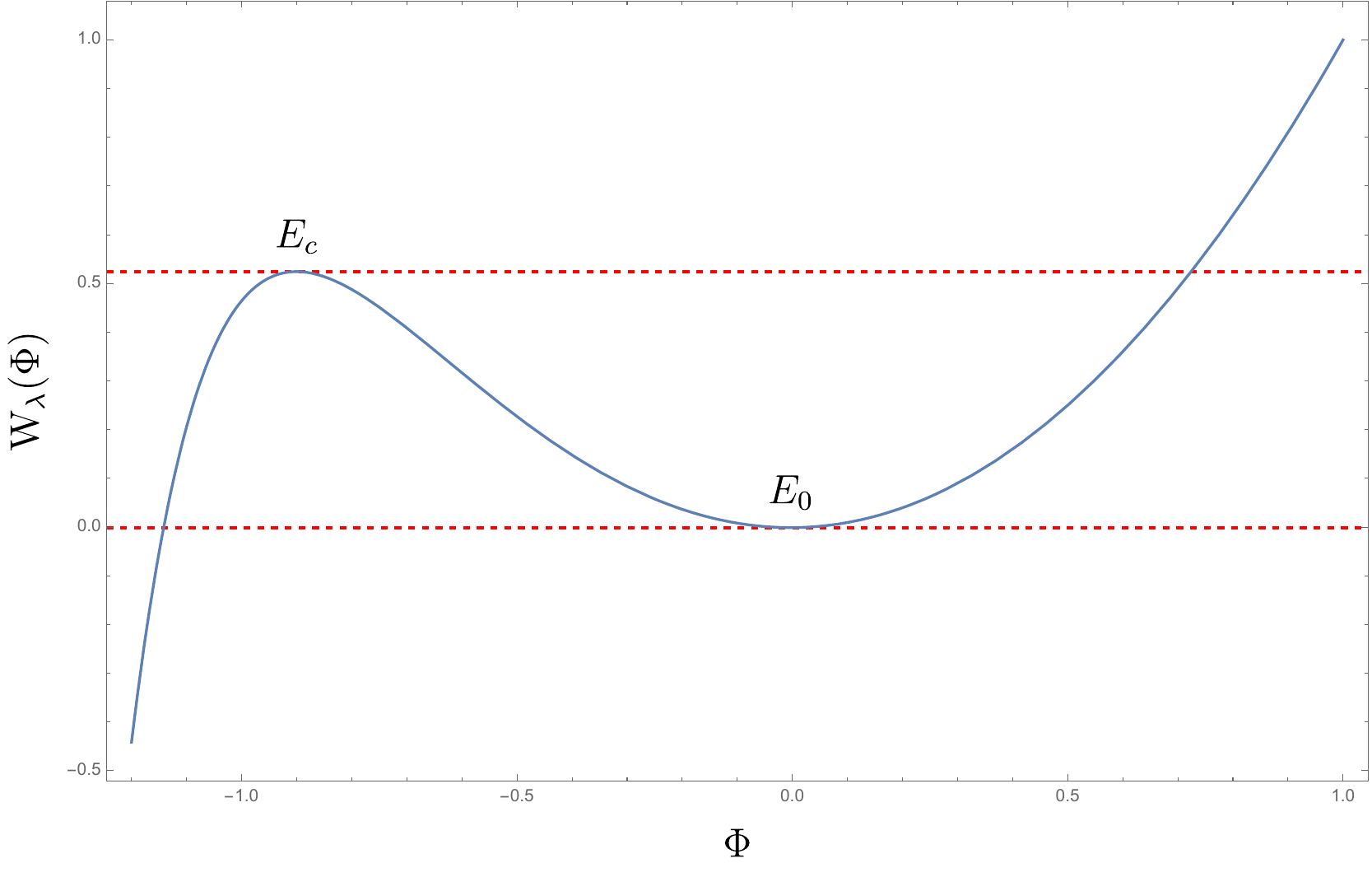}
    \caption{This figure illustrates the pre-potential $W_{\lambda}(\Phi)$ for $\alpha = 0$ and $\lambda = .001$. We label the energy of the zero temperature classical solution, which is set by the value of $W_{\lambda}(\Phi)$ at its outermost minimum, $E_0$. We also label the critical energy $E_c$. Solutions with $E>E_c$ have one horizon while solutions with $E_c>E>E_0$ have three.}
    \label{fig:groundstateenergy}
\end{figure}

Note that the equation $U_{\lambda}(\Phi_h) =0$ does not always have a real solution for all $\lambda$ and $\alpha$. As we turn up $\lambda$, there is a critical $\l = \l_c$ above which there no longer exists a real, local minimum for $W_{\lambda}$. A brief computation shows that $\l_c= (2\pi^2 e (1-\a)^2)^{-1}$ and $\Phi^c_h = -(2\pi (1-\a))^{-1}$. The ground state energy of the model with this critical $\l$ is $E_0 = -(4\pi^2(1-\a)^2)^{-1} = -\Phi_h^2$. As we will see in the next section, in the classical limit this critical $\lambda$ corresponds to the coupling (up to exponentially small corrections) where the ground state energy of the dual matrix model becomes complex.

In summary, for $\l < \l_c$ the smallest energy that is continuously connected to solutions of infinite energy is obtained by solving for $\Phi_h$ in the equation $\partial_{\Phi_h}E_{\rm ADM} = 0$ and gives a non-trivial function of $\l$. As we will see later on, this is actually a crude estimate of the critical $\l$ and $E_0(\l)$ and the exact version of the ground state energy $E_0$ as a function of $\l$ is obtained in a more intricate manner, following the matrix model techniques developed in \cite{Wit20MM, MaxTur21, TurUsaWen21}. Essentially we are asking here too much of the semiclassical theory in order to give the ground state energy. As discussed in the introduction, we will propose that other (complex) solutions play an important role in determining the exact ground state energy of the model.

\subsection{Lorentzian signature}

The more exciting feature of these solutions is their Lorentzian continuation. Asymptotically the geometry is still AdS$_2$, but once we go inside the black hole, the correction to the JT potential dominates. When $\l < \l_c$, $W_{\lambda}(\Phi)$ is non-monotonic and one can have three types of solutions. As we vary the AdM energy of the solution, we vary between these three possible types of solutions, which have differing numbers of horizons. There are two critical energies $E_c(\lambda)$ and $E_0(\lambda)$ at which the number of horizons changes.
\begin{figure}
    \centering
    \includegraphics[scale=.65]{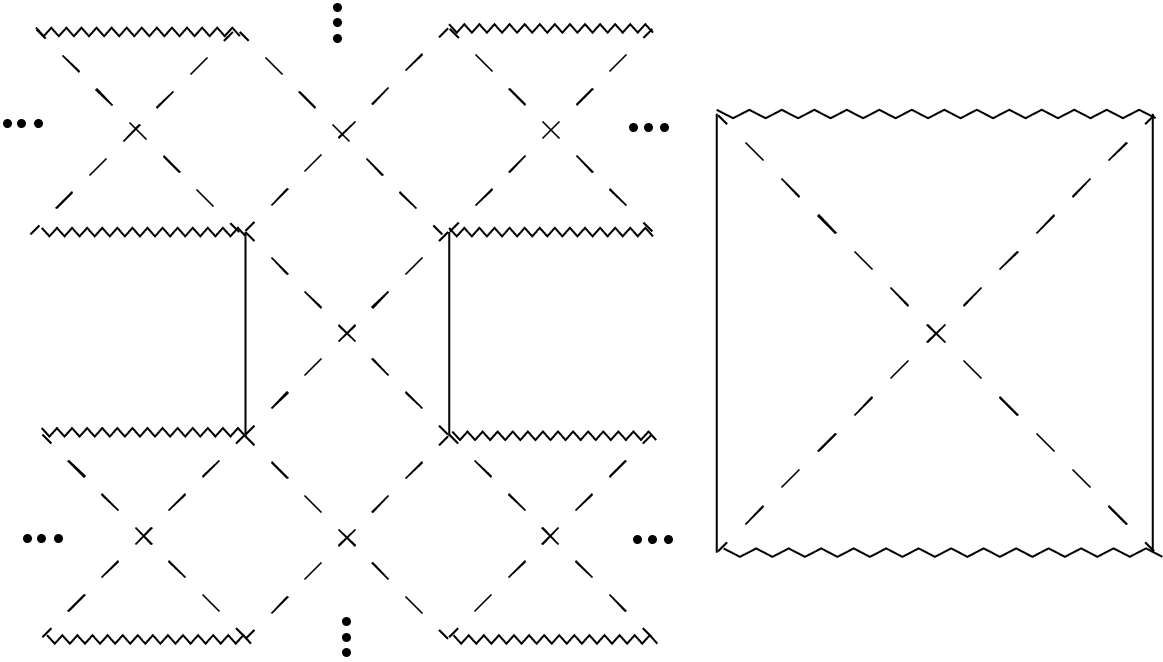}
    \caption{This figure illustrates the Penrose diagram for solutions with prepotential $W_{\lambda}(\Phi)$ with $\lambda_c>\lambda >0$. The diagram on the left shows the solution at low energies $E$ where $W_{\lambda}(\Phi) = E$ has three real roots. The diagram at the right where there is just one real real root. The jagged lines denote regions where both the curvature blows up and $\Phi \to - \infty$. The solid lines denote asymptotically AdS boundaries and the dashes lines are horizons. Note that we have drawn the singularities as horizontal but they will have an energy dependent shape.}
    \label{fig:penrose}
\end{figure}
\begin{enumerate}
    \item One horizon: The energy $E$ is chosen to obey $E> E_c$ or $E< E_0$ such that the solution to $W_{\lambda}(\Phi) = E$ has a single real zero in the complex dilaton plane. The Lorentzian solution has a spacelike singularity and the Penrose diagram looks qualitatively similar to that of AdS Schwarzschild in higher dimensions. We illustrate the Penrose diagram for this case on the right of Fig. \ref{fig:penrose}.
    \item Two horizons: The energy is such that there is one single zero and one double zero. The double zero is an example of a degenerate zero, much like an extremal horizon. This occurs when $E = E_c(\lambda)$ or when $E = E_0$, with the double zero occuring at the outer-horizon in the latter case and at the inner-horizon in the former case. 
    \item Three horizons: The solution has energy $E_0< E < E_c$ such that there are three real solutions to $W_{\lambda}(\Phi) = E$. In this case the solution is exotic, as its Penrose diagram extends to an infinite grid in $\mathbf{R}^2$, i.e. there are an infinite number of universes. This situation can only happen when $W_{\lambda}(\Phi)$ is non-monotonic. We illustrate the Penrose diagram for this case on the left of Fig. \ref{fig:penrose}. We will refer to the three real solutions to $W_{\lambda}(\Phi) = E$ as the inner, middle and outer solution, where the inner solution has the most negative dilaton value and the outer has the most positive.
\end{enumerate}
Clearly there exists a transition where one goes between three and one horizon with a degenerate intermediate two-horizon solution. Again, this transition can only occur when $W_{\lambda}(\Phi)$ is non-monotonic and so $\l < \l_c$. 

One of the most interesting features of these solutions is that they have spacelike singularities. In the three horizon case for $E_0<E<E_c$, the spacelike singularities are hidden by a Cauchy horizon and so will evade any simple probes from the boundary. The single horizon solutions, however, are much like the black holes we know in higher dimensions. In particular, the Ricci scalar $R \to +\infty$ when $\Phi \to -\infty$. Note that the curvature becoming large and positive is exactly what happens to the sectional curvature $R_{rtrt}$ in higher-dimensional black holes. A diffeomorphism invariant way of characterizing the singularity is to fire in light rays from both boundaries and ask at which boundary time $t_*$ do the two light rays meet at the singularity \cite{FidLuk04, FesLiu08}. This is given by the integral
\be 
t_{\star} = \int \d t = {\rm PV} \int_{-\infty}^{\infty} \frac{d\Phi}{G(\Phi)},
\ee
which is finite due to to the exponential decay of the integrand close to the singularity. Furthermore, calculating this integral for a fixed temperature solution, reveals that $t_{\star}$ can be positive and negative. For $t_{\star} < 0$, the singularity is bent downwards, while for $t_{\star} > 0$ it bends up \cite{FidLuk04}. This also implies that one can have so-called ``bouncing'' space-like geodesics in these geometries, just as in the higher dimensional case studied in \cite{FidLuk04}. Indeed, these bouncing geodesics will be important for understanding subtleties in the canonical approach discussed below. See Sec. \ref{sec:cansubtlety} for more discussion.

Interestingly, as we will show in the next section, the exact matrix model dual found via an expansion in a gas of defects seems to leave out solutions with $E < E_0$, at least to leading order in $1/G_N$. In other words, classically the ground state energy of the matrix model is $E = E_0$, up to non-perturbatively small corrections.

\subsection{Matrix model dual}

As is well-known, ordinary JT gravity has a matrix model dual \cite{SaaSheSta19}. In \cite{MaxTur21, Wit20MM, TurUsaWen21,EbeTur23} this was extended to non-trivial dilaton gravities consisting of the exponential form discussed above (with certain restrictions that we will come to in a bit) by doing an expansion in $\l$ and resumming that expansion. The thing one gains by doing an expansion in $\l$ is that each insertion of an exponential in the dilaton changes the dilaton equations of motion to 
\begin{align}
    \sqrt{g}(R + 2) = 4\pi(1-\a)G_N\delta^{(2)}(x - x_0).
\end{align}
The solution represents (to leading order in the genus) a disk with a deficit angle of amount $2\pi(1-\a)G_N$ at $x = x_0$. In \cite{MaxTur21,Wit20MM}, only sharp defects were considered with $\frac{1}{G_N} > (1-\a)>\frac{1}{2G_N}$. In \cite{EbeTur23, TurUsaWen21}, the answer was extended to all $\a$.

At order $\l^k$ in the $\l$ expansion, one has $k$ such defects and one needs to integrate over their positions, dividing out by diffeomorphism redundancy. This amounts to integrating over the moduli space of hyperbolic surfaces with $k$ conical defects. For $k< \frac{1}{G_N(1-\a)} = k_c$, this integral was done in \cite{EbeTur23}. As mentioned in the introduction, the proper treatment of this moduli space for $\a$ in the blunt range requires dealing with contact term ambiguities in the definition of the exponential operator. Geometrically, these ambiguities arise due to the possible merging of two defects into a single defect of twice deficit angle. 
\begin{figure}
    \centering
    \includegraphics[scale=.65]{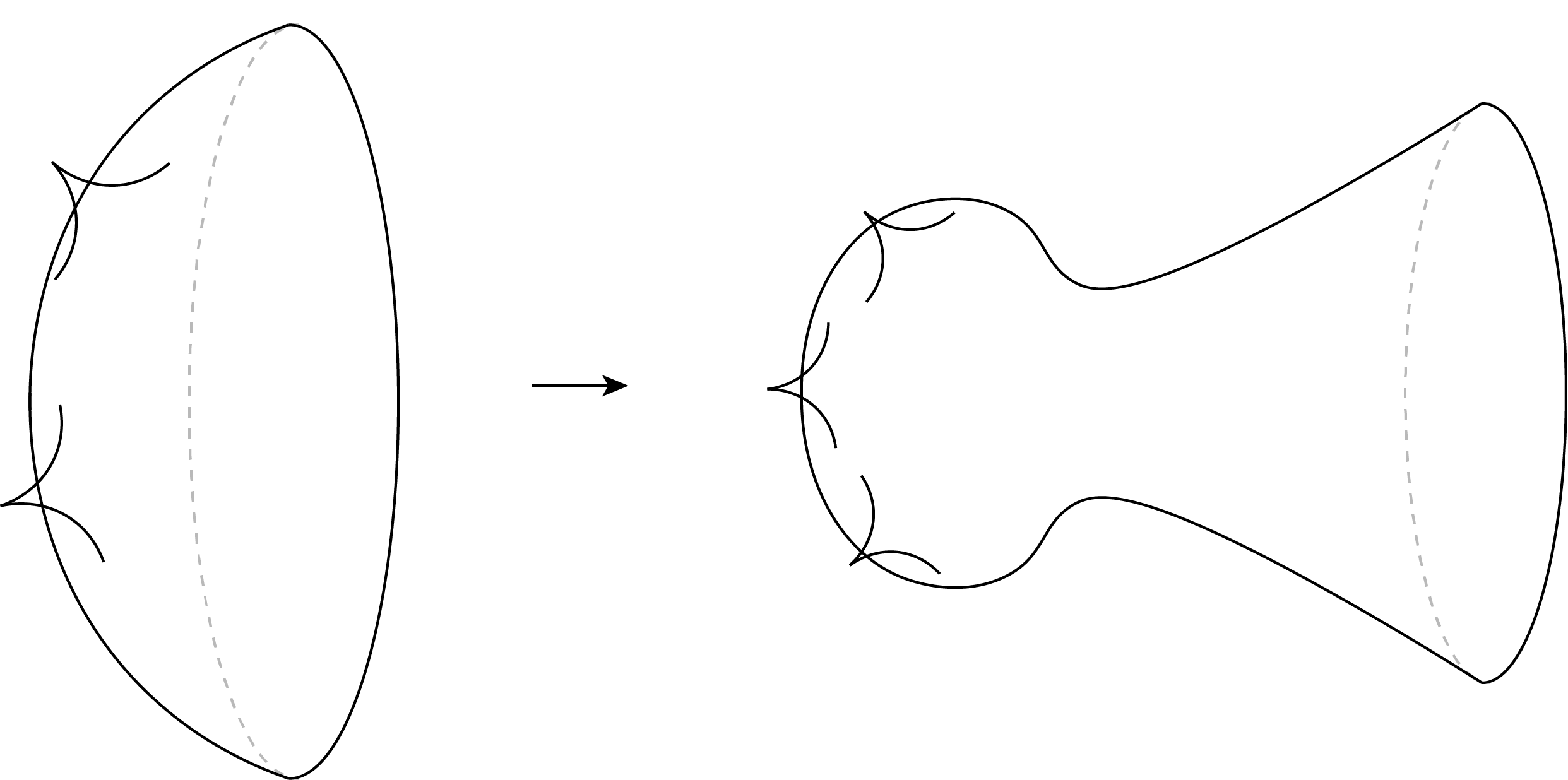}
    \caption{This figure illustrates the transition in the bulk geometry when the defect number, $L$, is above the bound $L > L_c = \lfloor \frac{1}{G_N (1-\alpha)} \rfloor$. We illustrate this phenomenon for the choice $\frac{1}{2G_N} > 1-\alpha > \frac{1}{3G_N}$ so that the transition number is $L_c = 2$. This transition number corresponds to when the net deficit angle of all the defects is more than $2\pi$. In the minimal string picture, where the defects correspond to insertions of the tachyon operator on the worldsheet, this phenomenon corresponds to the fact that there are a finite number of such primaries in the minimal model \cite{TurUsaWen21}.}
    \label{fig:nucleatinggeodesic}
\end{figure}

For $k>k_c$, the net deficit angle of all the defects together is greater than $2\pi$. In this case, one can always separate the region with defects from the asymptotic boundary using a closed geodesic \cite{Wit20MM,EbeTur23}.  We illustrate this in Fig. \ref{fig:nucleatinggeodesic} for the case $\frac{1}{2} >(1-\alpha)G_N > \frac{1}{3}$ so that a closed geodesic appears when the defect number $L \geq 3$. For defect number above this bound, one needs to compute the volumes of moduli space with a closed geodesic boundary and $k$ defects with deficit $2\pi (1-\a)G_N$. For $\a$ in the sharp range, this can be accomplished in a straightforward way from moduli space volumes for surfaces with $k$ geodesic boundaries via analytic continuation to complex geodesic boundary lengths \cite{Wit20MM,MaxTur21}. More generally one can use the recursion relations outlined in \cite{EbeTur23}.

The upshot is that at genus zero, these methods result in an exact expression for the density of states,
\be \label{exact_DOS}
\rho_{g = 0}(E) = \frac{e^{\S}G_N}{2\pi} \int_{E^{\rm exact}_0}^E \frac{du}{\sqrt{E - u}} \partial_u \mathcal{F}(u)
\ee
with the ground state energy $E^{\rm exact}_0$ given by the largest real zero of $\mathcal{F}$,
\begin{align}\label{eqn:E0exact}
    \mathcal{F}(E_0^{\rm exact}) = 0.
\end{align}
As mentioned in the introduction, in the orthogonal polynomial approach to the matrix model, this function $\mathcal{F}$ defines the so-called \emph{string equation}, $\mathcal{F}(u(x)) = x$, which is used to systematically compute higher-orders in the matrix integral. See \cite{Joh19} for a nice review of this technology. The function $\mathcal{F}(u)$ depends on the dilaton potential $W$; for the potential given by eq. \eqref{eqn:specprepot}, the function $\mathcal{F}$ is given by the expression \cite{EbeTur23,TurUsaWen21}
\be \label{minimal_string_F}
\mathcal{F}(u) = \sum_{L=0}^{\lfloor \frac{1}{(1-\a)G_N} \rfloor} \frac{\left(\l \pi (1-\a)\right)^{L}}{G_NL!} \left(\frac{2\pi(1 - L(1-\a)G_N)}{\sqrt{u}}\right)^{L-1} I_{L-1}\left(2\pi(1 - L(1-\a)G_N)\frac{\sqrt{u}}{G_N}\right),
\ee
where $I_L$ is the modified Bessel function of the first kind. To get this formula from \cite{TurUsaWen21}, we have re-scaled all of our energy variables $E \to E/G_N^2$ and $u \to u/G_N^2$ as well as the $(1-\a)$ variable. We are also working with a different normalization for our coupling $\lambda$. 

Note that the limit on $L$ is set by the floor of $\frac{1}{(1-\a)G_N}$. From the minimal string perspective, where this formula was first derived, the upper limit on the sum arises due to the fact that there are a finite number of primaries in the minimal model CFT. From the geometric perspective, however, the cutoff on the sum comes from the condition discussed above that the net deficit angle of all of the defects needs to be less than $2\pi$ or else the geometry nucleates a closed geodesic. It is interesting that the presence of a closed geodesic in the geometry signals an overcompleteness from the minimal string point of view; there are a finite number of tachyon vertex operator primaries, and as we add more insertions eventually we stop generating new states. In this work, we also find that the presence of a closed geodesic signals linear dependencies between naively orthogonal states.

Note, however, that the truncation of $\mathcal{F}$ at order $\l^{\lfloor \frac{1}{(1-\a)G_N} \rfloor}$ does not mean that all physical observables, such as the density of states or partition function truncates at a finite order in $\lambda$ as well. This is so because in these observables $E^{\rm exact}_0$ also makes an appearance, and $E_0^{\rm exact}$ does have an infinite expansion in $\l$, due to solving eq. \eqref{eqn:E0exact}. In Sec. \ref{sec:lengthambiguity}, we will argue that the corrections coming from $E^{\rm exact}_0$ are actually very interesting geometrically. Furthermore, note also that in the semiclassical limit, where $G_N \to 0$, the sum truncates at a very large order and in fact the truncation errors are non-perturbative in $1/G_N$. In equations we have
\begin{align}
    \mathcal{F}^{\rm semi-classical}(u) \approx \mathcal{F}^{\rm exact}(u) + \mathcal{O}\left(\lambda^{\frac{1}{G_N(1-\a)}}\right),
\end{align}
where 
\begin{align}\label{eqn:Fsemiclass}
\mathcal{F}^{\rm semi-classical}(u) \equiv \sum_{L=0}^{\infty} \frac{\left(\l \pi (1-\a)\right)^{L}}{G_NL!} \left(\frac{2\pi(1 - L(1-\a)G_N)}{\sqrt{u}}\right)^{L-1} I_{L-1}\left(2\pi(1 - L(1-\a)G_N)\frac{\sqrt{u}}{G_N}\right).
\end{align}
Interestingly, the modified string equation defined by eq. \eqref{eqn:Fsemiclass} can arise from using a different dilaton contour for the theory, as discussed in Sec. \ref{sec:altcont}.

\subsection{Re-summing the string equation}
The authors in \cite{TurUsaWen21} showed that one can in fact nicely re-sum the exact expression for $\mathcal{F}$ given in \eqref{minimal_string_F}. They noted that the modified Bessel functions in \eqref{minimal_string_F} have a simple form in terms of an inverse Laplace transform over an auxiliary variable $y$. Re-writing \eqref{minimal_string_F} in this manner allows one to simply re-sum the expansion in $\lambda$ to get
\begin{align}\label{eqn:Fexactresum}
    \mathcal{F}^{\rm exact}(u) = \frac{1}{2\pi i G_N^2} \int_{\mathcal{C}} dy \ \left(y - \sqrt{W(y) - u}\right)e^{2\pi \frac{y}{G_N}},
\end{align}
where the contour $\mathcal{C}$ is an inverse Laplace transform contour and so lies to the right of all non-analyticities in the integrand. The function $W(y)$ is just the dilaton pre-potential.\footnote{Note that this formula has been derived for general $W(y)$ such that $\delta W(y) \equiv W(y)-y^2$ has an inverse Laplace transform, $\mathcal{L}^{-1}(W)(a)$, which has support only in the range $a \in [0,2\pi/G_N]$.} Plugging this formula for $\mathcal{F}$ into the expression in eq. \eqref{exact_DOS}, we find
\begin{align}\label{eqn:exactdosinvlap}
    \rho_{g=0}(E) &= \frac{e^{S_0}}{8\pi^2 i G_N}\int_{E_0^{\rm exact}}^{E} \frac{du}{\sqrt{E- u}} \int_{\mathcal{C}} dy \frac{e^{2\pi \frac{y}{G_N}}}{\sqrt{W(y) - u}} \nonumber \\
    &= \frac{e^{S_0}}{2\pi G_N} \int_{\mathcal{C}} \frac{dy}{2\pi i} e^{\frac{2\pi y}{G_N}} \tanh^{-1} \left(\sqrt{\frac{E - E_0^{\rm exact}}{W(y) - E_0^{\rm exact}}} \right),
\end{align}
where in the second line we did the $u$ integral. Again, although we are focusing on a specific dilaton (pre-)potential, eq. \eqref{eqn:Fexactresum} is expected to be true for at least any dilaton potential whose deformation away from JT gravity admits an expansion in defects.\footnote{We expect, however, that the expression in \eqref{eqn:Fexactresum} is quite general. It is tantalizing that the formulae in these equations look strikingly similar to those in the discussion of the Wheeler de Witt wavefunctions of dilaton gravity, as in \cite{IliKruTurVer20, Hen85}. We suspect that using these formulae together with the formalism of \cite{JafKol19} could lead to a direct derivation of the disk density of states. By a direct derivation, we mean one that does not proceed by expanding in a gas of defects.} Interpreting $y$ as the dilaton value at the horizon, this formula can be read as imposing the smoothness condition for the geometry at the horizon. Similar ideas for computing the density of states in JT gravity were explored in \cite{JafKol19}. There the imposition of the smoothness condition was interpreted as the insertion of a defect operator in the bulk. The insertion of $e^{2\pi y/G_N}$ can be thought of as related to this defect operator. More generally, decomposing the path integral into sums over fixed horizon area states was discussed in \cite{CarTei95,DonHarMar18, AkeRat18}.

\section{New saddles and the exact density of states}\label{sec:exactdos}

In the previous section we met an exact formula for the genus zero density of states in our dilaton gravity theory. As discussed above, these functions admit a semi-classical expansion in the $G_N \to 0$ limit, yielding the ideal playground to see if we can reproduce them from the gravity theory. First we will analyze the expression in \eqref{eqn:exactdosinvlap} via a saddle point analysis. We will find saddle points in the $y$ integral at each solution to $W(y) = E$. We will then interpret these solutions in terms of novel dilaton gravity saddles with fixed energy boundary conditions. 

\subsection{Saddle point analysis of the exact density of states}

We start by analyzing the $y$ integral in the first line of eq. \eqref{eqn:exactdosinvlap} at fixed $u$ and then perform the $u$ integral. We want to compute the integral
\begin{align}\label{eqn:fprime}
     \mathcal{F}'(u) = \frac{1}{2\pi i G_N} \int_{\mathcal{C}} dy \frac{e^{2\pi \frac{y}{G_N}}}{\sqrt{W(y) - u}}.
\end{align}
The $y$ integral has saddle points obeying the equation
\begin{align}\label{eqn:F'saddle}
    \frac{2\pi}{G_N} - \frac{U(y)}{2(W(y) - u)} = 0 \implies u = W(y) + G_N \frac{U(y)}{4\pi}.
\end{align}
\begin{figure}
    \centering
    \includegraphics[scale=.44]{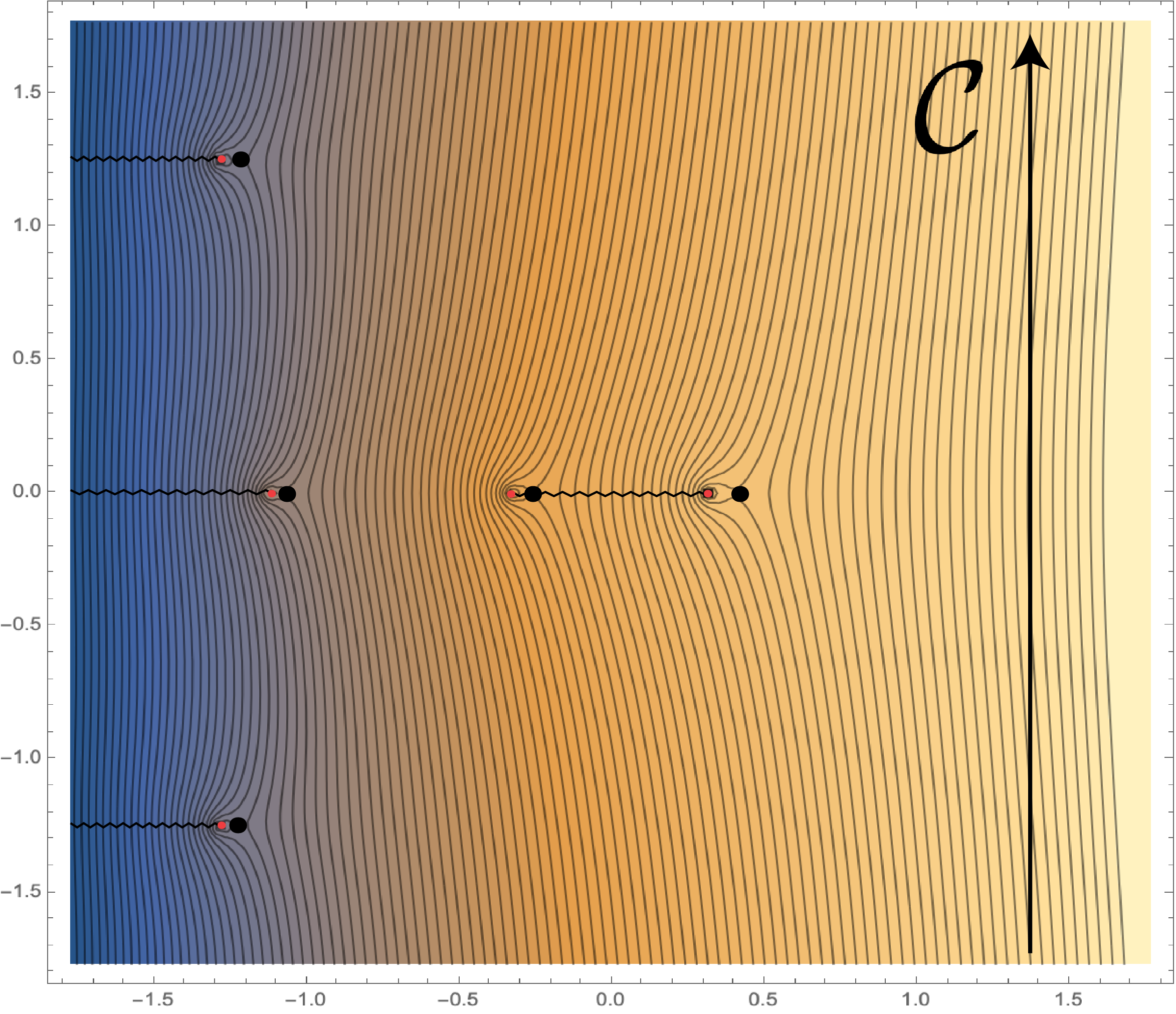}
    \caption{This figure illustrates the analytic structure of the integrand in eq. \eqref{eqn:fprime} in the complex $y$ plane. The contour of integration is pictured as the vertical arrow on the right. Each branch cut due to the square root in \eqref{eqn:fprime} is indicated by a black jagged line. The saddle points are indicated by black dots and the branch points by small red dots. The contour lines are lines of constant real part for the integral. The steepest descent contours are then orthogonal to these lines and pass through each black dot. More positive real part is in orange and more negative in blue.}
    \label{fig:steepestdescentcont}
\end{figure}
In other words, near each solution to $u = W(y)$ there is a saddle point. For our potential $W(y) = W_{\lambda}(y)$, we can even see easily which saddles contribute for a given $u$. As discussed in Sec. \ref{sec:dilgravrev}, for $E_c > u> E_0$, there are three real solutions to this equation and an infinite number of complex solutions, due to the exponential behavior of the potential. For the integrand  in eq. \eqref{eqn:fprime}, there are branch cuts emanating from each solution, with the two largest real solutions connected by a branch cut between each other. We illustrate this analytic structure in Fig. \ref{fig:steepestdescentcont}.

We can deform the contour to the steepest descent contours emanating from each saddle point.\footnote{Note that the factor of $i$ in the denominator of eq. \eqref{eqn:fprime} is cancelled due to the fact that the steepest descent contours run parallel to the imaginary axis near the saddle points.} We see that each saddle contributes, although the middle horizon (what would be the inner horizon in JT gravity) must be treated differently since the saddle point sits on a branch cut. Actually this saddle point contributes an amount which is ambiguous because we are evaluating $\mathcal{F}'$ on a Stokes line. We can break this ambiguity by continuing $G_N$ to be slightly complex. This saddle point then contributes something which is purely imaginary. 

One might be concerned that the middle horizon then contributes a complex number to the density of states. This is fixed due to the fact that to get the density of states we have to integrate $\mathcal{F}'(u)$ over $u$. This integral \emph{also} localizes to a saddle in the $G_N \to 0$ limit near the endpoint of the integral at $u=E$. For the term corresponding to the middle horizon, this integral is localized at a saddle point at $u_* = E - c G_N$ with $c>0$, but this turns out to be an unstable saddle point. Deforming the $u$ contour so that it picks up this saddle along the steepest descent contour leads to another factor of $\pm i$ from rotating the contour. Again, we lie on a Stokes line and so we need to continue $G_N$ slightly off the real axis to determine the sign in front of this second factor of $i$. One can check that the sign in front of these two $i$'s is directly correlated with the sign of the small imaginary part given to $G_N$ in order to move off the Stoke's line. The upshot is that together these two factors of $i$ give an overall minus sign. In the context of JT gravity, this minus sign is what gives the $-e^{-2\pi \sqrt{E}/G_N}$ in the expression for the density of states, $\rho_{JT}(E) \sim \sinh(2\pi \sqrt{E}/G_N)$.

Instead of going into more detail about the steepest descent analysis we have just mentioned, we will instead turn to a slightly simpler computable than the density of states which illustrates the key points of interest for us.

\subsection{Ground state energy}
Since computing the full density of states is rather complicated, we will instead focus for the rest of the paper on a simpler aspect of these formulae: the ground state energy of the matrix model. As described above, the ground state energy is fixed by the demand that the density of states has a square-root edge characteristic of a matrix model. This statement leads to the equation
\begin{align}
\mathcal{F}(E_0^{\rm exact}) = 0,
\end{align}
where again the density of states is given in terms of $\mathcal{F}$ by eq. \eqref{exact_DOS}.

Using a similar saddle point analysis to that described above, we can estimate $E_0^{\rm exact}$ in the $G_N \to 0$ limit, using eq. \eqref{eqn:Fexactresum}. Before picking up saddle points, however, we can first deform the contour so that it wraps all the branch-cuts emanating from each branch point. Again, each solution $y$ to $W(y) = u$ is a branch point in the complex $y$-plane. Furthermore, assuming there exists an outermost minimum to $W(y)$, then at small enough $u$ there are two branch points with the largest real part, as illustrated in Fig. \ref{fig:steepestdescentcont}. These set the small $G_N$ behavior of $\mathcal{F}$ for generic $u$. Thus, when the contribution due to the branch cut connecting these two points vanishes, then $\mathcal{F}$ vanishes up to corrections due to all the other solutions with more negative real part. One can easily see that the contribution from these two outermost solutions vanishes when the two solutions merge. This is because each branch point corresponds to a simple zero of $W(y)-u$ and so $\sqrt{W(y) - u}$ will in turn have a simple zero when the two solutions coincide at $u = E_0$. This means we have
\begin{align}
    \mathcal{F}(E_0) = \mathcal{O}(e^{-1/G_N}),
\end{align}
so that it vanishes up to corrections which are exponentially small in $1/G_N$. To find the exact ground state energy, $E^{\rm exact}_0$, it is then natural at small $G_N$ to expand around $E_0$. Doing so gives the equation, 
\begin{align}
E_0^{\rm exact} = E_0 + \delta E_0,\ \ \ \mathcal{F}(E_0) \vert_{\mathcal{O}(e^{-1/G_N})} \approx - \mathcal{F}'(E_0)\delta E_0.
\end{align}

So far, what we have said about the ground state energy holds for general $W(y)$ such that there exists an outermost minimum to the potential. For the specific choice, $W(y) = W_{\lambda}(y)$ with $\lambda_c>\lambda >0$, then the leading correction to $\mathcal{F}(E_0)$ will come from the innermost real solution to $W(y) = E_0$. We denote this solution by $y_i$. All the other complex saddles have more negative real part than than this solution and so are suppressed in the semi-classical limit.

To compute the contribution of this innermost real branch point at $y = y_i(E_0)$ to $\mathcal{F}(E_0)$, we just compute the piece of the deformed contour wrapping this branch cut. We then get the integral
\begin{align}
\mathcal{F}(E_0) \approx \frac{1}{G_N\pi} \int_{-\infty}^{y_i(E_0)} dy \left( \sqrt{E_0 - W_{\lambda}(y)}\right)\,e^{\frac{2\pi y}{G_N}} + \text{contributions from sub-leading branch points}.
\end{align}
Written in this form, the integral has a saddle point along the defining contour at the position $y_*$ defined by
\begin{align}
    W_{\lambda}(y_*) = E_0 - \frac{G_N W_{\lambda}'(y_*)}{4\pi}.
\end{align}
To leading order in $G_N$, including the one-loop correction, we find 
\begin{align}
    \mathcal{F}(E_0) \approx \frac{1}{2G_N} \left( \frac{G_N}{2\pi}\right)^{3/2} \sqrt{W_{\lambda}'(y_i)} e^{\frac{2\pi y_i}{G_N} + \frac{\pi^2}{2}}.
\end{align}
Furthermore, $\mathcal{F}'(E_0)$ can be calculated by noticing that when the two outermost branch points collide, then they form a simple pole in the expression for $\mathcal{F}'$ in eq. \eqref{eqn:fprime}. Picking up this pole gives
\begin{align}
    \mathcal{F}'(E_0) \approx \frac{\sqrt{2}}{ G_N \sqrt{W_{\lambda}''(y_o)}}e^{\frac{2\pi y_o}{G_N}} + \text{ exponentially sub-leading}.
\end{align}
This leads to 
\begin{align}\label{eqn:exactshift}
    \delta E_0 \approx (\mathrm{one-loop}) \times e^{\frac{2\pi \Delta y}{G_N}} = -\frac{1}{4}\sqrt{W_{\lambda}'(y_i)W_{\lambda}''(y_o)} \left(\frac{G_N}{2\pi}\right)^{3/2}e^{\frac{2\pi \Delta y}{G_N}+\frac{\pi^2}{2}}
\end{align}
where $\Delta y = y_i - y_o$ is the difference between the inner root and the outer root of the equation $W_{\lambda}(y) =E_0$. Note that this difference is negative so that the change in energy is indeed exponentially suppressed in $1/G_N$.


\subsection{Alternate dilaton contour and the semi-classical density of states}\label{sec:altcont}

It is tempting to ask if the function $\mathcal{F}^{\rm semi-classical}$ in \eqref{eqn:Fsemiclass}, which ignores the upper limit on the sum in the expression for $\mathcal{F}$ in eq. \eqref{minimal_string_F}, has a representation analogous to eq. \eqref{eqn:Fexactresum}.
\begin{figure}
    \centering
    \includegraphics[scale=.45]{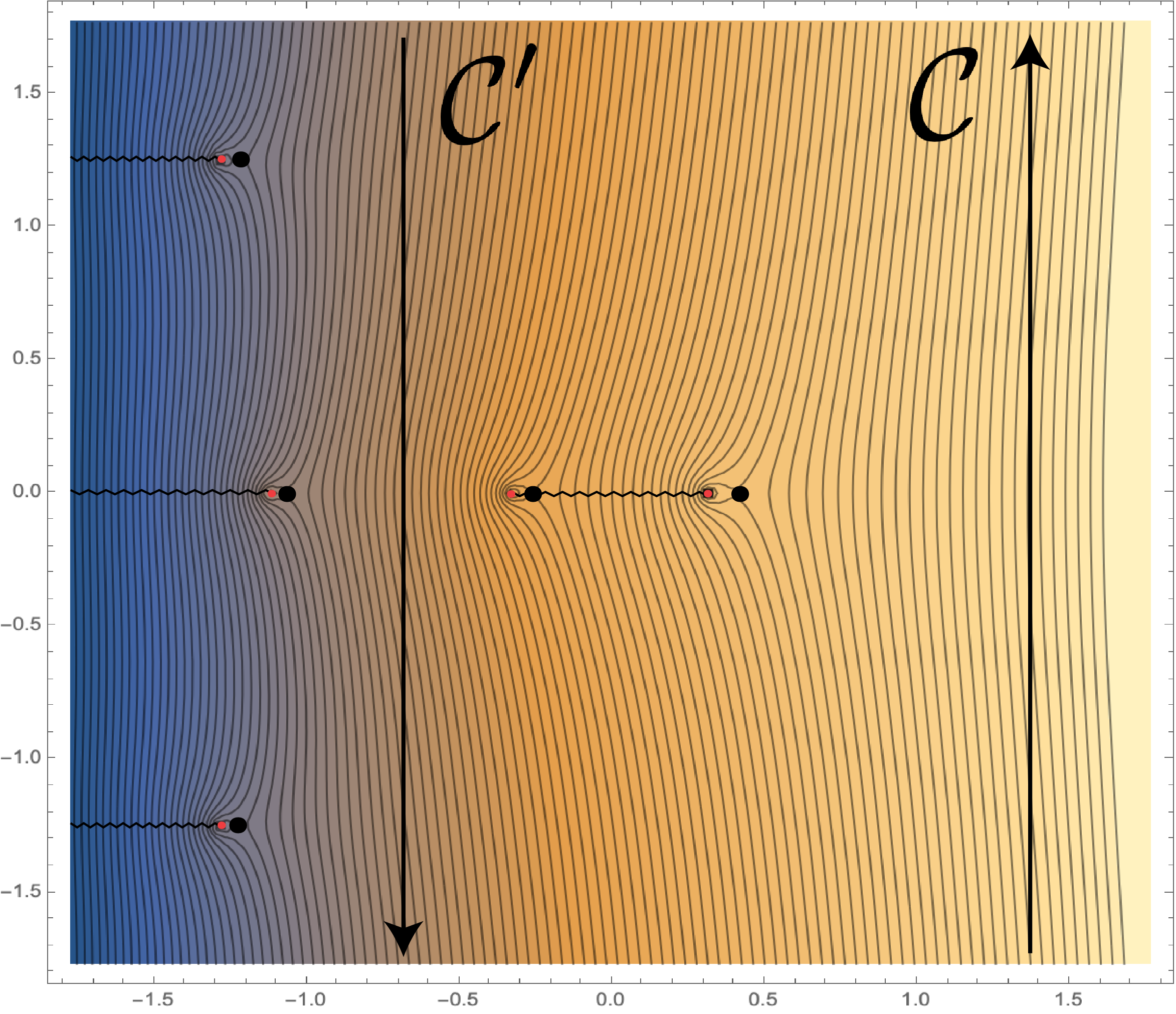}
    \caption{This figure illustrates the possible alternate contour for the $y$ variable that is used to compute the density of states in eq. \eqref{eqn:exactdosinvlap}. This contour excludes contributions from saddle points other than those associated with the two outermost horizons. This produces a result for the ground state energy that is consistent with that predicted by canonical quantization.}
    \label{fig:alternatecontour}
\end{figure}
For the dilaton potential $W_{\lambda}(y)$ with $0<\lambda<\lambda_c$, consider an energy $u$ small enough such that there are three real solutions to the equation $W_{\lambda}(y) = u$. Following the same steps presented in \cite{TurUsaWen21} for re-summing the exact function $\mathcal{F}^{\rm exact}$, one finds that if we modify $\mathcal{C} \to \mathcal{C} \cup \mathcal{C'}$, where $\mathcal{C}'$ is a vertical contour that lies between the middle and innermost solution but running in the opposite direction from $\mathcal{C}$, then the $\lambda$ expansion does not truncate at a finite order. In other words, we have
\begin{align}\label{eqn:Fsemiclassresum}
    \mathcal{F}^{\rm semi-class}(u) = \frac{1}{2\pi i G_N^2} \int_{\mathcal{C} \cup \mathcal{C}'} dy \ \left(y - \sqrt{W(y) - u}\right)e^{2\pi \frac{y}{G_N}}.
\end{align}
We illustrate this other contour in Fig. \ref{fig:alternatecontour}. Note that this contour effectively excludes any saddle point contributions from horizons other than the two outermost horizons.

From this representation of $\mathcal{F}^{\rm semi-classical}$ it is clear that the ground state energy $E_0$ as defined by $\mathcal{F}^{\rm semi-classical}(E_0) = 0$ is just given by the value when the two outermost roots of the equation $W(y) = E$ merge in the complex plane. This is because we can deform the contours to pick up the branch cut connecting the outer two horizons. This branch cut vanishes into a simple zero when the two roots collide at $y_0$ given by $W(y_0) = E_0$. 

Unfortunately, this contour is only good at lower energies for $\lambda >0$. For high enough energies, the inner contour $\mathcal{C}'$ can be pinched by the merging of the middle root and the inner-most root. These two roots collide when the Penrose diagram truncates from an infinite number of asymptotic regions to just one. This means that the formula for $\mathcal{F}^{\rm semi-classical}$ encounters a non-analyticity at this energy. It is tempting to interpret this non-analyticity in terms of a phase transition. In Sec. \ref{sec:nearphasetrans}, we will show that the exact answer as defined by equations \eqref{eqn:exactdosinvlap} does not see such a non-analyticity at this energy.

It is actually interesting to take this alternate contour seriously for a moment and see what it predicts for the density of states by plugging $\mathcal{F}^{\rm semi-classical}$ and $E_0$ into eq. \eqref{eqn:exactdosinvlap} for $\mathcal{F}^{\rm exact}$ and $E_0^{\rm exact}$, respectively. Examining the second line of eq. \eqref{eqn:exactdosinvlap}, we can re-write this expression in a more illuminating way as
\begin{align}\label{eqn:dosrewritten}
    \rho_{g=0}(E) \sim \int_{\mathcal{C} \cup \mathcal{C}'} \frac{dy}{2\pi i} e^{\frac{2\pi y}{G_N}} \left(\frac{1}{2}\log \left(W(y) - E\right) - \log \left( \sqrt{W(y) - E_0} + \sqrt{E - E_0}\right)\right).
\end{align}
From the first term, we see that there are branch cuts emanating from all solutions to $W(y) = E$ in the complex $y$ plane. From the second term, we see that there appear to also be branch cuts emanating from solutions to $W(y) - E_0$. In fact, since $E_0$ occurs at the outermost minimum, $y_0$, of $W$, then the square root, $\sqrt{W(y) - E_0}$, in the second term of eq. \eqref{eqn:dosrewritten} is regular at $y_0$ since $W(y) - E_0 \approx \frac{W''(y_0)}{2} (y- y_0)^2$ for $y \approx y_0$. 

Since all the other solutions to either $W(y) = E_0$ or $W(y) = E$ are not bounded by the alternate contour, we can then just deform $\mathcal{C} \cup \mathcal{C}'$ to pick up the branch cut in eq. \eqref{eqn:dosrewritten} running between the two outermost solutions to $W(y) - E$. We are then left with the pleasingly simple answer
\begin{align}\label{eqn:outertworootsdos}
    \rho^{\rm semi-classical}(E) \sim e^{\frac{2\pi}{G_N} W_{+}^{-1}(E)} - e^{\frac{2\pi}{G_N}W_{-}^{-1}(E)},
\end{align}
where $W^{-1}_{\pm}(E)$ denotes the outermost $(+)$ and second outermost $(-)$ roots of the equation $W(y) = E$. 

Since $E_0$ differs from $E_0^{\rm exact}$ by exponentially suppressed contributions, eq. \eqref{eqn:outertworootsdos} actually agrees with the exact density of states up to corrections that are exponentially small in $1/G_N$, at least in a specific range of energies. In particular, for the potential $W_{\lambda}$ formula \eqref{eqn:outertworootsdos} is true only for energies not too large or not too small, since these corrections can become unsuppressed in these limits. For energies $E$ too close to $E_0$, this formula breaks down due to edge effects near $E_0^{\rm exact}$ in the formula eq. \eqref{eqn:exactdosinvlap}. These edge effects are the main topic of this paper. For $E$ too large, this formula breaks down because the middle real root of $W_{\lambda}(y) = E$ can merge with the innermost real root as discussed above.


\subsection{Bulk interpretation of the saddles}\label{sec:bulkinterp}
Note that the saddle points in the integral defining $\mathcal{F}$ obey the same equations that defines the relationship between the dilaton at the horizon and the boundary AdM energy in the classical solutions, as in eq. \eqref{eqn:E=W}. This suggests that we should interpret these solutions as associated to bulk geometries with fixed energy boundary conditions. 

Fixed energy boundary conditions in JT gravity are given by imposing fixed $\partial_n \Phi$ and fixed $\Phi$ along the asymptotic boundary such that $\partial_n \Phi - \Phi = E$.\footnote{One may wonder how it is that doing the inverse Laplace transform of fixed length boundary conditions leads to the boundary conditions presented here. This point was nicely explained in \cite{GoeIliKruYan20}.} Such boundary conditions allow for complex values of the dilaton at the horizon. This is because the identification of $\tau \sim \tau + \beta$ is not fixed to be real anymore. These solutions are then defined by analytic continuations of the standard disk geometry into the complex dilaton plane, so that the dilaton takes values along a complex contour instead of along the real axis. 

Note that we can discuss these complex gometries just in the context of pure JT gravity with fixed energy boundaries. In JT gravity, to find the classical solutions with these boundary conditions, we just have to solve the equation $E = \Phi_h^2$. In that case, there is the standard disk saddle, with $\Phi_h$ given by $\Phi_h = \sqrt{E}$. The contribution to the path integral of this saddle gives a term like $e^{2\pi \sqrt{E}/G_N}$. There is, however, another saddle whose dilaton at the horizon takes on the value of the other root to this equation, $\Phi_h = - \sqrt{E}$.

\begin{figure}
    \centering
    \includegraphics[scale=1]{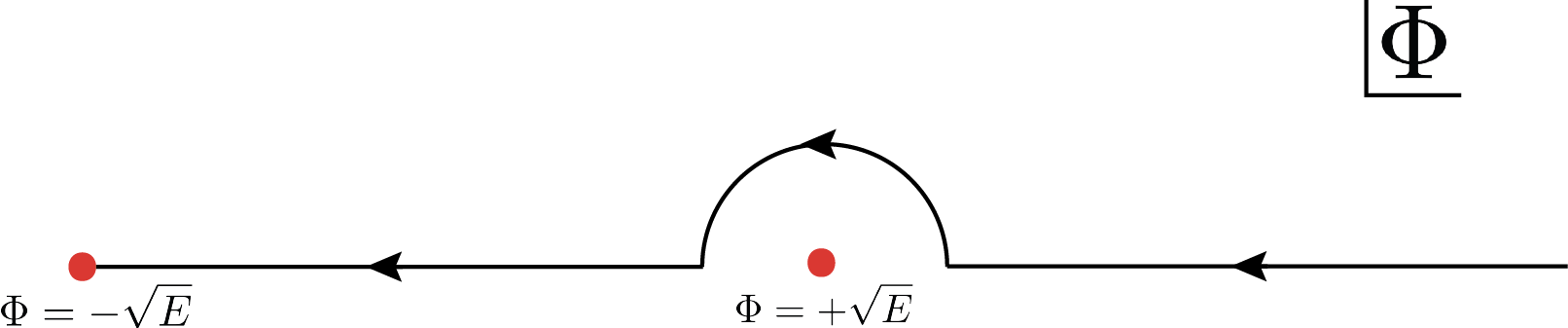}
    \caption{We illustrate the complex contour for the dilaton (radial coordinate) in the sub-leading saddle described in the main text. This defines a sub-leading saddle point geometry, which contributes the exponentially decaying in energy term to the density of states in JT gravity.}
    \label{fig:geocontour}
\end{figure}

To make sense of such a saddle, we can consider a complex geometry with metric given by
\begin{align}
    ds^2 = (W(\Phi) - E) d\tau^2 + \frac{d\Phi^2}{W(\Phi) - E}
\end{align}
but where the dilaton follows the complex contour running from $+\infty$ down to $-\sqrt{E}$, going around the other root at $+\sqrt{E}$ in the complex plane.  Of course, there is an ambiguity on how one actually chooses the contour. As usual, all contours which are deformable to each other without passing through non-analyticities in the action should be deemed equivalent. This proposed saddle is illustrated in Fig. \ref{fig:disksphere}, and we illustrate the proposed contour for JT gravity in Fig. \ref{fig:geocontour}.

Note that upon continuing around the pole in the metric at $\Phi = +\sqrt{E}$, the metric becomes that of the anti-sphere, i.e. the sphere with an overall minus sign in front of its metric. Such a solution to JT gravity was also considered in \cite{MahStaYan21}. Note that this metric violates the Kontsevich-Segal criterion for allowable metrics. Presumably, this saddle point, when its fluctuations have been properly accounted for, reproduces the sub-leading term in the $\sinh(2\pi \sqrt{E})$ density of states for JT gravity. We have not done a direct bulk analysis of this minus sign, but it would be interesting to understand how this works in more detail. One path forward would be to more directly derive the formulae in eq. \eqref{eqn:Fexactresum} and \eqref{eqn:exactdosinvlap} from the bulk. We hope to study this possibility in the future.

\subsection{A (near) phase transition}\label{sec:nearphasetrans}

As mentioned in Sec. \ref{sec:dilgravrev}, for $\lambda >0$ but not too large, there is a critical energy $E_c(\lambda)$ above which the classical solutions go from an infinite number of universes to a single universe. We can ask if the matrix model's thermodynamics see any signature of this transition. In particular, it is tempting to look for a phase transition in the model. 

As we crank up the energy, two innermost real roots of the equation $W_{\lambda}(y) = E$ merge and then go off into the complex plane. They merge precisely at critical energy. Since $\rho(E) \sim \mathcal{F}'(E)$ semi-classically, then we can just analyze $\mathcal{F}'$ as a function of energy. Just as discussed in the context of the ground state energy, when these two roots of the equation $W_{\lambda}(y) = E$ merge, they form a simple pole in \eqref{eqn:F'saddle}. Picking up this pole gives a contribution of the form
\begin{align}
    \mathcal{F}' \supset \frac{\sqrt{2}e^{\frac{2\pi y_i}{G_N}}}{G_N\sqrt{|W_{\lambda}''(y_i)|}}.
\end{align}
This contribution is finite and so there is no actual divergence in the density of states. Interestingly, the one-loop factor is enhanced for this saddle point by factors of $G_N$ because $W_{\lambda}'(y_i) = 0$ at this energy. One can understand this enhancement by examining the one-loop fluctuations for the saddle points defined by eq. \eqref{eqn:F'saddle}. This suggests that the horizon dilaton fluctuations about these sub-leading saddles is becoming strong at this point, since horizon area fluctuations appear to be diverging like $1/G_N$ near this critical point.  

Of course, this is not totally suprising: in the Penrose diagram, as we approach this critical point, two horizons are merging and developing an approximatedly dS$_2$ region which lies behind the outermost horizon. It would be interesting to understand if this mode could be related to the Schwarzian mode near asymptotic future of the nearly de Sitter region as discussed in \cite{MalTurYan19, CotJenMal19}. Note that from the point of view of the full density of states of the model, however, this is a non-perturbatively small effect which is swamped by the leading classical answer. It would be interesting to find other parameter regimes where this merger of the two inner branch points leads to a sharper signal in boundary observables.

\subsection{The critical coupling}\label{sec:critcoup}
In \cite{JohRos20,MaxTur21,Wit20MM}, the authors observed a special value of the coupling $\lambda = \lambda_c^{MM}$ for the potential $W_{\lambda}(\Phi)$ such that the exact string equation given in eq. \eqref{eqn:Fexactresum} no longer has a real solution to $\mathcal{F}^{\rm exact} (E_0) = 0$. The authors in \cite{JohRos20} understood this as a phase transition wherein the matrix model becomes non-perturbatively unstable for $\lambda > \lambda_c^{MM}$. It is natural that this transition should be associated to the critical coupling $\lambda_c = \frac{1}{2\pi^2 e (1-\alpha)^2}$ discussed in Sec. \ref{sec:dilgravrev} at which the dilaton potential $W_{\lambda}(\Phi)$ becomes monotonic. The reason is that for $\lambda > \lambda_c$ the theory no longer has a classical solution with $\beta = \infty$.

In the $G_N \to 0$ limit, it is easy to see that indeed the critical coupling $\lambda_c^{MM}$ at which the matrix model becomes unstable is (approximately) equal to the critical coupling $\lambda_c$ at which $W_{\lambda}(\Phi)$ becomes monotonic. The reason can be seen easily from the manipulations laid out in this section. In the $G_N \to 0$ limit, we have shown that the ground state energy is (up to non-perturbative corrections) associated with the value of $W_{\lambda}(\Phi)$ at the outermost minimum of the potential. When this outermost minimum no longer exists, there is no longer a way for $\mathcal{F}(E_0)$ to (almost) vanish. The outermost minimum vanishes from the real axis when the potential becomes monotonic and so $\lambda_c \approx \lambda_c^{MM}$, up to non-perturbatively small corrections.


\section{Canonical quantization in geodesic gauge}\label{sec:canquant}

At the disk level, JT gravity can be studied via the methods of canonical quantization. In fact, dilaton gravity with more general dilaton potentials can be quantized in this way as well. Canonical quantization of dilaton gravity for general dilaton potentials was discussed in \cite{IliKruTurVer20, Hen85, NanSakTri23, MarGegKun94}. To proceed, we first write a general metric in AdM gauge as
\begin{align}
ds^2 = -N^2 dt^2 + h(dx + N_{\perp} dt)^2,
\end{align}
where the time variable $t$ denotes a choice of Cauchy slice for the geometry and $x$ denotes the spatial coordinate on the slice. The dynamical variables on each slice can then be thought of as the dilaton and the scale factor $h = e^{2\sigma}$. There is a large gauge redundancy in describing the state using these variables, which we would like to fix. One natural choice is to fix the extrinsic curvature of the slice to be zero. This is the so-called geodesic gauge. Picking geodesic gauge has been discussed extensively in the context of JT gravity, where it was used to solve for the disk-level wavefunctions exactly \cite{HarJaf18,Yan18}. 

In dilaton gravity, the extrinsic curvature is canonically conjugate to the dilaton, whereas the induced metric on a geodesic slice is conjugate to the normal derivative of the dilaton \cite{GoeIliKruYan20}. By combining the Wheeler de-Witt and momentum constraints, one can get a classical expression for the momentum conjugate to the intrinsic metric which is
\begin{align}
P_{\ell} =\frac{1}{G_N} \sqrt{\left(\partial_u \Phi\right )^2 - W(\Phi) + E}
\end{align}
where $E$ is the AdM mass of the solution.\footnote{The factor of $1/G_N$ comes from the re-scaling of $\Phi$ and the couplings required for the classical limit.} We denote the conjugate momentum by $P_{\ell}$ because it is conjugate to the length of the geodesic slice, $\ell$, which is the only gauge invariant quantity for a one-dimensional metric. Here $\partial_u$ means derivatives with respect to a proper length variable along the slice. Using the ansatz \eqref{eqn:onshellmetric}, one can show that this is precisely the equation for a spacelike geodesic with momentum $P_{\ell}$, which is conserved due to the boost symmetry of the spacetime. 

\begin{figure}
    \centering
    \includegraphics[scale=.45]{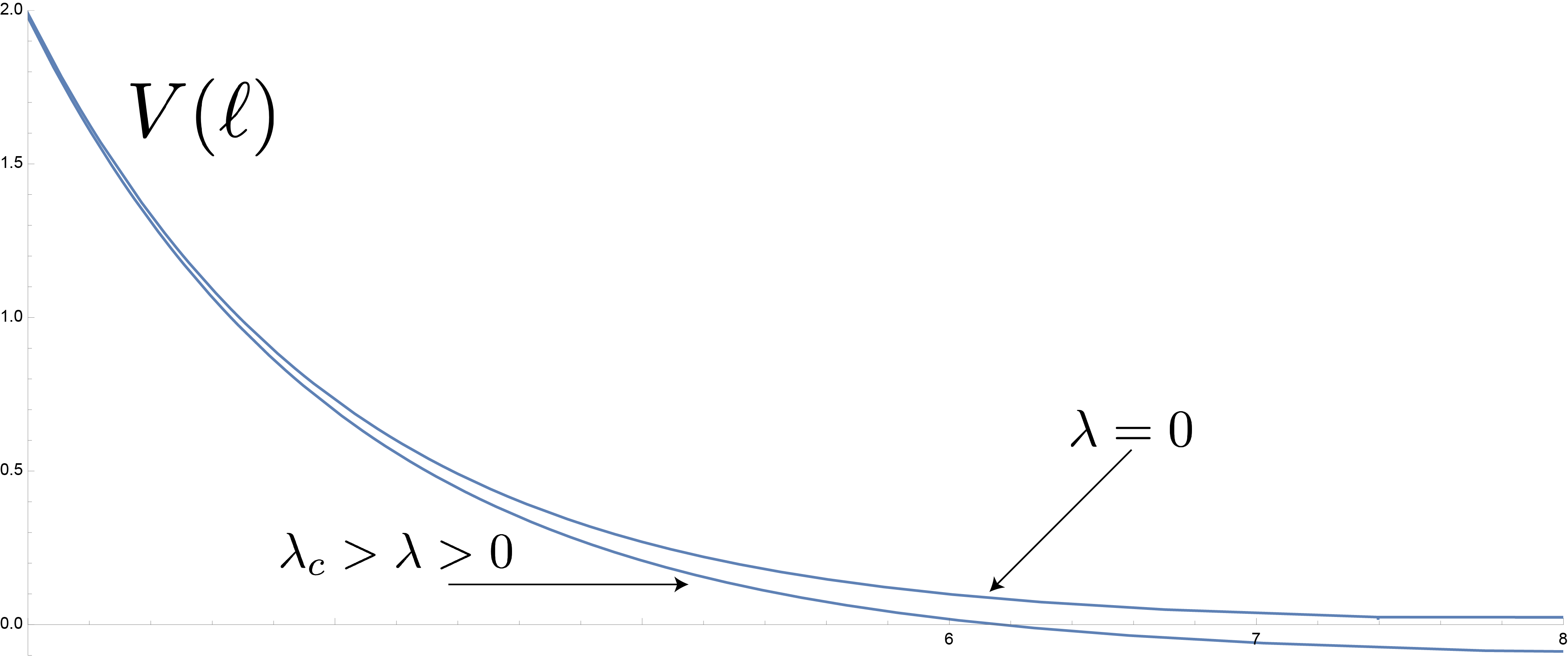}
    \caption{We plot the length potential for dilaton gravity with potential $W_{\lambda}(\Phi)$. For $\lambda = 0$, we just get the exponential potential of \cite{HarJaf18}. For the $\lambda_c>\lambda>0$, we have a mildly modified potential with a different asymptotic value at $\ell \to +\infty$. This corresponds to the shift in the ground state energy of the model from $E_0 = 0$ to $E_0$, set by the outermost minimum of $W_{\lambda}(\Phi)$. Note that the potentials agree at $\ell \to -\infty$ since this probes the asymptotically AdS region.}
    \label{fig:lengthpotential}
\end{figure}

We can solve this expression for the length in terms the momentum and energy, to get an expression
\begin{align}\label{eqn:lengthpotential}
\ell = \lim_{\Phi_b \to \infty} 2\left( \int_{W^{-1}(E-G_N P_{\ell}^2)}^{\Phi_b} \frac{d\Phi}{\sqrt{W(\Phi) - E + G_N^2 P_{\ell}^2}} - \log \Phi_b\right).
\end{align}
Just as in JT gravity, the $\log \Phi_b$ term is a universal renormalization of the length operator that makes it well-defined in the $\Phi_b \to \infty$ limit. Solving this for $V(\ell) \equiv E - G_N P_{\ell}^2$, we get an expression for the energy in terms of the momentum and potential
\begin{align}
E = G_N^2 P_{\ell}^2 + V(\ell).
\end{align} 
In JT gravity, this equation becomes
\begin{align}
E = G_N^2 P_{\ell}^2 + 4e^{-\ell},
\end{align}
as found in \cite{HarJaf18}. For the models we consider here which are deformed away from JT gravity by an exponential potential, the length potential is not analytically solvable, except for in certain regimes. We plot the potential analytically in Fig. \ref{fig:lengthpotential} for $\lambda$ in the range $\lambda_c > \lambda >0$ and for $\lambda = 0$ (pure JT). 

Regardless, one could imagine quantizing this theory in the obvious way by making $P, \ell$ into operators which obey the standard commutation relations. This way of quantizing the theory ignores certain gauge ambiguities which will be important for us.\footnote{There are also ordering ambiguities in solving eq. \eqref{eqn:lengthpotential} for the length potential, which could in principle be important, but following \cite{IliKruTurVer20} we will ignore these entirely.} What we will show in the remainder of this section is that this naive quantization procedure gets the correct answer for the ground state energy of the matrix model as well as the low energy behavior of the density of states, up to corrections which are exponentially small in $1/G_N$ (but enhanced at very low energies).

\subsection{Length potential at large positive and negative length}
While the length potential cannot be solved for analytically except for in a very specific class of examples, it can be solved for in certain limits. At large positive lengths, we can try to solve equation \eqref{eqn:lengthpotential} for $V(\ell)$. The only way to get a length which is arbitarily large is if we have tuned $E$ to be near the outermost minimum of the potential $W(\Phi)$. Let the value that $\Phi$ takes at this location be denoted by $\Phi_0$ so that
\begin{align}
W(\Phi_{0}) = E_0.
\end{align}
Focusing on the zero momentum geodesic for now, for such low energies we have
\begin{align}
\ell \approx \int_{W^{-1}(V(\ell))} \frac{d\Phi}{\sqrt{W'' (\Phi - \Phi_{0})^2/2 + (E_0 - V(\ell))}} \approx -\frac{2 \sqrt{2} }{\sqrt{W''(\Phi_0)}}\left( \log \sqrt{ \frac{V(\ell)-E_0}{\Lambda}}\right)
\end{align}
up to a constant, $\Lambda$, which is independent of $V(\ell)$. This constant can only be found numerically since it receives contributions from pieces of the $\Phi$ integral away from the endpoint. Solving for the length potential at large lengths then takes the form
\begin{align}
V(\ell) = E_0 + \Lambda e^{-\ell \sqrt{W''(\Phi_0)/2} }\left(1 + \mathcal{O}(1/\ell)\right).
\end{align}
This means that a quantization of this model along the lines described above leads to a prediction for the ground state energy of the dual matrix model as occuring at $E_0$ defined to be the value of the dilaton (pre)-potential at its outermost minimum. As discussed in Sec. \ref{sec:exactdos}, this answer almost matches with the exact answer in the semi-classical $G_N \to 0$ limit. It is off from the exact answer due to a non-perturbatively small shift in the ground state energy of the model. In the next section, we will interpret this shift as coming from a non-perturbatively small overlap between different states of fixed geodesic length. In other words, at non-perturbative in $1/G_N$ orders, geodesic length states of differing lengths are in fact not linearly independent, and so we should not expect canonical quantization in the geodesic length gauge to get the correct answer unless this linear dependancy is properly accounted for.

We can also solve for the potential at large negative $\ell$. Due to the large shift by $\log \Phi_b$ in the renormalization of $\ell$ in \eqref{eqn:lengthpotential}, the way to get large negative $\ell$ is when the turning point at the lower limit of the integral goes to $+\infty$. This is when $V(\ell)$ is very large. In this limit, the geodesic only probes the geometry at large $\Phi$ and so the dilaton potential limits back to JT gravity. Thus, we see that the length potential $V(\ell)$ has the behavior 
\begin{align}\label{eqn:largeellpot}
V(\ell) \sim 4e^{-\ell} \ \text{as } \ell \to -\infty,\ \text{and}\ V(\ell) \sim E_0 + \Lambda e^{-\ell \sqrt{W''(\Phi_0)/2} }\ \text{as}\ \ell \to +\infty.
\end{align}
This agrees with the numerically generated answer shown in Fig. \ref{fig:lengthpotential}. Given this answer for the predicted ground state of the model, in the specific example where $W(\Phi) = W_{\lambda}(\Phi)$, we can ask what happens if we start cranking up the coupling in the potential. As we discussed in Sec. \ref{sec:dilgravrev}, there is a critical $\lambda = \lambda_c$ above which the potential becomes monotonic. We can ask what happens to the length potential $V(\ell)$ at this critical coupling. Indeed, we see that the quantum mechanics defined by the potential $V(\ell)$ becomes ill-defined for very large $\ell$.\footnote{One other way to understand this is that for $\lambda > \lambda_c$ there are no longer geodesics in the classical Lorentzian solutions that are anchored to the two AdS boundaries for arbitrarily late times.} As mentioned in Sec. \ref{sec:critcoup}, this coupling $\lambda_c$ is associated with the critical coupling at which the matrix model becomes non-perturbatively unstable. It is clear that much about the exact matrix model's disk-level density of states can be learned just from analyzing the quantum mechanics defined by quantizing the theory in geodesic gauge. 

\subsection{Density of states and the WKB approximation}\label{sec:wkbdos}

Although the length potential defined by \eqref{eqn:lengthpotential} is not analytically tractable except for special dilaton potentials, in the semi-classical limit we can try to solve Schroedinger's equation via the WKB approximation. In that limit, the wavefunctions in the length basis for energy eigenstates are given by linear combinations of
\begin{align}
\psi_E(\ell) \supset \frac{1}{(E - V(\ell))^{1/4}} \exp \left( \frac{\pm i}{G_N} \int^{\ell}_{V^{-1}(E)} d\ell' \sqrt{E - V(\ell')}\right)
\end{align}
which is the answer in the classically allowable region. Using the results above, we see that at very large $\ell$ and fixed $E$, we get that the wavefunctions are 
\begin{align}
\psi_E(\ell) \sim \frac{1}{(E-E_0)^{1/4}} \exp \left( \pm i \ell \frac{\sqrt{E - E_0}}{G_N}\right).
\end{align}
At large $\ell$, we thus have that the general real solution takes the form\footnote{We will focus on real wavefunctions because this is what the Euclidean path integral will necessarily output for the length wavefunction in the energy eigenbasis, $\braket{\ell | E}$.}
\begin{align}\label{eqn:classicalregionexp}
\psi_E(\ell) =  \frac{1}{2(E-E_0)^{1/4}} \left( C(E) \exp \left( i \ell \frac{\sqrt{E - E_0}}{G_N}\right) + C^*(E) \exp \left( - i \ell \frac{ \sqrt{E - E_0}}{G_N}\right)\right).
\end{align}

In fact, due to an argument adapted from \cite{Lin22}, these wavefunctions are enough to in principle determine the disk-level density of states, assuming that fixed geodesic length states are delta-function orthogonal. The argument goes as follows: let the density of states be given by $\rho_0(E)$. Then the wavefunctions $\psi_E(\ell)$ should obey the orthogonality relations
\begin{align}\label{eqn:orthogonality}
&\int_{E_0}^{\infty} dE \rho_0(E) \psi^*_E(\ell) \psi_E(\ell') = \delta (\ell - \ell'), \nonumber \\
& \int_{-\infty}^{\infty} d\ell \psi_E(\ell) \psi_{E'}(\ell) = \frac{\delta (E - E')}{\rho_0(E)}.
\end{align}
Given the above orthogonality relations, the fact that the wavefunctions decay doubly exponentially in $\ell$ due to \eqref{eqn:largeellpot} tells us that the only way to get a delta function in energy out of the integral over $\ell$ is from the large, positive $\ell$ integral. This means that
\begin{align}
1/\rho_0(E) = |C(E)|^2.
\end{align}

Naively $C(E)$ is arbitrary. The trick, however, is then to view the disk-level partition function as a return amplitude in Euclidean time, from the $-\infty$ geodesic length state back to itself in $\beta$ time:
\begin{align}
Z(\beta) \sim \braket{-\infty| e^{-\beta H} | -\infty} = \int dE e^{-\beta E} \rho_0(E) \psi_E(-\infty) \psi_E(-\infty).
\end{align}
At large negative $\ell$, the wavefunctions decay doubly-exponentially in $\ell$. Stripping off this decay and absorbing it into a re-definition of $S_0$, we see that this formula produces the correct partition function provided that $\psi_E(\ell)$ is independent of $E$ at large, negative $\ell$. The demand that $C(E)$ be such that the transmission amplitude for $\ell \to -\infty$ is independent of $E$ is in fact enough to fix the form of $C(E)$, up to an energy independent constant.

We can then view solving for the density of states as a scattering problem, where waves come in from large positive $\ell$ and scatter off the potential. The above discussion says that there is a boundary condition at large negative $\ell$ which demands that the transmission coefficient is independent of $E$. Note that demanding this boundary condition amounts to doing the Euclidean gravitational path integral at the disk level. In particular, this argument should not be read as a new derivation of the density of states in gravity. Rather, one should view this calculation as performing the Eucliean path integral by slicing the geometry along geodesics.

In principle, one could try to solve for these transmission coefficients $C(E)$ in the semi-classical limit to arrive at the density of states. Instead, we will opt to determine their behavior at $E$ near the ground state, $E_0$, as we will need this formula later. It would be interesting to use WKB techniques of \cite{FesLiu08, KnoSch76} to try to compute $C(E)$ semi-classically for all energies. At low energies, however, the problem simplifies. This is due to the fact that at low energies the wavefunction really only feels the potential at large lengths of order $\log 1/(E- E_0)$. To understand this, we note that the wavefunction rapidly decays for $\ell$ in the classically forbidden region. Thus, the integral in eq. \eqref{eqn:orthogonality} over $\ell$ can be truncated to values of $\ell$ that are only in the classically allowed region, up to corrections that are non-perturbatively suppressed in $1/(E-E_0)$. In this region, since the lengths are long (greater than $\log 1/(E-E_0)$, then we can use the expression in eq. \eqref{eqn:largeellpot} for the potential.

The solution for these potentials is just the same as in the JT gravity case and are given by Bessel functions, namely 
\begin{align}
    \psi_E(\ell) \sim K_{\frac{2i\sqrt{(E-E_0)}}{G_N} \sqrt{\frac{2}{W''(\Phi_0)}}}\left(2 \sqrt{\frac{2\Lambda}{W''(\Phi_0)G_N}} e^{-\frac{\ell}{2} \sqrt{\frac{W''(\Phi_0)}2}} \right),
\end{align}
where again $\Lambda$ is a constant which needs to be determined numerically. These wavefunctions are just the same as those of JT gravity but with a re-scaled and shifted energy and so the density of states is just that of JT gravity with a re-scaled and shifted energy
\begin{align}\label{eqn:lowenergydensityofstates}
    \rho(E) \sim \sinh \left(2\pi \sqrt{\frac{2}{W''(\Phi_0)}} \frac{\sqrt{E - E_0}}{G_N} \right)
\end{align}
for $E$ close enough to $E_0$. Note that this formula just requires that we are at low enough energies so that the second equation in eq. \eqref{eqn:largeellpot} holds. Since the large $\ell$ approximation of $V(\ell)$ cares only about $\lambda$, we have in particular not made the assumption that $E-E_0 \sim G_N^2$. Thus, the formula in eq. \eqref{eqn:lowenergydensityofstates} should hold for a parametrically large range of energies. Intuitively this formula should be obvious; we have zoomed in on energies where the dilaton potential is well-approximated by a quadratic and so we have reduced the theory back to JT gravity approximately.

Of course, since we are focusing on a specific region in energy space and since our method can miss an overall, energy independent constant in the density of states, the normalization in this formula may not be correct to match onto higher energies. To find the correct normalization, we need to solve for the density of states at higher energies and match the two expressions in intermediate energies. At high enough energies, the semi-classical answer comes purely from the leading disk saddle point. Its contribution to the density of states gives
\begin{align}
   \rho(E) \sim e^{2\pi W^{-1}(E)/G_N},
\end{align}
where $W^{-1}(E)$ is the branch of $W^{-1}$ with the largest real part. At lower energies this gives
\begin{align}
\rho(E) \sim \exp \left( \frac{2\pi}{G_N} \left( \Phi_0 + \sqrt{\frac{2\left(E-E_0\right)}{W''(\Phi_0)} } \right) \right),
\end{align}
where again $\Phi_0$ is the value of the dilaton at the outermost minimum of the potential $W(\Phi)$. We then expect the density of states goes like 
\begin{align}\label{eqn:lowedoscanquant}
    \rho(E) \sim e^{\frac{2\pi}{G_N} \Phi_0}\sinh \left(2\pi \sqrt{\frac{2}{W''(\Phi_0)}} \frac{\sqrt{E - E_0}}{G_N} \right)
\end{align}
up to order one factors. Note that this formula agrees with the semi-classical density of states in eq. \eqref{eqn:outertworootsdos} predicted by using the alternate contour of Sec. \ref{sec:altcont}. In eq. \eqref{eqn:lowedoscanquant} we are effectively evaluating $\rho^{\rm semi-classical}(E)$ in eq. \eqref{eqn:outertworootsdos} at energies close enough to $E_0$ so that $W(y)$ can be approximated by a Gaussian around $y = y_0$.  Apparently, slicing the path integral in geodesic gauge leads to an answer that prefers the alternate contour. As we will argue in Sec. \ref{sec:lengthambiguity}, this is because the alternate contour excludes the geometries which cause a breakdown in geodesic gauge.

\subsection{A subtlety with geodesic gauge in $W_{\lambda}(\Phi)$ gravity}\label{sec:cansubtlety}
So far in this section, we have been agnostic about the form of $W(\Phi)$, except that we have demanded that $W$ has an outermost minimum at $\Phi_0$, where $W(\Phi_0) = E_0$. If we restrict our attention to the potential $W_{\lambda}(\Phi)$ for $\lambda_c>\lambda>0$, such that an outermost minimum exists, there is a subtlety with geodesic gauge already at the level of the Lorentzian classical solutions. As discussed in Sec. \ref{sec:dilgravrev}, for solutions with AdM energy $E>E_c$ (or $E<E_0$), the Lorentzian universe truncates to a single AdS boundary. We can solve for the geodesics anchored between two boundary points in such spacetimes. Just as in higher dimensional AdS black holes \cite{FidLuk04}, one can show that there are bouncing geodesics for $E>E_c$. This means that there can be more than one geodesic anchored to the same two boundary points in the same solution. The length of a geodesic anchored to fixed left and right boundary times then does not uniquely described the state; geodesic gauge has broken down. 

We will see in the next section that a similar phenomenon, albeit one that is inherently Euclidean, can explain the discrepancy between the exact answer for the ground state energy as given by eq. \eqref{eqn:E0exact} and the answer predicted by a canonical analysis. It would be interesting, however, to understand how to incorporate this Lorentzian breakdown into the framework of \cite{IliKruTurVer20} and to see if this might lead to a Lorentzian derivation of the exact ground state energy.

In this work, we have been interested in classical black hole interiors that look drastically different from those in pure JT gravity. For the purposes of canonical quantization, however, it may be less confusing to consider the theory with dilaton potential $W_{\lambda}(\Phi)$ where now $\lambda <0$. In this case, $W_{\lambda}(\Phi)$ looks qualitatively just like the JT pre-potential with a single unique minimum. In this case, geodesic gauge does not breakdown for any energy and we can proceed just as in JT gravity. Still, however, we find a discrepancy between the exact ground state energy from eq. \eqref{eqn:E0exact} and that predicted by canonical quantization in geodesic gauge. We will come back to this case in Sec. \ref{sec:lengthambiguity} and Sec. \ref{sec:discspec}. We will argue that indeed geodesic gauge still breaks down in these theories but in a way that requires the use of complex geometries. In this case, it is hard to imagine a way of deriving the exact ground state energy for $\lambda <0$ by only considering purely real, Lorentzian geometries. Instead, we can interpret this discrepancy as a modification of the inner product of the theory as suggested in \cite{IliLevLinMaxMez24}.


\section{Corrections to the length basis inner product}\label{sec:lengthambiguity}

In the previous section, we explained how a canonical analysis naively disagrees with the $E_0^{\rm exact}$ predicted by eq. \eqref{minimal_string_F}. It is important to understand what we got wrong in applying the canonical framework. What we will argue in this section is that we assumed that the geodesic length basis is delta function normalized, which is a false assumption. This non-trivial overlap comes from spacetime geometries which have the same topology as the disk but also contain a closed geodesic loop in the spacetime.
\begin{figure}
    \centering
    \includegraphics[scale=.6]{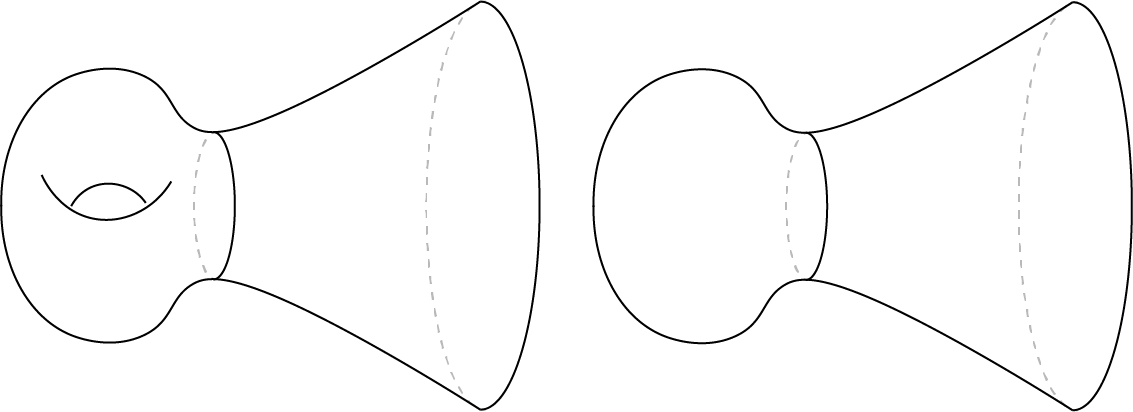}
    \caption{Two types of geometries that are summed over in more general theories of dilaton gravity. In JT gravity, however, only geometries like the one on the left appear, which have higher genus than the disk and so are topologically suppressed. In the theories considered in this work, one also needs to include contributions from geometries like that on the right, which appear at the disk level.}
    \label{fig:highergenus}
\end{figure}

In JT gravity, fixed geodesic length states have overlaps, denoted by $\braket{\ell | \ell'}$, that are defined by the Euclidean gravitational path integral with fixed geodesic length boundary conditions \cite{IliLevLinMaxMez24}. The geometries that contribute to such a path integral for $\ell \neq \ell'$ are non-perturbatively small in $S_0$. The non-zero overlap comes from spacetimes which contain a trumpet geometry. In JT gravity though, the trumpet can only be glued to other trumpets or a pair of pants, and so $\braket{\ell| \ell'}$ only receives contributions from spacetimes with higher topology. What we will see in this section is that in more general dilaton potentials the trumpet can instead be glued to a geometry which caps off and so has the topology of the disk. We will refer to such geometries with the topology of the disk but with a closed geodesic as \emph{pacifier spacetimes} for their resemblance to a baby's pacifier. 

We begin by discussing the trumpet geometry with closed geodesic length $b$ and fixed energy boundary conditions in a general model with potential $W$ such that $W$ has an outermost minimum. We will then discuss geometries with the topology of the disk that also end on a closed geodesic of length $b$ for the specific potential $W_{\lambda}$ with $0< \lambda < \lambda_c$. We will comment on these types of geometries for the case $\lambda <0$. Finally, we sew these two geometries together to form a pacifier and show how such geometries lead to a non-trivial inner-product between fixed length states.

As discussed in Sec. \ref{sec:dilgravrev}, the classical solutions of dilaton gravity are labeled by the energy or size of the thermal circle at infinity. Given these asymptotic parameters, the geometry is completely determined by the dilaton pre-potential via equations \eqref{eqn:onshellmetric} and \eqref{eqn:metricprepot}. Let the outermost minimum of this potential be at dilaton value $\Phi_{0}$ with associated energy $E_{0} = W(\Phi_{0})$. One can consider classical solutions with asymptotic energy $E < E_{0}$.\footnote{In fact, just such solutions were considered in \cite{AnnHar22, AnnGalHof18}.}

We have shown in this paper, however, that such geometries in the $G_N \to 0$ limit have energy below the ground state energy of the dual matrix model, since $E^{\rm exact}_0 = E_{0}$ up to non-perturbatively small corrections. Apparently, such classical geometries are excluded from the matrix model spectrum. This does not mean that the matrix model does not know about these geometries. Rather, they are present in the quantum effects of the model. As we now show, these types of solutions can contribute to the inner product of geodesic length states. An equivalent way of stating this issue is that for general dilaton potentials geodesic gauge is non-perturbatively ill-defined due to the possibility of nucleating a closed geodesic in the geometry. We explain this in detail now.

\subsection{Trumpet geometries}\label{sec:trumpetgeom}
In our model, we would like to find the dominant contribution to the path integral given two boundary conditions - one a geodesic circle of length $b$ and the other an asymptotic boundary of energy $E$. In fact, we will start with the problem at fixed boundary length/temperature and then inverse Laplace transform the answer back to fixed energy. 

We now show that there is a fixed temperature saddle which ends on these boundaries. Consider saddles with metric of the usual on-shell form
\begin{align}\label{eqn:trumpetmetric}
    ds^2 = (W(\Phi) - E) d\tau^2 + \frac{d\Phi^2}{W(\Phi)- E}, 
\end{align}
but where $E < E_0$. In Euclidean signature, the geodesic equation is equivalent to 
\begin{align}
\left(\partial_u \Phi(u) \right)^2 = W(\Phi(u)) - E - P_E^2,
\end{align}
where $P_E$ is the Euclidean momentum of the geodesic and $u$ is the proper length coordinate along the geodesic. We remind the reader that this equation is just that of a classical particle with coordinate $\phi$ in a potential $-W(\Phi)$ and with total energy $-E-P_E^2$.

Using this equation, we see that there is an unstable equilibrium point for this particle at $\Phi_0$, where $W(\Phi_0) = E_0$. This corresponds to a geodesic that just sits at $\Phi_0$ and so is a closed geodesic. The momentum of the geodesic that sits at this point is given by
\begin{align}
    P^2_E = E_0 - E > 0.
\end{align}
The length of this geodesic can then be related to $\beta$ and $E$ by the equation
\begin{align}\label{eqn:trumpetenergy}
   b = \beta \sqrt{E_0-E}.
\end{align}

We would now like to calculate the contribution of the trumpet to the gravity path integral. The on-shell action of the solution in \eqref{eqn:trumpetmetric} with energy given by \eqref{eqn:trumpetenergy} is quite simple. We have
\begin{align}
    S &= \frac{1}{2G_N} \int_{M} d^2 x \sqrt{g} \left(\Phi R + U(\Phi) \right)+ \frac{\Phi_b}{G_N} \int_{\partial M} \sqrt{h} K. \nonumber \\
    & = \frac{\beta}{2G_N} \int_{\Phi_0}^{\Phi_b} d\Phi (-\Phi W''(\Phi) + W'(\Phi)) + \frac{\Phi_b}{G_N} \int_{\partial M} \sqrt{h} K.
\end{align}
Integrating by parts in $\Phi$, using that $W(\Phi_0) = E_0$ and the expression for the on-shell value of $K$ in terms of $E$, we arrive at the total on-shell action
\begin{align}
     S= -\frac{\beta E_0}{G_N} + \frac{\beta E}{2G_N}= -\frac{1}{2G_N} \left(\beta E_0 + \frac{b^2}{\beta}\right).
\end{align}
In the second equality, we use the expression $\eqref{eqn:trumpetenergy}$ for $E$ as a function of $b, \beta$.

Since later we will be interested in low energies, it is important to also include the one-loop contribution from the boundary mode, which should become strongly coupled at low energies and temperatures. We do the one-loop calculation in Appendix \ref{app:schwloops}, and, assuming that the measure for the boundary mode is modified only up to $\beta$-independent constants from the JT gravity answer, we find that the trumpet with geodesic boundary of length $b$ and boundary of renormalized length $\beta$ contributes
\begin{align}
    Z_{trumpet}(b, \beta) \sim \frac{1}{\beta^{1/2}} e^{-\frac{\beta E_0}{2G_N} - \frac{b^2}{2\beta G_N}},
\end{align}
up to $\beta$-independent order one factors. Again, the determination of the order one factors would require a more careful analysis of the boundary particle measure that we leave for future work.

Inverse Laplace transforming, we land on the JT gravity answer but with an energy shifted by $E \to E - E_0$. This looks as if the ground state of the model has shifted to that predicted by canonical quantization. This answer in energy space then becomes
\begin{align}
    Z_{\mu, trumpet}(E,b) \sim \frac{1}{\sqrt{\pi (E-E_0)}} \cos\left(\frac{b \sqrt{E-E_0}}{G_N} \right),
\end{align}
up to energy-independent prefactors. As in JT gravity, this answer diverges at the shifted ground state energy, $E = E_0$.

\subsection{Disk geometries ending on a closed geodesic and sewing the pacifier}\label{sec:diskclosedgeo}
\begin{figure}
    \centering
    \includegraphics[scale=.8]{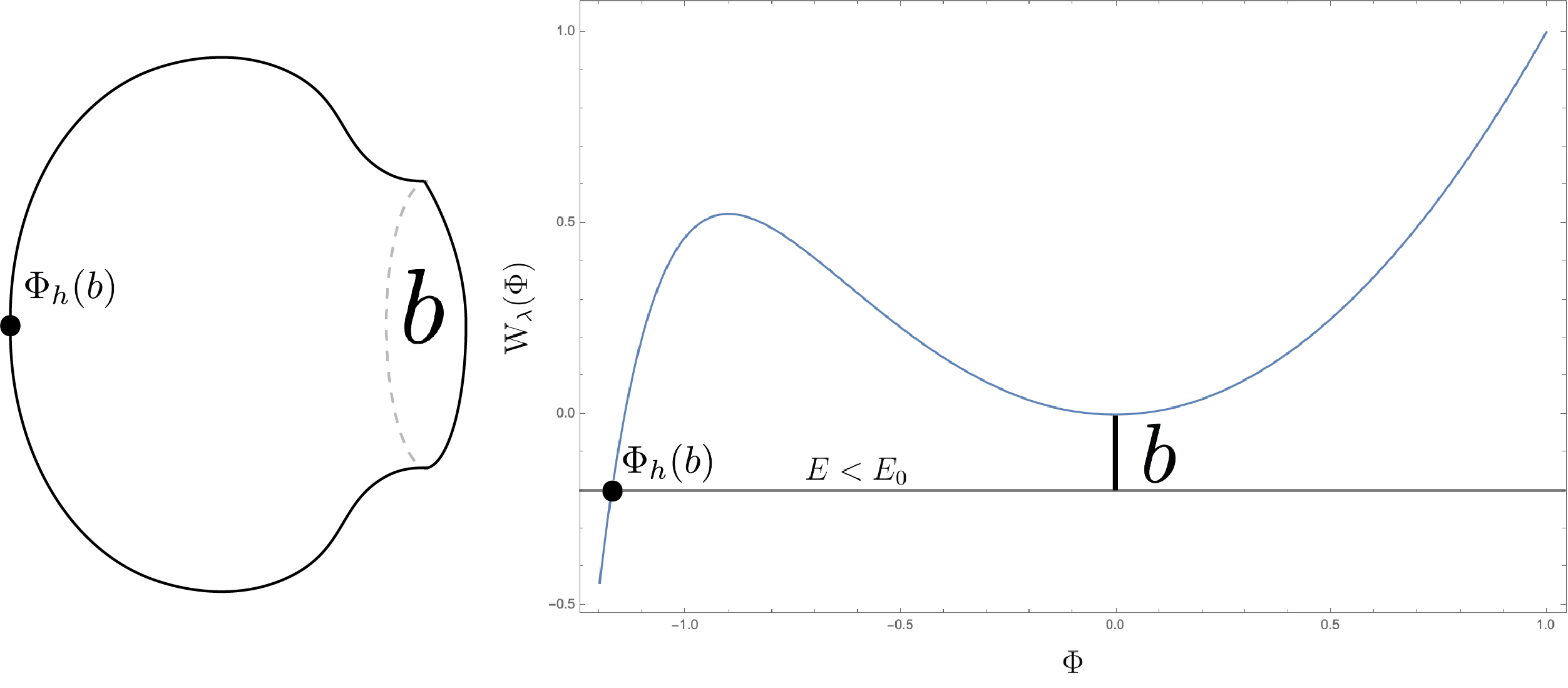}
    \caption{On the left, we illustrate a geometry with a closed geodesic of length $b$ and the topology of a disk. On the right we point out how this cap can be associated with a piece of a solution with energy $E<E_0$. We remind the reader that the difference between the black horizontal line and the curve of $W_{\lambda}(\Phi)$ is the metric component $g_{\tau \tau}$. The dilaton at the horizon $\Phi_h(b)$ lies on the third real branch of the function $W^{-1}$.}
    \label{fig:paccap}
\end{figure}
Given the answer for the trumpet with geodesic boundary of length $b$, we would like to sew in a disk geometry with a closed geodesic of length $b$. This is most explicit for the potential $W_{\lambda}(\Phi)$ with $0< \lambda < \lambda_c$. Again we look for solutions of the form in eq. \eqref{eqn:trumpetmetric} with $E< E_0$. In this energy range, the solution can cap off at the dilaton value given by $W_{\lambda}(\Phi) = E$, where now $\Phi$ is on the innermost real branch of $W_{\lambda}^{-1}(E)$. The equation relating $b$ to $\beta$ and $E$ in \eqref{eqn:trumpetenergy} still holds. We get one more equation by demanding that the geometry end at a smooth cap. Going through the usual steps, this equation reads
\begin{align}\label{eqn:diskgeotemp}
    \beta = \frac{4\pi}{U(\Phi_h)},
\end{align}
where $\Phi_h(b)$ and $E(b)$ are related via the equation 
\begin{align}
    W_{\lambda}(\Phi_h(b)) = E(b).
\end{align} 
These two equations together with equation \eqref{eqn:trumpetenergy} can be used to solve for $\beta(b), E(b)$ and $\Phi_h(b)$ in terms of $b$.

We plug this solution into the action to solve for the on-shell action. In this case, the only boundary term would come from the boundary at the closed geodesic, but this vanishes because $K = 0$ there. We are thus left with just the bulk term which gives
\begin{align}\label{eqn:diskwithgeoaction}
    S_{disk}(b) &= \frac{1}{2G_N} \int d^2 x \sqrt{g} \left(-\Phi W_{\lambda}''(\Phi) + W_{\lambda}'(\Phi)\right) \nonumber \\
    & =  \frac{2\pi \Phi_h(b)}{G_N} - \frac{\beta(b)}{G_N} (W_{\lambda}(\Phi_h(b))- E_0) \nonumber \\
    & = \frac{1}{G_N} \left(2\pi \Phi_h(b) + b \sqrt{E_0 - W_{\lambda}(\Phi_h(b))}\right)
\end{align}

For the potential $W_{\lambda}(\Phi)$, analytic formulae for $\beta(b)$ and $E(b)$ are not possible for general $b$. We can instead solve for them in various limits. In particular, in the limit that $ b\to 0$ then $E \to E_0$ but $\beta$ goes to a constant so that the second term in eq. \eqref{eqn:diskwithgeoaction} vanishes. The constant that $\beta$ approaches is set by $U(\Phi)$ at the innermost solution to the equation $W_{\lambda}(\Phi) = E_0$. As we increase $b$, $E(b)$ decreases and so $\Phi_h$ decreases.

In the opposite regime where $b$ becomes large, equation  \eqref{eqn:trumpetenergy} tells us that $E$ becomes more negative so that the equation $W_{\lambda}(\Phi_h) = E$ leads to $\Phi_h \approx -\frac{1}{2\pi (1-\alpha)} \log \frac{|E|}{\lambda}$, but then equation \eqref{eqn:diskgeotemp} tells us that $\beta \sim 1/|E|)$. Plugging this back into \eqref{eqn:trumpetenergy}, we see that in fact $b \sim 1/\sqrt{|E|}$ at large $E$, which is in contradiction with $b$ getting large. Therefore at large $b$ there seem to not exist saddles of the type we found at small $b$. Since we will mostly be interested in the small $b$ limit here, we just point out that we have not analyzed all possible classical saddles which have the topology of the disk and a closed geodesic of length $b$. Perhaps there are more exotic classical saddles at large $b$ where the dilaton at the horizon is complex, akin to what we discussed in Sec. \ref{sec:exactdos}. We leave further exploration of these geometries for future work. 

We now briefly comment on these geometries for $\lambda <0$. We can start again with a solution with energy $E(b) < E_0$. In this case, there is no obvious real geometry that ends on the geodesic of length $b$. That said, there are infinitely many complex geometries that end on the complex roots of the equation $W(\Phi(b)) = E(b)$. Just as the disk-sphere saddle in JT gravity, these can be understood by analytically continuing the dilaton contour. Using this contour to evaluate the on-shell action in fact will give the same answer as in eq. \eqref{eqn:diskwithgeoaction}, but now with $\Phi_h(b)$ complex.

Finally, note that in this analysis we have not analyzed the one-loop contributions to this disk geometry with a closed geodesic. Since we have not analyzed these factors, we should not expect to match the one-loop contribution to the exact answer for the shift in the ground state energy. We leave a more careful one-loop analysis for future work, perhaps using the recent techniques of \cite{BloMerYao23}.

\begin{figure}
    \centering
    \includegraphics[scale=.6]{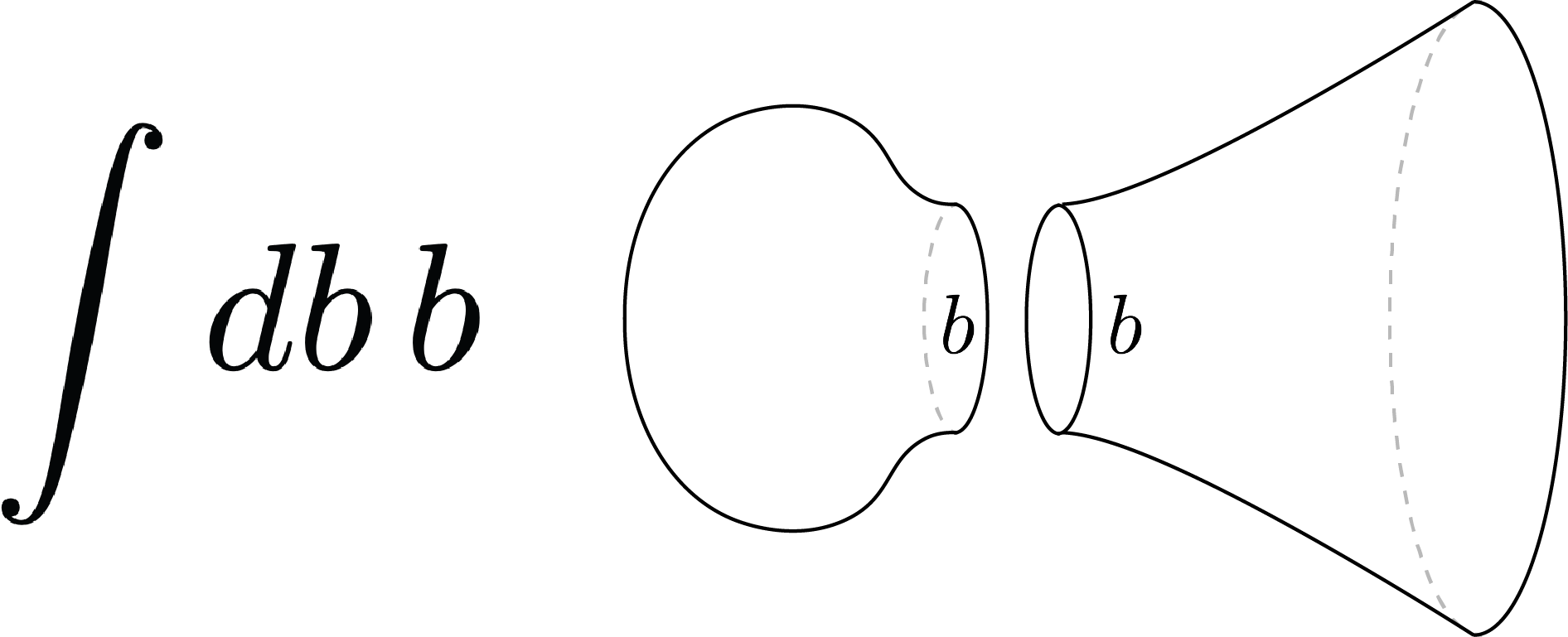}
    \caption{The cap with a closed geodesic of length $b$ is sewn to the exterior trumpet along the closed geodesic.}
    \label{fig:sewnpac}
\end{figure}
Nevertheless, having constructed the trumpet and the disk bounded by a closed geodesic, we now sew these two spacetimes to form the pacifier, akin to what is done in JT gravity. We glue the two spacetimes along the closed geodesic and then integrate over its length, $b$. There is of course a question of the measure for this integral. Since we expect there to again be a zero-mode corresponding to twisting the disk-gometry relative to the trumpet, we expect the measure to again be the same as in JT gravity.\footnote{Note that the measure for $b$ is an informed guess and we have not derived it.} All in all we get that these partially on-shell and partially off-shell geometries contribute
\begin{align}\label{eqn:rhopacifier}
    \rho_{\rm pacifier}(E) \sim \frac{1}{\sqrt{\pi (E-E_0)}}\int_0^{\infty} db b \ \cos \left(\frac{b \sqrt{E-E_0}}{G_N}\right)  \times Z_{disk}(b),
\end{align}
up to energy independent prefactors. Here, $Z_{disk}(b)$ is given by the exponential of the action in \eqref{eqn:diskwithgeoaction}. 

\subsection{The pacifier contribution to $\braket{\ell | \ell'}$}
Given eq. \eqref{eqn:rhopacifier} for the contribution to the density of states from pacifier-type spacetimes, we can ask how the pacifier affects the inner product in the geodesic length basis. First, we point out that the trumpet does in fact contribute to the length inner product. This is because for any two points on the asymptotic boundary of the trumpet, there are two geodesics connecting those points. In order to show this, we do an analysis of the geodesics on the trumpet in App. \ref{sec:trumpetgeodesics}. 

However, supposing access to the length basis wavefunctions $\psi_E(\ell)$ defined in Sec. \ref{sec:canquant}, we can easily get the length basis inner-product from expression eq. \eqref{eqn:rhopacifier}. To see this, we can use the fact that the fixed energy wavefunctions $\psi_E(\ell)$ discussed in Sec. \ref{sec:canquant} form a complete basis in length so that
\begin{align}
    \int_{-\infty}^{\infty} d\ell \psi_E^* (\ell) \psi_{E'}(\ell) = \frac{\delta(E - E')}{\rho_0(E)}
\end{align}
where $\rho_0(E)$ is the density of states predicted by canonical quantization. Using this relation, we see that 
\begin{align}
    \braket{\ell| \ell'} \supset \int_{E_0}^{\infty} dE\frac{1}{\sqrt{\pi (E-E_0)}}\int_0^{\infty} db b \ \cos \left(\frac{b \sqrt{E-E_0}}{G_N}\right)  \times Z_{disk}(b) \times \psi_E^*(\ell) \psi_E(\ell').
\end{align}

In JT gravity, at the level of the disk topology, this inner product is trivial, $\braket{\ell| \ell'} = \delta (\ell - \ell') + \mathcal{O}(e^{-S_0})$, since the only purely AdS$_2$ geometry with boundary conditions given by two fixed length geodesics ending on the same asymptotic time is the trivial geometry. In other words, at infinite $S_0$, the analog of $Z_{disk}(b)$ is $0$.\footnote{Or really, the analog of the wavefunction $Z_{disk}(b)$ averages to zero in JT gravity, when JT gravity is viewed as an ensemble of dual quantum theories. This fact that in a specific member of the JT ensemble, $Z_{disk}(b)$ is a random-number perhaps suggests a connection between fixed members of the ensemble and deformed dilaton gravity with random couplings, as was suggested in \cite{BloIliKru22, BloKru22}. In fact, the deformations considered in \cite{BloIliKru22, BloKru22} are of the exponential type considered here, but with complex $\alpha$, since they correspond to opening up geodesic holes in the spacetime.} If, however, one allows surfaces of higher topology, this inner product can acquire off-diagonal contributions. This fact was used to argue that at late times a black hole in JT gravity may tunnel into a white hole \cite{StaYan22}. See \cite{IliLevLinMaxMez24} for further discussion of this point. This means that if one works at large $S_0$ in JT gravity, the length basis is a ``good'' basis in the sense that length states of different length $\ell$ are independent. We have seen here that with more general dilaton potentials, the length basis is a bad basis already at the disk level since $Z_{disk}(b)$ is non-zero.

\subsection{Pacifiers and the non-perturbative shift in the ground state energy}

We now show that the expression for $\rho_{\rm pacifier}(E)$ in eq. \eqref{eqn:rhopacifier} contributes an enhancement to the full density of states near $E \approx E_0$. We interpret this as a non-perturbatively small shift in the ground state energy of the model due to the pacifier spacetime. We compare with the exact answer found Sec. \ref{sec:exactdos}.

It is clear that the expression in \eqref{eqn:rhopacifier} diverges like $1/\sqrt{E - E_0}$ at small $\delta E \equiv E - E_0$ so that 
\begin{align}
    \rho_{\rm pacifier}(E) \approx \frac{c(\lambda,\alpha)}{\sqrt{E - E_0}} + \mathcal{O}((E- E_0)^0).
\end{align}
We would like to compute the coefficient $c(\lambda,\alpha)$. This requires us to do the $b$ integral in expression \eqref{eqn:rhopacifier} at small $\delta E$. Since we are in the semi-classical, $G_N \to 0$ limit, we could can look for a saddle in $b$. 

Treating each branch of the cosine in \eqref{eqn:rhopacifier} separately and using the expression for $S_{disk}(b)$ in \eqref{eqn:diskwithgeoaction}, we find that the saddle point equation for $b$ is 
\begin{align}
    2\pi \Phi_h'(b) \pm i \sqrt{E - E_0} + \sqrt{E_0 -W(\Phi_h(b))} - \frac{b U(\Phi_h) \Phi_h'(b)}{2\sqrt{E_0 - W(\Phi_h(b))}} = 0.
\end{align}
Using the expression for $\beta$ given in \eqref{eqn:diskgeotemp} together with equation \eqref{eqn:trumpetenergy}, we get that the first and last terms cancel. So the saddle point equation for $b$ just sets 
\begin{align}
    E = W(\Phi_h(b)). 
\end{align}
This is not surprising since integrating over $b$ should just enforce a jump condition that $\Delta \partial_n \phi =0$ at the junction where the two geometries were sewed. Such a jump condition indeed enforces that $E = W(\Phi_h(b))$. After setting $b$ on-shell, we see that our geometry just becomes that associated to the complex saddle of the type discussed in Sec. \ref{sec:bulkinterp}.

In the end, we have 
\begin{align}
    \rho_{\rm pacifier} \sim \frac{1}{\sqrt{E-E_0}} \exp \left( \frac{2\pi W^{-1}(E_0)}{G_N}\right) \times \left(\text{one-loop}\right),
\end{align}
where the one-loop factor comes from fluctuations about the pacifier geometry. Combining this with the prediction from canonical quantization for the low-energy density of states given in \eqref{eqn:lowedoscanquant}, we get that the non-perturbative shift in the ground state energy takes the form
\begin{align}
\delta E \sim (\text{one-loop}) \times e^{2\pi \left(W^{-1}_{inner}(E_0) - W^{-1}_{outer}(E_0)\right)/G_N}
\end{align}
where $W^{-1}_{inner/outer}(E_0)$ is the inner/outer-most root of the equation $W(\Phi) = E_0$. This agrees with the expression for the shift in the ground state energy from evaluating the exact density of states in eq. \eqref{eqn:exactshift}, up to one-loop factors. Again, we have not done a thorough analysis of the one-loop factors for the pacifier geometries, but this is obviously an important calculation for the future.



\section{Classical dilaton gravity from a gas of very blunt defects}\label{sec:bluntexpansion}
So far we have explored the classical solutions of dilaton gravity and their contributions to the path integral in the $G_N \to 0$ limit. As explained in Sec. \ref{sec:intro} and outlined in \cite{MaxTur21,Wit20MM,EbeTur23}, an alternative route can be taken for the class of dilaton potentials we consider, which are deformed away from JT gravity by an exponential in the dilaton - $e^{-2\pi (1-\alpha) \phi}$ (or linear combinations thereof with different powers $\alpha$). This route amounts to expanding perturbatively in these deformations and integrating insertions of exponential operators over the spacetime.

More explicitly, for a dilaton potential of the form 
\begin{align}
    U(\Phi) = 2 \Phi + 2\pi (1-\alpha) \lambda e^{-2\pi (1-\alpha) \Phi}
\end{align}
one can expand the partition function at small $\lambda$. At order $L$ in the $\lambda$ expansion, the gravity partition function receives a correction of the form
\begin{align}\label{eqn:defectexpansion}
     Z_{grav} \vert_{\mathcal{O}(\lambda^L)} \sim \frac{\left(2\pi (1-\alpha) \lambda\right)^L }{L!} \int D\Phi Dg \  \int  \left(\prod_{i=1}^L d^2 x_i\right) \exp \left(-I_{JT} - \sum_{i=1}^L 2\pi (1-\alpha) \Phi(x_i) \right)
\end{align}
with $I_{JT}$ given by 
\begin{align}\label{eqn:unrescaledaction}
I_{JT} = -\frac{S_0}{2\pi} \left( \frac{1}{2} \int_M \sqrt{g} R + \int_{\partial M} \sqrt{h} K\right) - \frac{1}{2} \int_M \sqrt{g} \Phi(R+2) - \int_{\partial M} \sqrt{h} \Phi(K-1).
\end{align}

Note that for the moment, we have a different normalization for the action than what we discussed in eq. \eqref{eqn:bulkaction} in Sec. \ref{sec:intro}. The normalization in \eqref{eqn:unrescaledaction} matches the convention of \cite{TurUsaWen21}. To compare with previous sections, we would like to work in a saddle point approximation to the dilaton gravity path integral. To do this, it is helpful to re-scale $\Phi \to \Phi/G_N$, with $G_N$ some small, dimensionless number. We can think of $G_N$ as controlling the size of the boundary value of the dilaton, which in JT gravity controls for example the Schwarzian coupling. As described in the introduction, to get classical dilaton gravity, we should then re-scale the couplings
\begin{align}
    &\lambda 2\pi (1-\alpha) \to \frac{\lambda 2\pi (1-\alpha)}{G_N}, \nonumber \\
    & (1-\alpha) \to G_N (1-\alpha).
\end{align}
In the $G_N \to 0$ limit, this corresponds to making the defect insertions very blunt so that they do not perturb the geometry very much. We will see that the saddle points of dilaton gravity are built up out of many (order $1/G_N$) blunt defects. As we now show, the smooth geometries that dominate the semi-classical dilaton gravity path integral can then be thought of as made up of a dense gas of very blunt defects. 

In this spirit, we introduce another field, which is the defect density field
\begin{align}
\rho_0(x) = \frac{1}{L} \sum_{i=1}^L \delta^2 (x-x_i)/\sqrt{g}.
\end{align}
We would then like to focus on terms in the defect expansion in \eqref{eqn:defectexpansion} with fixed defect number $L$, where $L$ scales with $1/G_N$ as 
\begin{align}
    L \sim \kappa/G_N
\end{align}
with $\kappa$ fixed. For such terms, we find that \eqref{eqn:defectexpansion} becomes
\begin{align}
\frac{(\lambda 2\pi (1-\alpha))^L}{G_N^L L!}\int D\Phi Dg \int \prod_i d^2 x_i \sqrt{g_i} \exp \left(\frac{1}{2G_N}\int_M \sqrt{g} \left(\Phi R + 2\Phi- 4\pi (1-\alpha)\kappa \rho_0(x) \Phi(x)\right)\right)
\end{align}
where $\kappa \equiv L G_N$ is fixed as we take $G_N \to 0$ and $L \to \infty$. For now, we are ignoring the Gibbons-Hawking boundary term in the action.

To proceed, we need to deal with the measure $\int \prod_i d^2 x_i$. We follow a standard procedure given in \cite{LivNovViv17} and insert the identity by writing
\begin{align}
\int D\rho(x) \  \delta \left(\rho(x) - \rho_0(x)\right) =1
\end{align}
and then exchanging the order of integration. Writing the delta function in terms of an auxiliary density, $\hat{\rho}(x)$, we get
\begin{align}
\int D\hat{\rho}(x) \int \prod_i\left( d^2 x_i \sqrt{g_i}\right) \exp \left(i \int_M d^2 x \sqrt{g} \ L\hat{\rho}(x) \left(\rho(x) - \frac{1}{L}\sum_i \frac{\delta^2 (x-x_i)}{\sqrt{g}}\right)\right) 
\end{align}
This becomes
\begin{align}
&\int D\hat{\rho}(x)  \exp \left(i \int_M d^2 x \sqrt{g} \ L\hat{\rho}(x)\rho(x) \right) \left(\int d^2 y \sqrt{g} e^{-i\int d^2 x \hat{\rho}(x) \delta^2 (x-y)}\right)^L \nonumber \\
&=\int D\hat{\rho}(x)  \exp \left(i \int_M d^2 x \sqrt{g} \ L\hat{\rho}(x)\rho(x) + L\log \left(\int d^2 y \sqrt{g}\  e^{-i\hat{\rho}(y)} \right)\right) 
\end{align}
Evaluating this integral by saddle point to eliminate the $\hat{\rho}$ integral, we get that the saddle point equation is
\begin{align}
\rho(x) = \frac{e^{-i\hat{\rho}(x)}}{\int d^2 y \sqrt{g} \ e^{-i\hat{\rho}(y)}}
\end{align}
Plugging this solution back into the action we end up with the integral
\begin{align}
=\frac{(\lambda 2\pi (1-\alpha))^L}{G_N^L L!}\int D\Phi Dg D\rho_{=} \exp \left(\frac{1}{2G_N}\int_M \sqrt{g} \left(\Phi R + 2\Phi- 4\pi (1-\alpha)\kappa \rho(x) \Phi(x) - 2\kappa \rho(x) \log \rho(x)\right)\right)
\end{align}
where by $D\rho_{=}$ we mean integrating over all unit normalized $\rho$, $\int d^2 x \sqrt{g} \rho(x) = 1$. 

The term in the action proportional to $\rho \log \rho$ came from the path integral measure upon changing from an integral over $x_i$'s to an integral over $\rho$. This term has a simple interpretation in terms of the statistical entropy. It counts the number of microscopic configurations of defects for a given density profile, $\rho$. This entropic term creates a statistical pressure for the defects to not be on top of each other. On the other hand, due to the coupling to the dilaton, there is a pressure to bring the defects together, toward the minimum of the dilaton. As advertised, and shown explicitly below, these two forces balance to give stable solutions.

To deal with the normalization condition on the $\rho$ field, we can insert a Lagrange multiplier to enforce this condition, and so the path integral at fixed defect number becomes
\begin{align}\label{eqn:fixedLdensityaction}
=&\frac{(\lambda 2\pi (1-\alpha))^L}{G_N^L L!} \int D\Phi Dg D\rho \,dz \times \nonumber\\
&\exp \left(\frac{1}{2G_N}\int_M \sqrt{g} \left(\Phi R + 2\Phi- 4\pi (1-\alpha)\kappa \rho(x) \Phi(x) - 2\kappa \rho(x) \log \rho(x)+ 2i z \kappa (\rho(x) -1)\right)\right).
\end{align}
This is the JT path integral with a large number, $L$, of defects on the background, ignoring higher order contributions from the measure. We now want to find saddle points for this integral.

\subsection{Classical solutions for the defect density at fixed boundary length and fixed defect number}
Varying the action in \eqref{eqn:fixedLdensityaction} with respect to the $\rho, z$ and $\Phi$ gives
\begin{align}\label{eqn:denseom}
&2\pi (1-\alpha) \Phi + \log \rho + 1 + iz = 0, \nonumber \\
&\int d^2 x \sqrt{g} \rho(x) = 1, \nonumber \\
& R = -2 + 4\pi (1-\alpha) \rho(x).
\end{align}
The solution to the first two of these equations is just
\begin{align}
\rho(x) = Ce^{-2\pi (1-\alpha)\Phi(x)}
\end{align}
with $C$ determined by the normalization condition
\begin{align}\label{eqn:normcondition}
\int d^2 x \sqrt{g} \rho(x) = \int d^2 x \sqrt{g} Ce^{-2\pi (1-\alpha)\Phi(x)} =1.
\end{align}
Before we solve for the metric, we have already learned something interesting. Classically the defects bunch up exponentially toward the region of more negative dilaton.\footnote{This exponential decay of the density is exactly analogous to the Debye screening effect in charged metals at finite temperature. We thank Steve Shenker for emphasizing this point to us.}

To solve for the metric, we follow the steps laid out in Sec. \ref{sec:dilgravrev} and \cite{Wit20PT} and again look for solutions of the form 
\begin{align}\label{eqn:metricansatz}
    ds^2 = A(r)d\tau^2 + \frac{dr^2}{G(r)}
\end{align}
where we make the identification $\tau \to \tau + \beta$. Varying the action with respect to $G$ and then using a gauge freedom to set $G = A$, we get
\begin{align}\label{eqn:dilatoneom}
   \Phi'(r) A'(r) = 2\Phi(r) -4\pi (1-\alpha) \kappa \rho(r) \Phi(r).
\end{align}
Plugging in the ansatz in \eqref{eqn:metricansatz} and varying the action in \eqref{eqn:fixedLdensityaction} with respect to $A$ further gives
\begin{align}
\Phi'' = 0.
\end{align}
We see that just as before we can set $r = \Phi$.

Assuming that the horizon occurs at $\Phi = \Phi_h$ and using the form of the metric in \eqref{eqn:metricansatz}, the normalization condition in \eqref{eqn:normcondition} implies
\begin{align}
C = \frac{2\pi (1-\alpha)e^{2\pi (1-\alpha) \Phi_h}}{\beta}.
\end{align}
Plugging in for $\rho$ in \eqref{eqn:dilatoneom} and integrating twice, the metric component $A(\Phi)$ then takes the form
\begin{align}
    A(\Phi) = W_{\lambda(\kappa)}(\Phi) - E.
\end{align}
with $E$ an integration constant, which one can check is the AdM mass of the solution. Here we have defined the fixed-$\kappa$ pre-potential as 
\begin{align}\label{eqn:fixedkappaprepot}
    W_{\lambda(\kappa)}(\Phi) = \Phi^2 -\lambda(\kappa) e^{-2\pi (1-\alpha)\Phi},
\end{align}
where we have defined an effective coupling at fixed $\kappa$ to be
\begin{align}\label{eqn:lambdakappa}
   \lambda(\kappa,\beta) \equiv \frac{\kappa C}{\pi (1-\a)} = \frac{2\kappa}{\beta}e^{2\pi (1-\alpha) \Phi_h(\kappa,\beta)}.
\end{align}
One should think of eq. \eqref{eqn:fixedkappaprepot} as the effective dilaton pre-potential induced by a $\kappa/G_N$ number of defects. Note that the effective coupling of the theory depends on the asymptotic boundary length $\beta$. This dependence is a result of the non-local condition on the theory that the defect density be normalized to one, eq. \eqref{eqn:normcondition}.

Demanding that this metric is smooth at the horizon enforces that
\begin{align}\label{eqn:betaphih}
    \beta =\frac{(2\pi - 2\pi (1-\alpha) \kappa)}{\Phi_h}.
\end{align}
A way to interpret this equation is that the dilaton at the horizon $\Phi_h$ is the generator of rotations and so $2\pi - \beta \Phi_h$ is the conical deficit of the geometry.\footnote{It may be confusing to the reader that we are demanding that the horizon be smooth but yet are interpreting the solution in terms of defects, which are not smooth. This is because we have taken the scaling limit of large defect number and small $(1-\alpha)$, where the blunt defects condense and approximate a smooth geometry.} While in pure JT gravity without defects the solution had a full angle of rotation by $2\pi$, now the defects have reduced this angle by an amount $2\pi (1-\alpha) \kappa$ which in the original, un-rescaled variables is just $2\pi (1-\alpha) L$. Furthermore, note that the energy is related to the dilaton at the horizon, $\Phi_h$, by 
\begin{align}\label{eqn:enatfixedbeta}
    \Phi_h^2 -\frac{2\kappa \Phi_h}{2\pi - 2\pi (1-\alpha) \kappa} = \left( \frac{2\pi (1-(1-\alpha)\kappa)}{\beta}\right)^2 - \frac{2\kappa}{\beta}= E.
\end{align}

We would now like to plug everything in and compute the on-shell action. The total action including boundary terms at fixed, renormalized boundary length $\beta$ is then 
\begin{align}
   \frac{1}{2G_N}\left[ \beta \int_{\Phi_h}^{\Phi_b} d\Phi \left( -\Phi A'' + 2 \Phi - 4\pi (1-\alpha)\kappa \rho \Phi - 2\kappa \rho \log \rho + 2iz \kappa \rho \right) - 2i z\kappa \right] + \frac{\Phi_b}{G_N} \int du \sqrt{h} (K-1) .
\end{align}
One way to evaluate the on-shell action is to integrate by parts on $A$. The boundary term from this integration by parts will kill the other boundary term proportional to the extrinsic curvature. Using that $A'(\Phi_h) \beta = 4\pi$ as well as the equations of motion, we then get that the total on-shell action is
\begin{align}\label{eqn:canonshell}
    &S_{on-shell}(\kappa,\beta) = \frac{1}{G_N}\left( (\pi - \pi (1-\alpha) \kappa)\Phi_h(\beta) + \kappa \log \frac{\beta}{2\pi (1-\alpha)}\right).
\end{align}
Up to one-loop contributions to this answer, which we briefly discuss in App. \ref{app:schwloops}, this exactly matches the answer found by \cite{EbeTur23} if we reintroduce $G_N$'s by sending $(1-\alpha) \to (1-\alpha)/G_N$ and $\kappa = LG_N$. This formula has an interpretation in terms of an ``effective'' defect with opening angle $(1-(1-\alpha)\kappa)$. Indeed, \cite{EbeTur23} found the answer given by \eqref{eqn:canonshell} by doing the exact integral over the moduli space of hyperbolic disks with $L$ defects, which they found localizes to the configuration where all $L$ defects sit on top of each other. The picture in \cite{EbeTur23} then seems different from what we have here since in our picture the defects are spread out over the geometry (albeit with a density decaying exponentially in $\Phi$). We would like to understand the origin of this difference more clearly.


\subsection{Behavior of solutions at large $\kappa$}\label{sec:throatgeometries}
\begin{figure}
    \centering
    \includegraphics[scale=.5]{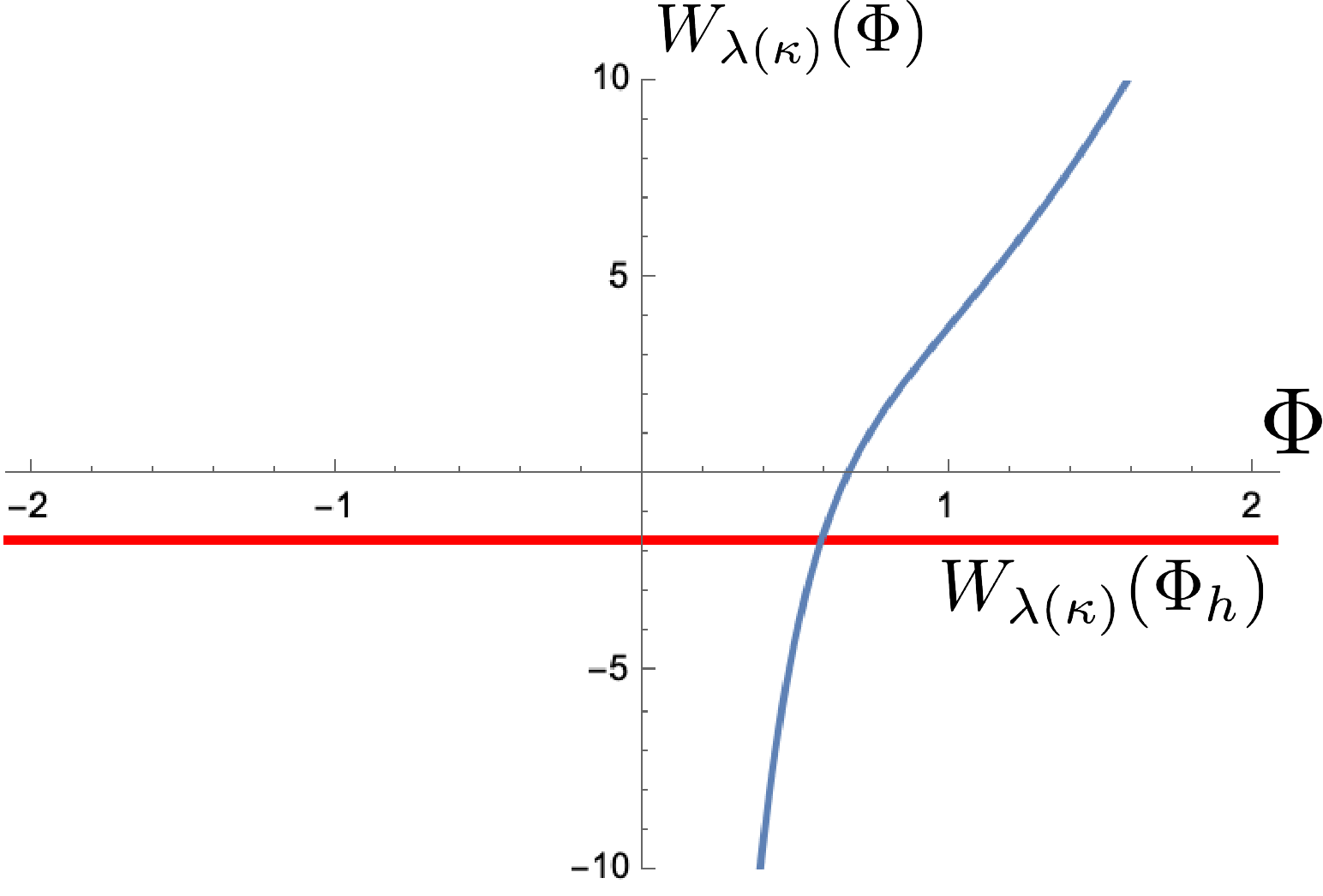}
    \includegraphics[scale=.5]{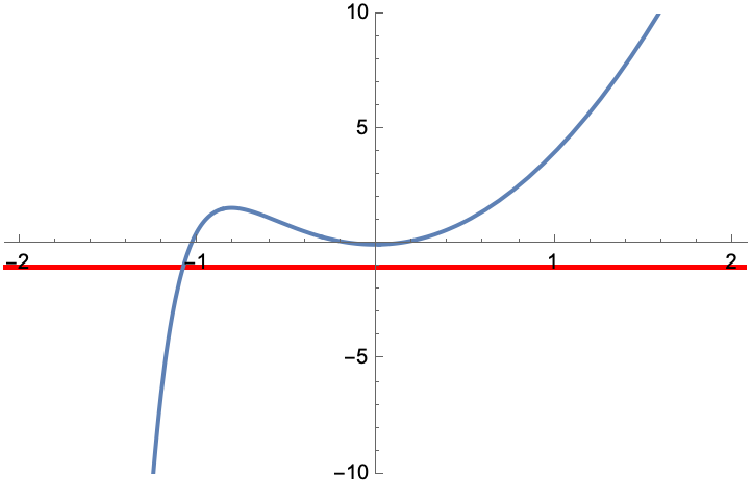}
    \includegraphics[scale=.5]{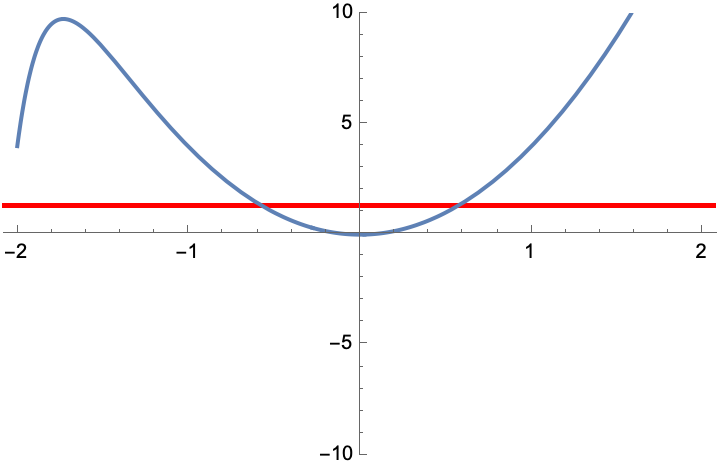}
    \caption{We plot the effective dilaton potential $W_{\lambda(\kappa)}(\Phi)$ for increasing values of $\kappa$ as one moves clockwise from the upper left. There is a critical $\kappa$ where the potential becomes non-monotonic. At this point, the energy of the solution $W_{\lambda(\kappa)}(\Phi_h)$ is below the outermost minimum of the potential and so the geometry develops a throat. At larger $\kappa$, the energy goes above the outermost minimum and the throat pinches off.}
    \label{fig:Wkappa}
\end{figure}
\vspace{-3mm}
As discussed in Sec. \ref{sec:dilgravrev}, something special happens when the number of defects is so large that their net angle deficit is bigger than $2\pi$. Indeed, this number of defects is precisely the order at which the string equation truncates in the $\lambda$ expansion, as shown in eq. \eqref{minimal_string_F}.  Intuitively, this is because the defects have soaked up all the opening angle in the geometry. In response, a closed geodesic nucleates in the geometry akin to what happens in the pacifier spacetimes of Sec. \ref{sec:lengthambiguity}. We will show that indeed one can think of the pacifier as also arising from a dense gas of defects with total defect number larger than $\kappa_c$. To understand this, we would like to see how our density picture sees the development of a throat for increasing defect number.

In our scaling limit, the net deficit angle for the defects becomes greater than $2\pi$ when $2\pi (1-\alpha) L G_N > 2\pi$ or in terms of the re-scaled defect number, $\kappa$, we have
\begin{align}
\kappa > \kappa_c \equiv 1/(1-\alpha).
\end{align} 
In our representation in terms of a smooth density of defects, near $\kappa = \kappa_c$ the spacetime geometry changes dramatically. To see this, we can plot the effective dilaton pre-potential, $W_{\lambda(\kappa)}(\Phi)$, which appears in the metric of our fixed $\kappa$ saddle, which we reproduce here:
\begin{align}
    &W_{\lambda(\kappa)}(\Phi) = \Phi^2 - \frac{2\kappa}{\beta}e^{2\pi (1-\alpha) (\Phi_h(\beta) - \Phi)}, \nonumber \\
    &\Phi_h(\beta) = \frac{2\pi (1-(1-\alpha) \kappa)}{\beta}.
\end{align}
According to these formulae, we can think of dialing $\kappa$ as dialing the coupling in the dilaton potential $W_{\lambda}(\Phi)$ discussed in Sec. \ref{sec:dilgravrev} where here the effective coupling is again
\begin{align}\label{eqn:lambdakappa}
    \lambda(\kappa) = \frac{2\kappa}{\beta}e^{2\pi (1-\alpha) \frac{2\pi (1-(1-\alpha)\kappa)}{\beta}}.
\end{align} In Sec. \ref{sec:dilgravrev}, we found that there was a critical coupling $ \lambda_c = \frac{1}{e 2\pi^2 (1-\alpha)^2}$ above which the pre-potential $W_{\lambda}(\Phi)$ becomes monotonic. From this, we have the following characterization of the fixed-$\kappa$ pre-potential:
\begin{align}
    &W_{\lambda(\kappa)}(\Phi)\ \text{is monotonic if } \lambda(\kappa) > \lambda_c, \nonumber \\
    &W_{\lambda(\kappa)}(\Phi)\ \text{is non-monotonic if } \lambda(\kappa) < \lambda_c.
\end{align}

As we increase $\kappa$ from near zero, one can show that $\lambda(\kappa)$ begins increasing and eventually passes $\lambda_c$. At very large $\kappa$, however, the exponential in eq. \eqref{eqn:lambdakappa} begins to take over and $\lambda(\kappa)$ rapidly goes back below $\lambda_c$. Note that all the while the dilaton at the horizon, $\Phi_h$, decreases according to eq. \eqref{eqn:fixedkappaprepot}. As one can check, once $\lambda(\kappa)$ drops below $\lambda_c$ from above, $\Phi_h$ is in fact less than the dilaton value, $\Phi_{outermin}$, at the outer-most minimum of $W_{\lambda(\kappa)}$. In fact, when $\lambda(\kappa)$ reaches $\lambda_c$ for the second time, one can show that the AdM energy of the solution, $E(\kappa)$, is below $W_{\lambda(\kappa)}(\Phi_{outermin})$. This means the geometry has a throat. As $\kappa$ increases further, the energy of the solution increases until $\kappa$ reaches $\kappa_{\text{pinch}}$ where the throat degenerates and the geometry pinches off. The geometry can be continued in $\kappa$ beyond this point but it becomes complex. We are not sure how seriously to take the saddle when the throat has degenerated. We suspect that by this point the saddle has left the integration contour. The evolution of $W_{\lambda(\kappa)}(\Phi)$ as $\kappa$ is increased is illustrated in Fig. \ref{fig:Wkappa}.

In short, as we crank up the defect number fixing the boundary temperature, the density of defects builds up near the horizon, creating a throat that eventually pinches off. As discussed in the next subsection, we expect that this pinching off effectively truncates the integral over $\kappa$ in eq. \eqref{eqn:kaction} rendering the full answer finite, although this deserves further study.


\subsection{Finding the saddle point in $\kappa$}

Having computed the contribution to the fixed boundary length path integral at fixed defect number, $\kappa$, we would like to see how the sum over defect number reproduces the classical answer. As we will now show, the sum over defect number has a saddle point in defect number, $\kappa_*$, which depends upon the dilaton coupling $\lambda$ in such a way that the on-shell effective pre-potential, defined in eq. \eqref{eqn:fixedkappaprepot}, is equal to the actual pre-potential of the theory at fixed $\lambda$:
\begin{align}
    W_{\lambda(\kappa_*)}(\Phi) = W_{\lambda}(\Phi), \ \ \lambda(\kappa_*) = \lambda.
\end{align}

To see this, we go back to the full sum over number of defects which defines the full gravitational path integral which is
\begin{align}\label{eqn:fullpathintegral}
=&\sum_L\frac{(\lambda \pi (1-\alpha))^L}{G_N^L L!} \int D\Phi Dg D\rho \,dz \times \nonumber\\
&\exp \left(\frac{1}{2G_N}\int_M \sqrt{g} \left(\Phi R + 2\Phi- 4\pi (1-\alpha)\kappa \rho(x) \Phi(x) - 2\kappa \rho(x) \log \rho(x)+ 2i z \kappa (\rho(x) -1)\right)\right).
\end{align}

If we focus on terms in the sum where $L$ scales with $G_N$ like $L = \kappa/G_N$, then we can replace the sum with an integral over $\kappa$.
At small $G_N$, the large $L$ contribution to the disk partition function then becomes 
\begin{align}\label{eqn:kaction}
    Z_{grav} &\sim \frac{1}{G_N}\int^{??}_{\sim G_N} d\kappa \int D\Phi Dg D\rho \,dz \times \exp \left(  \frac{\kappa}{G_N} \log \left(\frac{\lambda \pi (1-\a)}{\kappa}\right) + \frac{\kappa}{G_N} \right) \times\nonumber\\
    &\exp \left(\frac{1}{2G_N}\int_M \sqrt{g} \left [\Phi R + 2\Phi- 4\pi (1-\alpha)\kappa \rho(x) \Phi(x) - 2\kappa \rho(x) \log \rho(x)+ 2i z \kappa (\rho(x) -1) \right]\right).
\end{align}
where we used Stirling's approximation on the $L!$ in \eqref{eqn:fullpathintegral}. We have left the upper limit on $\kappa$ ambiguous for now but come back to this point momentarily. The lower limit is set by the value at which the density approximation becomes untrustworthy which is $\kappa$ of order $G_N$. 

Differentiating the action with respect to $\kappa$, we see that we should supplement the equations in eq. \eqref{eqn:denseom} with the equation
\begin{align}\label{eqn:kappaderiv}
    \log \left( \frac{\lambda \pi (1-\a)}{\kappa}\right) -\int d^2x \sqrt{g} \left( 2\pi (1-\a) \rho \Phi + \rho \log \rho\right) + iz \left(\int d^2 x \sqrt{g} \rho - 1\right) = 0.
\end{align}
Using the three equations in eq. \eqref{eqn:denseom} and again defining the effective coupling constant $\lambda (\kappa) \equiv \frac{C \kappa}{\pi (1-\a)}$ where $\rho(x) = C e^{-2\pi (1-\a) \Phi(x)}$, we see that eq. \eqref{eqn:kappaderiv} implies
\begin{align}\label{eqn:kappaeom}
    \lambda (\kappa_*) = \lambda,
\end{align}
as we wanted to show. This confirms that we should think of these smooth on-shell geometries as a dense collection of defects with total defect number set by eq. \eqref{eqn:kappaeom}.

We can be a bit more explicit for $\kappa_*$ by using the equation for $\lambda(\kappa)$ in \eqref{eqn:lambdakappa}. Further use of eq. \eqref{eqn:betaphih} allows us to solve for $\kappa_*$ in terms of $\lambda, \beta$ as
\begin{align}\label{eqn:ksaddleexplicit}
    \kappa_* = \frac{-\beta \mathcal{W} \left(-2\pi^2 \lambda (1-\alpha)^2 e^{-4\pi^2 (1-\alpha)/\beta}\right)}{4\pi^2 (1-\alpha)^2},
\end{align}
where $\mathcal{W}$ is the Lambert W-function, defined by the equation $\mathcal{W}(x) e^{\mathcal{W}(x)} = x$.

Interestingly, we see that for high temperatures where $1/\beta \gtrsim \log(1/G_N)$ then $\kappa \lesssim G_N$. In this regime, it seems that we leave the large $L = \kappa/G_N$ approximation. This happens because at such high temperatures the Euclidean disk geometry is so close to the regular JT gravity solution that you need only a few defects to account for the difference. For these temperatures and energies, the difference between the JT metric at the horizon and the exact metric is polynomially suppressed in $G_N$. Note, that by the discussion in Sec. \ref{sec:dilgravrev}, we expect these geometries to have a space-like singularity in the interior. Apparently then just a few exterior defects can modify the black hole interior drastically. 

\begin{figure}
    \centering
    \includegraphics[scale=.7]{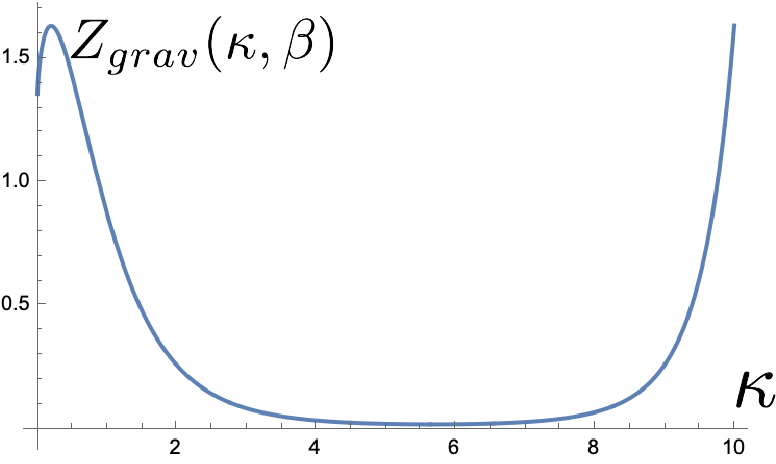}
    \caption{This figure shows a plot of the integrand in eq. \eqref{eqn:kaction} for $\lambda \approx .01, \alpha = 0$ and $\beta \approx 60$. This shows that the integrand has two saddle points at small $G_N$: one at small $\kappa$ which is stable and another, unstable saddle point at larger $\kappa$. The integral of this function is not convergent on the full real line and so must get regulated. We speculate on how this happens in the main text.}
    \label{fig:kaction}
\end{figure}

One final point is that the saddle point equation for $\kappa_*$ in \eqref{eqn:ksaddleexplicit} in fact has one other real solution in $\kappa$, since the Lambert W-function can take multiple real values on its different branches. This can be seen clearly in Fig. \ref{fig:kaction} for particular values of $\lambda, \beta$ and $\alpha$. Geometrically, these solutions with larger $\kappa$ will be the solutions which we discussed in the previous subsection, Sec. \ref{sec:throatgeometries}, which have $\Phi_h$ on the third branch of the inverse of the effective dilaton potential, $W_{\lambda(\kappa)}^{-1}$. One can check numerically however that this second saddle point $\kappa_*^{(2)}$ always appears to occur above the critical number, $\kappa_{\mathrm{pinch}}$, where the throat in the geometry pinches off as discussed in the previous subsection. This suggests to us that this other, sub-leading saddle point in the $\kappa$ integral will not contribute to the actual answer. Furthermore, we can see from Fig. \ref{fig:kaction} that the $\kappa$ integral in Eq. \eqref{eqn:kaction} does not converge unless there is an upper limit on the integral. Again, we suspect that this cutoff is provided by the effects discussed in the previous subsection. This point clearly deserves further study. 

\subsection{The Weil-Peterson volume for a large number of very blunt defects and a single closed geodesic}

Using these ideas, it is tempting to try to compute the Weil-Peterson volumes for a geometry with a large number of blunt defects and a single geodesic boundary.\footnote{The setup gets more complicated for higher numbers of geodesic boundaries but we suspect something akin to these techniques might generalize.} To understand this, we will look for saddle points with fixed defect number $\kappa$ on a geometry which is bounded by a geodesic of length $b$. We will see that this is only possible if $\kappa > \kappa_c$, as demanded by a simple application of the Gauss-Bonnet theorem. 

To find solutions with a closed geodesic, we take inspiration from the discussion in Sec. \ref{sec:diskclosedgeo} and we look for solutions with the AdM energy below the outermost minimum of $W_{\lambda(\kappa)}$. As discussed in Sec. \ref{sec:diskclosedgeo}, the closed geodesic will then sit at constant $\Phi = \Phi_*$, where $W_{\lambda(\kappa)}$ has this outermost minimum at $\Phi_*$. The equations of motion just as before give the metric component in terms of the dilaton
\begin{align}\label{eqn:fixedbmetric}
    A(\Phi) = \Phi^2 - \frac{2 \kappa}{2\pi (1-\alpha)} C e^{-2\pi (1-\alpha) \Phi} - E(b,\kappa)
\end{align}
where the time coordinate has periodicity $\tau \to \tau + \beta$, which will be determined by the equations below. Before doing that, we need to determine $C$ from the normalization condition
\begin{align}
    \beta \int_{\Phi_h}^{\Phi_*} d\Phi \rho(\Phi) = 1 \implies C = \frac{2\pi(1-\alpha)}{\beta (e^{-2\pi (1-\alpha) \Phi_h}-e^{-2\pi (1-\alpha)\Phi_*})},
\end{align}
where $\Phi_h$ is the value of the dilaton at the cap so that \begin{align}\label{eqn:energyconstraint}
    E(b,\kappa) = W_{\lambda(\kappa)}(\Phi_h) = \Phi_h^2 - \frac{2 \kappa}{\beta(1-e^{2\pi (1-\alpha) (\Phi_h - \Phi_*)})}.
\end{align} 
Smoothness at the cap demands
\begin{align}\label{eqn:smoothness}
   \beta \Phi_h +  \frac{2\pi (1-\alpha) \kappa}{1-e^{2\pi (1-\alpha) (\Phi_h - \Phi_*)}} = 2\pi.
\end{align}

We have two further conditions from the fact that $\Phi_*$ occurs at the minimum of the dilaton potential and another coming from demanding that the proper length of this geodesic is $b$. These two equations read
\begin{align}\label{eqn:phisminimum}
    2\Phi_* + \frac{4\pi (1-\alpha) \kappa e^{2\pi (1-\alpha) (\Phi_h-\Phi_*)}}{\beta(1-e^{2\pi (1-\alpha) (\Phi_h-\Phi_*)})} = 0
\end{align}
and
\begin{align}\label{eqn:lengthisb}
    \Phi_*^2 - \frac{2\kappa e^{2\pi (1-\alpha) (\Phi_h-\Phi_*)}}{\beta (1-e^{2\pi (1-\alpha)(\Phi_h - \Phi_*)})} - E(b,\kappa)= \frac{b^2}{\beta^2}.
\end{align}

Note that if we take eq. \eqref{eqn:phisminimum}, mulitply through by $\beta$ and subtract it off from eq. \eqref{eqn:smoothness}, we get the constraint
\begin{align}\label{eqn:GB}
    \beta (\Phi_h - \Phi_*) = 2\pi - 2\pi (1-\alpha) \kappa.
\end{align}
From this we wee that if $\Phi_* > \Phi_h$ then $\kappa > \kappa_c$ as expected. Equation \eqref{eqn:GB} follows simply from Guass-Bonnet applied to these geometries. Note that eq.'s \eqref{eqn:energyconstraint}, \eqref{eqn:smoothness}, \eqref{eqn:smoothness} and \eqref{eqn:phisminimum} comprise four equations for the four unknowns $\Phi_h, \Phi_*, E, \beta$ which in principle allow us to solve for these parameters in terms of the fixed quantities $\kappa$ and $b$. Unfortunately, solving these equations analytically is not possible except in certain limits.

One interesting effect that one can see numerically is that for fixed $\kappa > 1/(1-\alpha)$, as we crank up $b$, there are not solutions with arbitrarily large $b$. Note that this is precisely what happened when we analyzed $Z_{disk}(b)$ in Sec. \ref{sec:lengthambiguity}. The reason is that as we crank up $b$, the value of the dilaton at the horizon increases even though the energy of the solution, $E(b,\kappa)$, is decreasing. This leads to an increase in the effective coupling constant in $\lambda(\kappa)$. At a critical value of $b$, the effective coupling constant $\lambda(\kappa,b)$ hits the critical coupling constant $\lambda_c$ discussed above and the metric in eq. \eqref{eqn:fixedbmetric} becomes monotonic as a function of $\Phi$. This means it no longer contains a real closed geodesic.


To compute the WP volume for a hyperbolic geometry with $L = \kappa/G_N$ number of defects and a single geodesic boundary of length $b$, we then just need to compute the Shannon entropy of this solution since there are no boundary terms present in the problem. The density is 
\begin{align}
    \rho(\Phi) = Ce^{-2\pi (1-\alpha) \Phi}
\end{align}
where $C$ is given above. Introducing the notation
\begin{align}
    r = e^{2\pi (1-\alpha) (\Phi_h - \Phi_*)}
\end{align}
we have
\begin{align}
    C = \frac{2\pi(1-\alpha)}{\beta} \frac{e^{2\pi (1-\alpha) \Phi_h}}{1-r}.
\end{align}
Note that by construction $r<1$ so $C>0$ as it should be. Then the Shannon term becomes
\begin{align}
    &-\int d^2 x \sqrt{g} \rho \log \rho = -\log C+2\pi (1-\alpha) C \beta \int_{\Phi_h}^{\Phi_*} d\Phi e^{-2\pi (1-\alpha) \Phi} \Phi \nonumber \\
    & =-\log C  + \frac{1-r + 2\pi (1-\alpha )\Phi_h - r2\pi (1-\alpha) \Phi_*}{1-r} \nonumber \\
    & = \log \left(\frac{\beta(r) (1-r)}{2\pi (1-\alpha)}\right) + \frac{r}{1-r} \log r + 1.
\end{align}
It would be nice to compare this with the results of \cite{EbeTur23} in certain limits to check agreement. This would require solving their recursion relations for the volumes in the very blunt limit. We save this comparison for future work.

\section{Discussion \& Speculations}\label{sec:discspec}

We have proposed that new instantonic corrections in classical dilaton gravity are necessary to match with the exact answer found in \cite{EbeTur23,TurUsaWen21}. These exact answers were derived by expanding in a gas of exponential operators, which are viewed either as a gas of defects or as tachyon vertex operators in the minimal string picture. This expansion at first glance obscures the appearance of smooth, classical geometry. We have shown how this smooth geometry in fact arises from a dense gas of very blunt defects. We now end with some speculations and comments about what all of this means for the interior black hole geometry.

\subsection{Firewalls from pacifiers?}
We have shown that instantonic corrections due to the inclusion of pacifier spacetimes can lead to a breakdown of geodesic gauge. In \cite{StaYan22,IliLevLinMaxMez24}, such a breakdown of geodesic gauge was observed due to the inclusion of wormhole effects. The authors of \cite{StaYan22} proposed that such effects can be interpreted in terms of a tunneling process by which a black hole can turn into a white hole. It would of course be fascinating if a similar conclusion could be drawn with regards to the pacifier geometry. One difference between these phenomena, however, is that the wormhole effects of \cite{StaYan22} become enhanced at very late times of order $e^{S_0}$. The effects we discuss here are enhanced at low energies instead of long times. This suggests that they may modify the interior description of someone jumping into a very low temperature black hole. In particular, at low temperatures, there appears to be a growing probability that an observer falls into the geometry associated to the pacifier instead of the standard disk. Such a geometry, when continued from its $\mathbf{Z}_2$ symmetric geodesic time slice, has a space-like singularity. Instead of falling smoothly through an inner horizon, the observer hits a singularity that cuts off the geometry. Should the space-like singularity in this case be interpreted as a type of firewall?

\subsection{Defect expansion in Lorentzian signature}
Our understanding of how the classical saddle points can be recast as a dense gas of blunt defects is currently purely Euclidean. It is tempting, however, to consider analogous calculations in Lorentzian signature. For example, a transition amplitude might contain contributions from Lorentzian geometries bounded by two time slices. We then want to compute the contribution from
\be 
\exp\left( \i \int \sqrt{-g}\left[ \Phi R + 2 \Phi + \l e^{-2\pi (1-\a) \Phi}\right] \right)
\ee
to the path integral for the geometry. Following the same route as before, we can try to expand the exponential,
\be 
e^{\i I_{\rm JT}} \left(1 + \i \l \int \sqrt{-g} \l e^{-2\pi (1-\a) \Phi} + \dots \right)
\ee
To order $\l$ we get a Lorentzian action
\be 
\frac{1}{2}\int \sqrt{-g} \left[ \Phi R + 2\Phi + 2\i \a \Phi \delta^{(2)}(x-x_0) \right].
\ee
Integrating now over real dilaton instead of complex, since we always implicitly rotate the dilaton contour in Lorentzian signature (even in JT gravity), we get that this restricts us to surfaces with defects that have \emph{complex} curvature singularities. Proceeding as before and introducing a density for such defects in the small $G_N$ limit, we will find that the normalization condition on the density tells us that in fact the density of defects is itself complex. We are not sure how to interpret this fact except that likely this means the defect picture will be less useful in Lorentzian signature.

\subsection{A simpler theory close to the singularity}

It is interesting to explore the theories we have discussed in this paper by zooming in the large, negative dilaton region. This region lies close to the singularity and is therefore insensitive to the AdS asymptotics. At large negative dilaton, we can write the action 
\be 
I \approx \frac{1}{2} \int \d^2 x \sqrt{-g} \left( \Phi R + \l 2\pi (1-\a) e^{-2\pi (1-\a) \Phi} \right).
\ee
Taking this action seriously in its own right, we note that this theory does not have a good semi-classical limit. The reason is that $\lambda$ can always be absorbed into a constant shift of $\Phi$ which in turn can be absorbed into a re-definition of $S_0$. This means that, ignoring non-perturbative corrections, there is no good small $\lambda$ perturbation expansion. In the cases studied in the bulk of this work, the asymptotic AdS scale provided the other scale relative to which we could define a good weakly-coupled limit. Not surprisingly, the theory near the singularity is inherently strongly coupled.

It is interesting to re-write this theory using a different set of fields. Going to conformal gauge $\d s^2 = e^{\rho} \d x^+ \d x^-$ and defining $\Phi = \frac{1}{2\pi (1-\a)} (\rho + \chi)$, we get
\be 
I = \int \d x^+ \d x^- \left( \frac{2}{\a} \partial_+ \rho \partial_- \rho + \frac{2}{2\pi(1-\a)} \partial_+ \chi \partial_- \rho + \l e^{-\chi} \right).
\ee
Sicne the action for $\rho$ is quadratic we can integrate it out to get an action 
\be 
I = \int \d x^+ \d x^- \left( - \frac{1}{2\pi (1-\a)} \partial_+ \chi \partial_- \chi + \l e^{-\chi} \right)
\ee
This looks like a time-like Liouville action. We can alternatively view this as Liouville but with an inverted exponential potentialy. Note that can write $x^{\pm} = x \pm t$ and then continue to Euclidean by taking $t = -i \tau$. Such a continuation would give us the wrong sign kinetic term.

It is tempting to proceed using a canonical WdW analysis on mini-supserspace where $\chi = \chi_0$ is a constant function. Doing so leads us to solve the equation
\be 
\left[-\frac{1}{2\a} \partial_{\chi_0}^2 - \l e^{-\chi_0} \right]\psi_E(\chi_0) = E \psi_E(\chi_0).
\ee
This is a Schroedinger equation with a potential that is unbounded from below and so does not make much sense a priori. Interestingly, we can define a self-adjoint extension that makes the problem well-defined as was done in \cite{FreSch03}. Solving this problem in this manner amounts to putting a boundary condition at large negative $\chi_0$, which is related to the near singularity region. It would be fascinating to try to understand what this mini-superspace wavefunction corresponds to in the dual matrix model. Presumably this wavefunction corresponds to inserting an operator in the matrix model since we can think of it as a boundary condition in the theory. If so, what aspect of the matrix model corresponds to the boundary conditions needed to make this equation well-defined?

\subsection{Future directions}
In Sec. \ref{sec:bluntexpansion}, we argued that one could compute the Weil-Peterson volumes of a geometry with a single, closed geodesic boundary of length $b$ and many blunt defects using the density approximation. It of course would be interesting to solve the recursion relations of \cite{EbeTur23} in this limit and compare with our results. 

Also, in this work we have not properly treated the one-loop factors. This would require a thorough analysis of fluctuations about our saddle point solutions. It would be interesting to use the re-writing of general dilaton gravity in terms of a Poisson sigma model, a la the recent work of \cite{BloMerYao23}, to understand these fluctuations better. We plan to study this in the future. 

Finally, we end this discussion with a brief return to the case of $W_{\lambda}(\Phi)$ gravity with $\lambda < 0$. In this case, as mentioned in Sec. \ref{sec:cansubtlety}, geodesic gauge is a good gauge in Lorentzian signature in that the classical, Lorentzian solutions to $W_{\lambda}(\Phi)$ gravity at a given boundary time are specified uniquely by the geodesic length. That said, in Sec. \ref{sec:diskclosedgeo} we still found a breakdown of geodesic gauge, although this breakdown was due to the inclusion of complex geometries. As discussed in Sec. \ref{sec:altcont}, these other geometries can be excluded by an alternate choice of the defining contour for the dilaton. For the theory with $\lambda >0$, this alternate contour only made sense in a range of energies. For the theory with $\lambda < 0$, however, there is only a single minimum to the potential $W_{\lambda}(y)$, and so this alternate contour makes sense for all energies, since the inner contour never gets ``pinched'' by another root. This suggests a choice in defining the theory for $\lambda <0$ where one can consistently choose to include (or not) the other complex solutions to $W_{\lambda}(y) = u$. It would be interesting to match the density of states associated to this alternate contour for $\lambda <0$ with the density of states derived through geodesic slicing, as discussed in subsection Sec. \ref{sec:wkbdos}. It is tempting then to interpret this alternate dilaton contour as a modification of the defining inner product of the theory as in \cite{IliLevLinMaxMez24}. It would be interesting to show that if one demands that the geodesic length states of $\lambda <0$ gravity are delta function orthogonal, then we can derive the density of states predicted by the alternate contour of Sec. \ref{sec:altcont}.


\section*{Acknowledgements}
We thank Dionysios Anninos, Daniel Harlow, Luca Iliesiu, David Kolchmeyer, Henry Lin, Juan Maldacena, Edgar Shaghoulian, Arvin Shahbazi-Moghaddam, Jon Sorce, Sandip Trivedi,  Joaquin Turiaci, Mykhaylo Usatyuk and Wayne Weng for discussions. We thank Daniel Harlow, Jon Sorce, Joaquin Turiaci and Mykhaylo Usatyuk for comments on the draft. We also particularly thank Steve Shenker for initial collaboration and many stimulating discussions about the ideas in this work. JK is supported by NSF grant PHY-2207584. AL is supported through the Packard Foundation as well as the Heising-Simons Foundation under grant no. 2023-4430.

\newpage 
\appendix 

\section{Boundary particle fluctuations}\label{app:schwloops}

In this appendix, we analyze the boundary particle fluctuations about a general solution to the dilaton equations of motion. We also apply this formula to a background with a fixed but large number of defects, which induces an effective dilaton pre-potential $W_{\lambda(\kappa)}(\Phi)$. 

We proceed by fixing the background geometry. We then analyze the boundary mode fluctuations about this solution. The background metric at fixed temperature is of the form
\begin{align}
    ds^2 = (W(\phi) - E(\beta)) d\tau^2 + \frac{d\phi^2}{W(\phi)-E(\beta)}
\end{align}
where in order to avoid confusion we have temporarily relabeled the radial coordinate from uppercase to lowercase $\Phi$ to $\phi$, since we have effectively integrated out the dilaton. 

We would like to compute the extrinsic curvature of a cutout of this solution. To do so, it is convenient to inroduce a tortoise-like coordinate for this geometry defined by
\begin{align}
   r(\phi) = -\int_{\phi}^{\infty} \frac{d\phi'}{W(\phi') - E},
\end{align}
where $r$ goes to zero at the boundary and $-\infty$ at the horizon. Inverting this equation gives a function $\phi(r)$. In the tortoise coordinates, the metric is conformal to the flat cylinder,
\begin{align}
    ds^2 = \left(W(\phi(r)) - E\right) \left( d\tau^2 + dr^2\right) \equiv e^{2\omega} \left( d\tau^2 + dr^2 \right).
\end{align}
For a boundary cutout with embedding coordinates $x^{\mu}(u) = (\tau(u), r(u))$ one can easily show that the extrinsic curvature is given by 
\begin{align}
    K = e^{-\omega}K_{flat} + \partial_n \omega,
\end{align}
where $n^{\mu} = \frac{e^{-\omega}}{\sqrt{r'^2 + \tau'^2}}(r'(u), -\tau'(u))$.
Furthermore, the fixed length boundary conditions impose the relation between $\tau'$ and $r'$
\begin{align}\label{eqn:boundarymetric}
    e^{\omega(r)}\sqrt{r'^2 + \tau'^2} = \frac{1}{\varepsilon}.
\end{align}

In terms of the original $r$ coordinate, we have that 
\begin{align}
    &\omega(r) = \frac{1}{2} \log \left( W(\phi(r)) -E\right) \implies \partial_{r} \omega = U(\phi(r))/2.
\end{align}
Using these conditions, a short computation shows that
\begin{align}
    \partial_n \omega = -\varepsilon \tau' U(\phi(r))/2 \approx \varepsilon \tau'(u)/r(u),
\end{align}
and used that $\phi(r) \sim -1/r$ at small $r$. We now use the relation in equation \eqref{eqn:boundarymetric} to write $r'(u)$ in terms of $\tau'(u)$. Using that $W(\phi(r)) \approx 1/r^2$ at small $r$, i.e. the geometry asymptotes to AdS$_2$, we find the relation
\begin{align}
r(u) = \varepsilon \left(\tau'(u) + \varepsilon^2 \left(\frac{\left(\tau''(u)\right)^2}{2\tau'(u)} - \frac{E}{2} (\tau'(u))^3 \right)+ \mathcal{O}(\varepsilon^3)\right).
\end{align}
Using that $K_{flat}$ is given by 
\begin{align}
    K_{flat} = \frac{-r'\tau'' + r'' \tau'}{\left((r')^2 + (\tau')^2\right)^{3/2}}
\end{align}
we find that the extrinsic curvature term in the action is
\begin{align}
   \frac{\Phi_b}{ G_N} \int \sqrt{h} \left(K-1\right) = -\frac{1}{2G_N} \int_0^{\beta} du \left(\left(\frac{\tau''(u)}{\tau'(u)}\right)^2 - E\,\tau'(u)^2\right) + \mathcal{O}(\varepsilon).
\end{align}

Following the usual steps, we can get the fluctuation determinant by expanding around the saddle point. We can expand the reparameterization mode $\tau(u)$ as
\begin{align}
    \tau(u) = u + \varepsilon(u). 
\end{align}
Further expanding $\varepsilon(u)$ in Fourier modes, we have
\begin{align}
    \varepsilon(u) = \sum_{|n|\geq 1} e^{-2\pi i n \frac{u}{\beta}} \left( \varepsilon_n^{(R)} + i \varepsilon_n^{(I)}\right)
\end{align}
with $\varepsilon^{(R,I)}$ the real and imaginary parts respectively. As discussed in \cite{SaaSheSta19}, the real and imaginary parts are constrained $\varepsilon_n^{(R)} = \varepsilon_{-n}^{(R)}$ and $\varepsilon_{n}^{(I)} = -\varepsilon_{-n}^{(I)}$ by reality of $\varepsilon$. Plugging this in, we get that the action is
\begin{align}\label{eqn:flucaction}
    \frac{\Phi_b}{G_N}\int \sqrt{h} \left( K -1\right) = \frac{-(2\pi)^4}{2G_N \beta^3} \sum_{|n|\geq 1} \left(n^4 - \frac{\beta^2 E(\beta)}{(2\pi)^2}n^2\right) \left( \left(\varepsilon_n^{(R)}\right)^2 + \left(\varepsilon_n^{(I)}\right)^2\right) 
\end{align}
Note that the sum over $n$ skips $n = 0$ since this mode holds the geometry fixed, due to its $U(1)$ symmetry. Thus, we do not integrate over this mode instead treating it as gauge. We now make an assumption that the measure for these fluctuations takes the form
\begin{align}
   \prod_n \frac{1}{\beta^2} \left(n^3 - \frac{\beta^2 E(\beta)}{(2\pi)^2} n\right) d\varepsilon_n^{(R)} d\varepsilon_n^{(I)}
\end{align}
up to possible order one, $\beta$-independent factors. Note that this is a mild assumption since to derive this we just need to assume that the boundary mode fluctuation measure does not care about the geometry of the solution deep in the bulk. The measure then results from using the pure AdS$_2$ measure discussed in equation (126) of \cite{SaaSheSta19} for the trumpet of length $b$, but where we have set $b^2 = -\beta^2 E(\beta)$. Integrating the action above over the fluctuations gives us the resulting factor of $1/\beta^{1/2}$ due to the gauged $U(1)$ zero mode.

\subsection{Fixed defect number fluctuations}
It is interesting to apply the above calculation to the geometry with an $L = \kappa/G_N$ number of defects. The same manipulations as above go through with $E(\beta)$ now related to $\beta$ by eq. \eqref{eqn:enatfixedbeta} as
\begin{align}\label{eqn:enatfixedbeta2}
    E(\beta) = \left(\frac{2\pi (1-(1-\alpha)\kappa)}{\beta} \right)^2 - \frac{2\kappa}{\beta}.
\end{align}
Plugging equation \eqref{eqn:enatfixedbeta2} into eq. \eqref{eqn:flucaction}, we see that this is almost the fluctuation action associated to a $\kappa/G_N$ number of defects, although not quite due to the extra term in \eqref{eqn:enatfixedbeta} of the form $-2\kappa/\beta$. We do not fully understand the meaning of this extra term. We suspect we may be missing contributions due to fluctuations in the defect density very near the boundary. We leave an exploration of such contributions for future work. Furthermore there are subtleties in this case regarding whether or not we should integrate over the $n=1$ mode in the fluctuation action or not. We suspect we should not since otherwise we get the incorrect factors of $1/\beta$ to match with the answers of \cite{TurUsaWen21, EbeTur23}.

\section{Geodesics on the trumpet}\label{sec:trumpetgeodesics}

\begin{figure}
    \centering
    \includegraphics[scale=1.2]{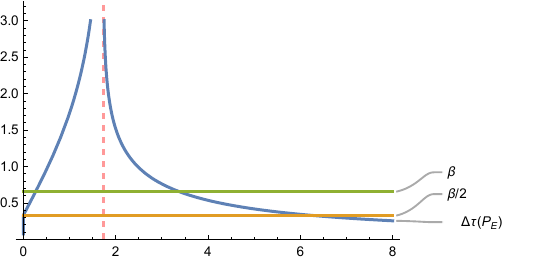}
    \includegraphics[scale=.6]{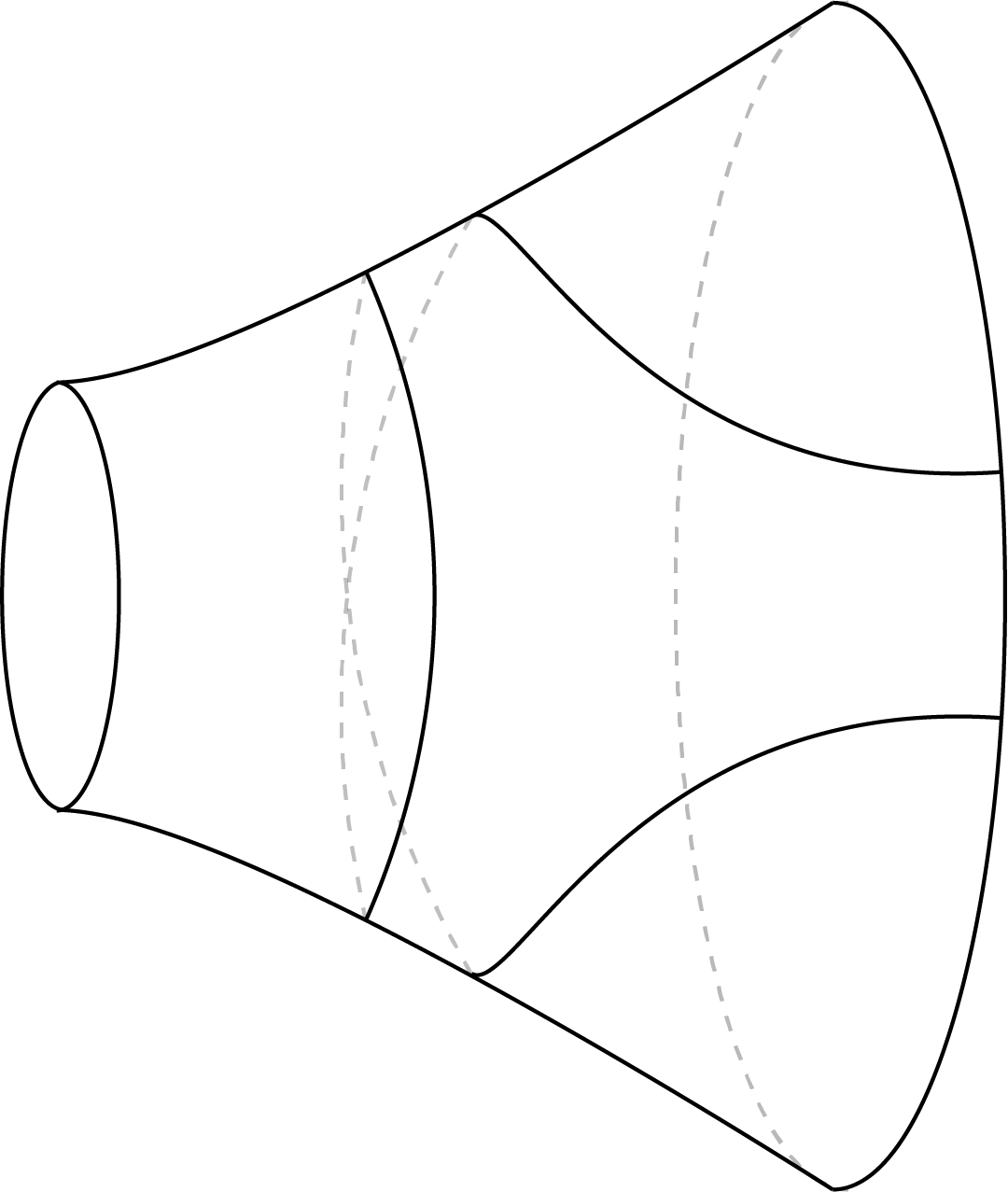}
    \caption{We plot $\Delta \tau(P_E)$ as a function of $P_E$. Note that the expression for $\Delta \tau(P_E)$ is antisymmetric in $P_E$, so it is obvious how to extend the above plot to negative $P_E$. The dashed red vertical line is the critical momentum at which $\Delta \tau$ blows up. On the right, we illustrate an example of a possible geodesic on the trumpet with $\Delta \tau > \beta$.}
    \label{fig:taupe}
\end{figure}

In this Appendix, we analyze boundary anchored geodesics on the trumpet geometries discussed in Sec. \ref{sec:trumpetgeom}. As we will show, these geometries have two boundary anchored geodesics for every pair of boundary points, just as in JT gravity. As discussed in Sec. \ref{sec:trumpetgeom}, in Euclidean signature the geodesic equation is equivalent to 
\begin{align}
\left(\partial_u \Phi(u) \right)^2 = W(\phi(u)) - E - P_E^2,
\end{align}
where $P_E$ is the Euclidean momentum of the geodesic and $u$ is the proper length coordinate along the geodesic. Since the trumpet geometries have $E< E_0$, the geodesics we will be interested in have $P_E^2> E_0 - E$ since the geodesics where $P_E^2 < E_0 -E$ all intersect the geodesic boundary, $b$.

Since there is a closed geodesic sitting at constant $\Phi = \Phi_0$, if we tune the momentum of the boundary anchored geodesic to be close to $P_E^2 \approx E_{0} - E$ then we will have a geodesic which winds many times around the geometry. If we back away from this limit, we can find geodesics which end on the same two boundary points if 
\begin{align}
    \Delta \tau_1 = \beta - \Delta \tau_2
\end{align}
where $\Delta \tau$ is given by
\begin{align}\label{eqn:lengthtime}
\ell = 2 \int_{\Phi_*}^{\infty} \frac{\d \Phi}{\sqrt{W(\Phi) - E-P_E^2}},\quad \tau(\Phi_*) - \tau(\infty) = P_E \int_{\Phi_*}^{\infty} \frac{\d\Phi}{(W(\Phi) - E)\sqrt{W(\Phi) - E - P_E^2}}.
\end{align}
Here $\Phi_*$ is the minimum value of the dilaton that the geodesic bounces off of so that
\begin{align}
    \Delta \tau(\ell) = 2 \left( \tau(\Phi_*) - \tau(\infty)\right).
\end{align}
Since later we will be interested in cutting the geometry along $\ell$, we will only focus on geodesics with $\Delta \tau < \beta$.

For the specific potential $W_{\lambda}(\Phi)$ with $\lambda < \lambda_c$, we plot $\Delta \tau$ as a function of $P_E$ in Fig. \ref{fig:taupe} for a generic choice of $\lambda, \alpha$ and $E$, namely $\alpha = 1/2$, $E = -3$ and $\lambda = .01$. There are a few important properties of this curve to point out.
\begin{itemize}
    \item Note that the expression for $\Delta \tau$ in eq. \eqref{eqn:lengthtime} is manifestly antisymmetric in $P_E$. 
    \item As expected, as we increase $P_E$ from zero, there is a critical $P_E$ where $E + P_E^2 = E_{0}$ at which $\Delta \tau = \infty$.
    \item For $P_E$ where $\Delta \tau(P_E)>\beta$, the geodesic winds multiple times around the geometry and so self-intersects. The geometries that contribute to the inner product $\braket{\ell|\ell'}$ are such that $\ell$, $\ell'$ do not intersect. Thus, the geodesics we are interested in lie below this line in the plot. We illustrate one of these winding geodesics in Fig. \ref{fig:taupe}
    \begin{figure}
    \centering
    \includegraphics[scale=1.2]{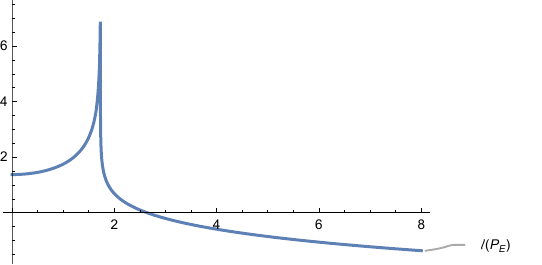}
    \caption{We plot $\ell(P_E)$ as a function of $P_E$.}
    \label{fig:lenpe}
\end{figure}
    \item Two geodesics with momentum $P_1$ and $P_2$ end on the same boundary points if $\Delta \tau_1 - \Delta \tau_2 = \beta$. By anti-reflecting the curve in fig. \ref{fig:taupe} across the vertical axis, we see that for every geodesic with $\Delta \tau_1 < \beta/2$ and $P_E>0$ there exists a geodesic with $P_E<0$ and $\beta >\Delta \tau_2 > \beta/2$ such that they are anchored to the same boundary points. In other words, for every two boundary points, there are two bulk geodesics anchored to those two points and which do not intersect the closed geodesic.
\end{itemize}  
We can also plot the renormalized length of the geodesics as a function of momentum $P_E$ as in Fig. \ref{fig:lenpe}.
Again we see that the length diverges at the critical momentum from winding many times around the geometry.

\bibliographystyle{ourbst}
\bibliography{Refs}

\end{document}